\documentclass[12pt]{article}

\newcommand{\VersionInformation}{}  
\InputIfFileExists{Debug}{}{}

\usepackage{amsmath}
\usepackage{amsthm}
\usepackage{amsfonts}          
\usepackage{amssymb}           
\usepackage[mathscr]{eucal}
\usepackage{graphicx}
\usepackage{color}
\usepackage{hypbmsec}
\usepackage{fancyvrb}
\usepackage{sepnum}

\usepackage{rotating}
\usepackage{slashbox}
\usepackage[isu,bf]{caption}
\newlength{\xtrawidth}
\newlength{\xtraheight}
\usepackage[all]{xy}           

\ifx\NoBackreferences\EMPTYMACRO
  \usepackage[backref,linktocpage,bookmarks]{hyperref}
\else
  \usepackage[linktocpage,bookmarks]{hyperref}
\fi

\ifx\DEBUG\EMPTYMACRO
\else
  \usepackage{showlabels}
\fi

\def\clap#1{\hbox to 0pt{\hss#1\hss}}

\makeatletter
  \def\adots{\mathinner{\mkern2mu\raise\p@\hbox{.}
      \mkern2mu\raise4\p@\hbox{.}\mkern1mu
      \raise7\p@\vbox{\kern7\p@\hbox{.}}\mkern1mu}}
\makeatother

\newcommand{\comma}[1]{\ensuremath{\sepnum{{.}}{{,}}{}{#1}}}

\newcommand{\eqdef}{=}

\newcommand{\R}{\ensuremath{{\mathbb{R}}}}
\newcommand{\C}{\ensuremath{{\mathbb{C}}}}

\newcommand{\Z}{\mathbb{Z}}
\newcommand{\CP}{\ensuremath{\mathop{\null {\mathbb{P}}}}\nolimits}

\newcommand{\ibar}{\ensuremath{{\bar{\text{\it\i\/}}}}}
\newcommand{\jbar}{\ensuremath{{\bar{\text{\it\j\/}}}}}

\newcommand{\betabar}{\ensuremath{{\bar{\beta}}}}
\newcommand{\sbar}{\ensuremath{{\bar{s}}}}
\newcommand{\zbar}{\ensuremath{{\bar{z}}}}

\newcommand{\FS}{\ensuremath{{\text{FS}}}}

\newcommand{\ptset}{\ensuremath{\{\text{pt.}\}}}

\DeclareMathOperator{\diff}{d\!}
\DeclareMathOperator{\Span}{span}

\DeclareMathOperator{\Mat}{Mat}

\DeclareMathOperator{\Vol}{Vol}
\DeclareMathOperator{\dVol}{dVol}
\DeclareMathOperator{\diag}{diag}

\newcommand{\Xt}{{\ensuremath{\widetilde{X}}}}

\newcommand{\ZZZ}{\ensuremath{{\Z_3\times\Z_3}}}
\newcommand{\Lsheaf}{\ensuremath{\mathscr{L}}}
\newcommand{\Osheaf}{\ensuremath{\mathscr{O}}}

\newcommand{\Yt}{{\ensuremath{\widetilde{Y}}}}

\newcommand{\dP}[1]{\ensuremath{dP_{#1}}}

\newcommand{\Kahler}{K\"ahler}

\newcommand{\cyclperm}{\ensuremath{(\text{cyc})}}

\newcommand{\Qt}{{\ensuremath{\widetilde{Q}}}}
\newcommand{\QtF}{{\ensuremath{\widetilde{Q}}_F}}
\newcommand{\QtFsub}[1]{{\ensuremath{\widetilde{Q}}_{F,{#1}}}}

\newcommand{\Pt}{{\ensuremath{\widetilde{P}}}}
\newcommand{\Rt}{{\ensuremath{\widetilde{R}}}}

\newcommand{\CY}{\text{CY}}
\newcommand{\Npoints}{\ensuremath{N_p}}

\newcommand{\CC}{C\nolinebreak\hspace{-.05em}\raisebox{.4ex}{\tiny\bf +}\nolinebreak\hspace{-.10em}\raisebox{.4ex}{\tiny\bf +}}
\newcommand{\blitzpp}{blitz\nolinebreak\hspace{-.05em}\raisebox{.4ex}{\tiny\bf +}\nolinebreak\hspace{-.10em}\raisebox{.4ex}{\tiny\bf +}}

{\end{list}}

{\everymath{\displaystyle\everymath{}}\array}%
{\endarray}

\newtheorem{theorem}{Theorem}

\newtheorem{definition}{Definition}

\begin{document}
\begin{titlepage}
  \vspace*{-2cm}
  \VersionInformation
  \hfill
   \parbox[c]{5cm}{
     \begin{flushright}
       arXiv:0712.3563 [hep-th]
       \\
       UPR 1192-T
     \end{flushright}
   }
  \vspace*{\stretch1}
  \begin{center}
     \Huge 
     Calabi-Yau Metrics for Quotients and Complete Intersections
  \end{center}
  \vspace*{\stretch2}
  \begin{center}
    \begin{minipage}{\textwidth}
      \begin{center}
        \large Volker Braun${}^1$, 
        Tamaz Brelidze${}^1$,
        Michael R.~Douglas${}^2$, and
        \\
        Burt A.~Ovrut${}^1$
      \end{center}
    \end{minipage}
  \end{center}
  \vspace*{1mm}
  \begin{center}
    \begin{minipage}{\textwidth}
      \begin{center}
        ${}^1$ 
        Department of Physics,
        University of Pennsylvania,        
        \\
        209 S. 33rd Street, 
        Philadelphia, PA 19104--6395, USA
      \end{center}
      \begin{center}
        ${}^2$ 
        Rutgers University,
        Department of Physics and Astronomy,
        \\
        136 Frelinghuysen Rd.,
        Piscataway, NJ 08854--8019, USA
      \end{center}
    \end{minipage}
  \end{center}
  \vspace*{\stretch1}
  \begin{abstract}
    \normalsize 
    We extend previous computations of Calabi-Yau metrics on
    projective hypersurfaces to free quotients, complete
    intersections, and free quotients of complete intersections. In
    particular, we construct these metrics on generic quintics,
    four-generation quotients of the quintic, Schoen Calabi-Yau
    complete intersections and the quotient of a Schoen manifold with
    $\Z_3\times\Z_3$ fundamental group that was previously used to
    construct a heterotic standard model. Various numerical
    investigations into the dependence of Donaldson's algorithm on the
    integration scheme, as well as on the K\"ahler and complex
    structure moduli, are also performed.
  \end{abstract}
  \vspace*{\stretch5}
  \begin{minipage}{\textwidth}
    \underline{\hspace{5cm}}
    \centering
    \\
    Email: 
    \texttt{vbraun}, \texttt{brelidze}, \texttt{ovrut@physics.upenn.edu};
    \texttt{mrd@physics.rutgers.edu}.
  \end{minipage}
\end{titlepage}

\tableofcontents

\section{Introduction}
\label{sec:Intro}

A central problem of string theory is to find compactifications which
can reproduce real world physics, in particular the Standard Model.
The first and still one of the best motivated ways to achieve this are
heterotic string compactifications on Calabi-Yau
manifolds~\cite{Candelas:1985en}. In particular, the so-called
``non-standard embedding'' of $E_8\times E_8$ heterotic strings has
been a very fruitful approach towards model building.

For a variety of reasons, the most successful models of this type to
date are based on non-simply connected Calabi-Yau threefolds. These
manifolds admit discrete Wilson lines which, together with a non-flat
vector bundle, play an important role in breaking the heterotic $E_8$
gauge theory down to the Standard Model~\cite{Ovrut:2003zj,
  Buchbinder:2002pr, Donagi:2000zf, Donagi:2000fw, Donagi:2004ia, 
  Donagi:2004qk}.  In the process,
they project out many unwanted matter components. In particular, one
can use this mechanism to solve the doublet-triplet splitting
problem~\cite{Donagi:2004ub, Donagi:2004su}. 
Finally, the non-simply connected threefolds have many fewer
moduli as compared to their simply connected covering spaces~\cite{Braun:2005fk}. 
In recent work~\cite{Braun:2005bw, Braun:2005nv, Braun:2005ux, Bouchard:2005ag}, 
three generation models with a variety of desirable features were
introduced. These are based on a certain quotient of the Schoen
Calabi-Yau threefold, which yields a non-simply connected Calabi-Yau
manifold with fundamental group $\Z_3\times \Z_3$.

Ultimately, it would be desirable to compute all of the observable
quantities of particle physics, in particular gauge and Yukawa
couplings, from the microscopic physics of string
theory~\cite{Braun:2006me, Braun:2005xp}. There are many issues which
must be addressed to do this.  Physical Yukawa couplings, for example,
depend on both coefficients in the superpotential and the explicit
form of the \Kahler{} potential.  In a very limited number of specific
geometries~\cite{Candelas:1987rx, Candelas:1990rm, Greene:1993vm, 
Donagi:2006yf}, the former can be computed using
sophisticated methods of algebraic geometry, topological string theory
and the like. For the latter, we generally have only the qualitative
statement that a coefficient is ``expected to be of order one''.
Doing better and, in particular, extending these calculations to
non-standard embedding, multiply connected compactifications has been
an outstanding problem~\cite{Candelas:1985en}.

In recent work~\cite{DonaldsonNumerical, Douglas:2006hz}, a plan has
been outlined to analyze these problems numerically, at least in the
classical limit. The essential point is that, today, there are good
enough algorithms and fast enough computers to calculate Ricci-flat
metrics and to solve the hermitian Yang-Mills equation for the gauge
connection directly.  Given this data, one can then find the correctly
normalized zero modes of fields, determine the coefficients in the
superpotential and compute the explicit form of the \Kahler{}
potential. Some progress in this direction was made
in~\cite{DonaldsonNumerical, Douglas:2006hz, Douglas:2006rr,
  MR2161248, MR1064867}, and also~\cite{Headrick:2005ch,
  Doran:2007zn, MR2154820}.

In the present work, we take some significant steps forward in
computing Calabi-Yau metrics.  Some of the steps are technical and
computational improvements, which we will discuss.  But the primary
new ingredient is the ability to solve for Ricci-flat metrics on
non-simply connected Calabi-Yau manifolds.  While in broad conceptual
terms the procedure is similar to the simply connected case, in
practice the problem of finding and working with a complete basis of
holomorphic sections of a line bundle, as used in Donaldson's method,
is now quite intricate. To solve this, we systematically use the
techniques of Invariant Theory~\cite{MR1255980}.

We begin, in \autoref{sec:quintic}, by extending Donaldson's algorithm
to the computation of Calabi-Yau metrics for generic quintic
threefolds.  This formalism is first applied to the simple Fermat
quintic, reproducing and extending the results
of~\cite{Douglas:2006rr}. We then numerically calculate the Calabi-Yau
metrics, and test their Ricci-flatness, for a number of random points
in the complex structure moduli space. All of these manifolds are, of
course, simply connected.  We then proceed to non-simply connected
manifolds or, equivalently, to covering spaces that admit fixed point
free group actions. In \autoref{sec:Invariants}, we outline the
general idea and review some of the Invariant Theory, in particular
the Poincar\'e series, Molien formula and the Hironaka decomposition,
that we will use.  This formalism will then be applied in
\autoref{sec:Z5Z5} to the subset of quintics that admit a $\Z_5 \times
\Z_5$ fixed point free group action.  Using the Molien formula and the
Hironaka decomposition, we determine the $\Z_5 \times \Z_5$ invariant
sections on such quintics and, hence, on the $\Z_5 \times \Z_5$
multiply connected quotient space. Using these, we can extend
Donaldson's algorithm and compute the Calabi-Yau metric, and test its
Ricci-flatness, on the quotient. As a by-product of this process, we
note that there are now two ways to compute a Calabi-Yau metric on the
covering space; first, as a point in the quintic moduli space
employing the methods of \autoref{sec:quintic} and second, by using
Donaldson's algorithm for invariant sections only. These two methods
are compared in \autoref{sec:Z5Z5}.  Note that only the second
approach descends to the $\Z_5 \times \Z_5$ quotient threefolds.

In \autoref{sec:Schoen}, we describe Schoen threefolds. We show how to
compute Calabi-Yau metrics on these simply connected complete
intersection manifolds and, as always, test their
Ricci-flatness. Schoen manifolds which admit a fixed point free $\Z_3
\times \Z_3$ group action are then discussed in
\autoref{sec:Z3Z3}. Proceeding as in \autoref{sec:Z5Z5}, the Molien
formula and the Hironaka decomposition are generalized to complete
intersections.  These are then used to find all $\Z_3 \times \Z_3$
invariant sections. These descend to the $\Z_3 \times \Z_3$ quotient,
and are used to compute the Calabi-Yau metric on this multiply
connected threefold.

In addition, we explicitly check algebraic independence of primary
invariants, which are defined in \autoref{sec:Invariants}, for
quintics in \autoref{sec:thetaQuintic}.  Finally some portions of the
code used in this paper are presented in
\autoref{sec:QuinticImplementation}.

\section{The Quintic}
\label{sec:quintic}

\subsection{Parametrizing Metrics}
\label{sec:MetricParam}

Quintics are Calabi-Yau threefolds $\Qt\subset \CP^4$.  As
usual, the five homogeneous coordinates
$\left[z_0:z_1:z_2:z_3:z_4\right]$ on $\CP^4$ are subject to the
identification
\begin{equation}
  \label{eq:P4Rescaling}
  \left[z_0:z_1:z_2:z_3:z_4\right]
  = 
  \left[\lambda z_0:\lambda z_1:\lambda z_2:\lambda z_3:\lambda z_4\right]
  \quad \forall \lambda \in \C-\{0\}
  .
\end{equation}
In general, a hypersurface in $\CP^4$ is Calabi-Yau if and only if 
it is the zero locus of a degree-$5$ homogeneous 
polynomial\footnote{Hence the name quintic.}
 \begin{equation}
 \label{eq:generic_quintic}
  \Qt( z) = 
  \sum_{n_0+n_1+n_2+n_3+n_4=5} 
  c_{(n_0,n_1,n_2,n_3,n_4)} 
  z_0^{n_0}
  z_1^{n_1}
  z_2^{n_2}
  z_3^{n_3}
  z_4^{n_4}
  .
\end{equation}
Note that, abusing notation, we denote both the threefold and its defining 
polynomial by $\Qt$. There are  $\binom{5+4-1}{4}=126$ degree-5
monomials, leading to 126 coefficients $c_{(n_0,n_1,n_2,n_3,n_4)} \in \C$.
These can be reduced by redefining the $z_{i}$-coordinates under 
$GL(5,\C)$. Hence, the number of complex structure moduli of a generic
$\Qt$ is $126-25=101$. A particularly simple point in this moduli
space is the so-called Fermat quintic $\QtF$, defined as the 
zero-locus of 
\begin{equation}
  \label{eq:FermatQuintic}
  \QtF(z)=z_0^5+z_1^5+z_2^5+z_3^5+z_4^5
  .
\end{equation}
We will return to the Fermat quintic later in this section. 

In general, the metric on a real six-dimensional manifold is a
symmetric two-index tensor, having $21$ independent
components. However, on a Calabi-Yau (more generally, a \Kahler{})
manifold the metric has fewer independent components. First, in
complex coordinates, the completely holomorphic and completely
anti-holomorphic components vanish,
\begin{equation}
  \label{eq:KahlerGeometry1}
  g_{ij}(z,\zbar) = 0
  ,\quad
  g_{\ibar \jbar}(z,\zbar) = 0
  .
\end{equation}
Second, the mixed components are the derivatives of a single
function
\begin{equation}
  \label{eq:KahlerGeometry2}
  g_{i\jbar}(z,\zbar) = g_{\ibar j}^*(z,\zbar)
  = 
  \partial_i \bar\partial_\jbar K(z,\zbar)
  .
\end{equation}
The hermitian metric $g_{i\jbar}$ suggests the following definition of
a real $(1,1)$-form, the \Kahler{} form
\begin{equation}
  \label{eq:KahlerForm}
  \omega = 
  \frac{i}{2} 
  g_{i\jbar} \diff z_i \wedge \diff \zbar_\jbar
  =
  \frac{i}{2} \partial \bar\partial K(z,\zbar)
  .
\end{equation}
The \Kahler{} potential $K(z,\zbar)$ is locally a real function, but
not globally; on the overlap of coordinate charts one has to patch it
together by \Kahler{} transformations
\begin{equation}
  K(z,\zbar) 
  \sim
  K(z,\zbar) + f(z) + \bar{f}(\zbar)
  .
\end{equation}
The metric eq.~\eqref{eq:KahlerGeometry2} is then globally defined.

The $5$ homogeneous coordinates on $\CP^4$ clearly come with a natural
$SU(5)$ action, so a naive ansatz for the \Kahler{} potential would be
invariant under this symmetry. However, the obvious $SU(5)$-invariant
$|z_0|^2+\cdots+|z_4|^2$ would not transform correctly under the
rescaling eq.~\eqref{eq:P4Rescaling} with $\lambda =
\lambda(z)$. Therefore, one is led to the unique\footnote{Unique up to
  an overall scale, of course. The scale is fixed by demanding that
  $\omega_\FS$ is an integral class, $\omega\in H^2(\CP^4,\Z)$. To
  verify the integrality, observe that the volume integral over the
  curve $[1:t:0:0:0]$ in $\CP^4$ is
  \begin{equation}
    \int_\C \frac{i}{2} \partial \bar\partial 
    K_\FS\big([1:t:0:0:0]\big) 
    = 
    \int_\C \frac{1}{\pi} \partial_t \bar\partial_{\bar t}
    \ln (1+t\bar t)
    \frac{i}{2}
    \diff t \diff \bar t
    = 
    1
    .
  \end{equation}
}
$SU(5)$ invariant \Kahler{} potential
\begin{equation}
  \label{eq:K_FS}
  K_\FS =\frac{1}{\pi} \ln \sum_{i=0}^4 z_i \zbar_\ibar
  .
\end{equation}
One can slightly generalize this by inserting an arbitrary hermitian
$5\times 5$ matrix $h^{\alpha\betabar}$,
\begin{equation}
  \label{eq:K_FS_h}
  K_\FS= \frac{1}{\pi} \ln \sum_{\alpha,\betabar=0}^4
  h^{\alpha\betabar} z_\alpha \zbar_\betabar
  .
\end{equation}
Any \Kahler{} potential of this form is called a Fubini-Study
\Kahler{} potential (giving rise to a Fubini-Study metric). At this
point the introduction of an arbitrary hermitian $h^{\alpha\betabar}$
does not yield anything really new, as one can always diagonalize it
by coordinate changes. However, strictly speaking, different
$h^{\alpha\betabar}$ are different Fubini-Study metrics.

The above \Kahler{} potential is defined on the whole $\CP^4$ and,
hence, defines a metric on $\CP^4$. But this induces a metric on the
hypersurface $\Qt$, whose \Kahler{} potential is simply the
restriction. Unfortunately, the restriction of the Fubini-Study metric
to the quintic is far from Ricci-flat. Indeed, not a single Ricci-flat
metric on any proper Calabi-Yau threefold is known. One of the reasons
is that proper Calabi-Yau metrics have no continuous isometries, so it
is inherently difficult to write one down analytically. Recently,
Donaldson presented an algorithm for numerically approximating
Calabi-Yau metrics to any desired degree~\cite{DonaldsonNumerical}. 
To do this in the quintic context, take a ``suitable'' 
generalization, that is, one
containing many more free parameters of the Fubini-Study metric
derived from eq.~\eqref{eq:K_FS_h} on $\CP^4$. Then restrict this
ansatz to $\Qt$ and numerically adjust the parameters so as to
approach the Calabi-Yau metric.  An obvious idea to implement this is
to replace the \mbox{degree-$1$} monomials $z_\alpha$ in
eq.~\eqref{eq:K_FS_h} by higher degree-$k$ monomials, thus introducing
many more coefficients in the process. However, note that the degree
$k$ is the \Kahler{} class
\begin{equation}
  k\in H^{1,1}(\CP^4,\Z) \simeq \Z
  .
\end{equation}
The reason for this is clear, for example, if we multiply $K_\FS$ in
eq.~\eqref{eq:K_FS} by $k$. Then
\begin{equation} 
  k \, K_\FS=
  \frac{k}{\pi}  \ln \sum_{i=0}^4z_i\zbar_i =  
   \frac{1}{\pi}\ln \sum_{i_1, \dots, i_k=0}^4
  z_{i_1} \cdots z_{i_k} 
  \zbar_{\ibar_1} \cdots \zbar_{\ibar_k}
  .
\end{equation}
Hence, if we want to keep the overall volume fixed, the correctly
normalized generalization of eq.~\eqref{eq:K_FS_h} is
\begin{equation}
  K(z,\zbar)=  
  \frac{1}{k \pi}
  \ln \sum_{
    \begin{smallmatrix}
      i_1, \dots, i_k=0 \\ 
      \jbar_1, \dots, \jbar_k=0
    \end{smallmatrix}
  }^4
  \,
  h^{(i_1, \dots, i_k), (\jbar_1, \dots, \jbar_k)}
  \underbrace{z_{i_1} \cdots z_{i_k}}_{
    \text{degree $k$}
  }
  \,
  \underbrace{  \zbar_{\jbar_1} \cdots \zbar_{\jbar_k}  }_{
    \text{degree $k$}
  }.
\end{equation}
Note that the monomials $z_{i_1} \cdots z_{i_k}$, where $i_1, \dots,
i_k=0,\dots,4$ and integer $k\geq 0$, are basis vectors for the space
of polynomials $\C[z_0,\dots,z_k]$. For fixed total degree $k$, they
span the subspace $\C[z_0,\dots,z_k]_k$ of dimension
\begin{equation}
  \hat{N}_k =  \binom{5+k-1}{k}
  .
\end{equation}
Some values of $\hat{N}_k$ are given in \autoref{tab:N_k_quintic}. In
particular, the matrix of coefficients $h$ now must be a hermitian
$\hat{N}_k\times \hat{N}_k$ matrix.

However, there is one remaining issue as soon as $k\geq 5$, namely,
that the monomials will not necessarily remain independent when
restricted to $\Qt$. In order to correctly parametrize the degrees of
freedom on $\Qt$, we have to pick a basis for the quotient
\begin{equation}
  \label{eq:QuinticCoordRing}
  \C\left[ z_0, \dots, z_4 \right]_k
  \Big/
  \big\langle \Qt(z) \big\rangle
\end{equation} 
for the degree-$k$ polynomials modulo the hypersurface equation. Let us 
denote this basis by $s_\alpha$,
$\alpha=0,\dots,N_k-1$. It can be shown that for any quintic 
\begin{equation}
  \label{eq:N_k_equation}
  N_k = 
  \begin{cases} 
    \hat{N}_k =  \binom{5+k-1}{k} 
    & 0 \leq k < 5 \\[1ex]
    \hat{N}_k - \hat{N}_{k-5} =
    \binom{5+k-1}{k} - \binom{k-1}{k-5} & k \geq 5 
    .
  \end{cases}
\end{equation}
Some values of the $N_k$ are listed in \autoref{tab:N_k_quintic}.
\begin{table}
  \centering
  \renewcommand{\arraystretch}{1.3}
  \begin{tabular}{c|cccccccc}
    $k$ &
    1 & 2 & 3 & 4 & 5 & 6 & 7 & 8
    \\ \hline \strut
    $\hat{N}_k$ &
    5& 15& 35& 70& 126& 210& 330& 495
    \\
    $N_k$ &
    5& 15& 35& 70& 125& 205& 315& 460
  \end{tabular}
  \caption{The number of homogeneous polynomials $\hat{N}_k$ and 
    the number of remaining polynomials $N_k$ after imposing the 
    hypersurface constraint, see eq.~\eqref{eq:N_k_equation}.}
  \label{tab:N_k_quintic}
\end{table}
For any given quintic polynomial $\Qt(z)$ and degree $k$, computing an
explicit polynomial basis $\{s_\alpha \}$ is straightforward. As an
example, let us consider the the Fermat quintic defined by the
vanishing of $\QtF(z)$, see eq.~\eqref{eq:FermatQuintic}. In this
case, a basis for the quotient eq.~\eqref{eq:QuinticCoordRing} can be
found by eliminating from any polynomial in $\C[ z_0, \dots, z_4 ]_k$
all occurrences of $z_0^5$ using $z_0^5=-(z_1^5+z_2^5+z_3^5+z_4^5)$.
   
Using the basis $s_\alpha$ for the quotient ring, one finally arrives at
the following ansatz
\begin{equation} 
  \label{eq:generalKahlerpotential}
  K_{h,k}= 
  \frac{1}{k \pi} 
  \ln \sum_{\alpha ,\betabar=0}^{N_k-1}
  h^{\alpha \betabar} s_\alpha \sbar_\betabar 
  =
  \frac{1}{k \pi} 
  \ln \|s\|^2_{h,k}
\end{equation}
for the Kahler potential and, hence, the approximating metric. Note
that they are formally defined on $\CP^4$ but restrict directly to
$\Qt$, by construction. Obviously, this is not the only possible ansatz 
for the approximating metric, and the reason for this particular 
choice will only become clear later on. However, let us simply mention 
here that there is a rather simple iteration scheme~\cite{Douglas:2006rr,  
MR2161248, MR1916953} involving $K_{h,k}$ which will converge 
to the Ricci-flat metric in the limit $k\rightarrow \infty$. Note that, in contrast
to the Fubini-Study \Kahler{} potential eq.~\eqref{eq:K_FS_h}, the
matrix $h^{\alpha \betabar}$ in eq.~\eqref{eq:generalKahlerpotential}
cannot be diagonalized by a $GL(5,\C)$ coordinate change on the
ambient $\CP^4$ for $k\geq 2$.

Let us note that there is a geometric interpretation of the
homogeneous polynomials, which will be important later on. Due to the
rescaling ambiguity eq.~\eqref{eq:P4Rescaling}, the homogeneous
polynomials are not functions on $\CP^4$, but need to be interpreted
as sections of a line bundle. The line bundle for degree-$k$
polynomials is denoted $\Osheaf_{\CP^4}(k)$ and, in particular,
the homogeneous coordinates are sections of $\Osheaf_{\CP^4}(1)$. In
general, the following are the same
\begin{itemize}
\item Homogeneous polynomials of degree $k$ in $n$ variables.
\item Sections of the line bundle $\Osheaf_{\CP^{n-1}}(k)$.
\end{itemize}
Moreover, the quotient of the homogeneous polynomials by the quintic,
eq.~\eqref{eq:QuinticCoordRing}, is geometrically the restriction
of the line bundle $\Osheaf_{\CP^4}(k)$ to the quintic
hypersurface. That is, start with the identification above,
\begin{equation}
  H^0\big( \CP^4, \Osheaf_{\CP^4}(k) \big) = \C[z_0,z_1,z_2,z_3,z_4]_k
  .
\end{equation}
After restricting the sections of $\Osheaf_{\CP^4}(k)$ to $\Qt$, they
satisfy the relation $\Qt(z)=0$. Hence, the
restriction is 
\begin{equation}
  \begin{split}
    H^0\big( \Qt, \Osheaf_\Qt(k) \big) 
    =&~ 
    \C[z_0,z_1,z_2,z_3,z_4]_k
    \Big/ 
    \left\langle \Qt(z) \right\rangle_k
    \\
    =&~ 
    \C[z_0,z_1,z_2,z_3,z_4]_k
    \Big/ 
    \Big(
    \Qt\, 
    \C[z_0,z_1,z_2,z_3,z_4]_{k-5}
    \Big)
    .
  \end{split}
\end{equation}
More technically, this whole discussion can be represented by the 
short exact sequence
\begin{equation}
  \vcenter{\xymatrix{
      0 
      \ar[r] 
      &
      H^0\big( \CP^4, \Osheaf_{\CP^4}(k-5) \big)
      \ar[r]^{\times \Qt(z)} 
      \ar@{=}[d]
      &
      H^0\big( \CP^4, \Osheaf_{\CP^4}(k) \big)
      \ar[r]^{\text{restrict}} 
      \ar@{=}[d]
      &
      H^0\big( \Qt, \Osheaf_\Qt(k) \big)
      \ar[r] 
      \ar@{=}[d]
      &
      0
      \\
      0 
      \ar[r]
      &
      \C\left[ z_0, \dots, z_4 \right]_{k-5}
      \ar[r]^{\times \Qt(z)}
      &
      \C\left[ z_0, \dots, z_4 \right]_{k}
      \ar[r]
      &
      \C[ z_0, \dots, z_4 ]_{k}
      \big/
      \left\langle \Qt(z) \right\rangle_k
      \ar[r]
      &
      0 
    }}
\end{equation}

\subsection{Donaldson's Algorithm}
\label{sec:Donaldson}

Once we have specified the form for the \Kahler{} potential, our
problem reduces to finding the ``right'' matrix
$h^{\alpha\betabar}$. This leads us to the notion of T-map and
balanced metrics, which we now introduce. First, note that
eq.~\eqref{eq:generalKahlerpotential} provides a way to define an
inner product of two sections. While it makes sense to evaluate a
function at a point, one cannot ``evaluate'' a section (a homogeneous
polynomial) at a point since the result would only be valid up to an overall
scale\footnote{In other words, at any given point one can only decide
  whether the section is zero or not zero.}. However, after picking 
$\|s\|_{h,k}^2$, see eq.~\eqref{eq:generalKahlerpotential}, one can
cancel the scaling ambiguity and define
\begin{equation} 
  \label{eq:LbMetric}
  (S,S')(p)
  = 
  \frac{  S(p)\, \bar{S}'(p)  }{ \|s\|_{h,k}^2 (p) }
  = 
  \frac{
    S(p) \, \bar{S}'(p)
  }{
    \sum_{\alpha,\betabar}  h^{\alpha\betabar} \, s_\alpha(p) \, \sbar_\betabar(p)
  } 
  \quad 
  \forall p\in \Qt
\end{equation}
for arbitrary sections (degree-$k$ homogeneous polynomials) $S$, $S'
\in H^0\big( \Qt, \Osheaf_\Qt(k) \big)$. Note that the $s_0$, $\dots$,
$s_{N_k-1}$ are a basis for the space of sections, so there are always
constants $c^\alpha\in \C$ such that
\begin{equation}
  S = \sum_{\alpha=0}^{N_k-1} c^\alpha s_\alpha
  .
\end{equation}
The point-wise hermitian form $(~,~)$ is called a metric on the line
bundle $\Osheaf_\Qt(k)$. Given this metric, we now integrate
eq.~\eqref{eq:LbMetric} over the manifold $\Qt$ to define a
$\C$-valued inner product of sections
\begin{equation}
  \begin{split}
    \big\langle 
    S, S'
    \big\rangle
    =&~ 
    \frac{N_k}{\Vol_\CY(\Qt)}
    \int_\Qt (S,S')(p) \dVol_\CY 
    \\ 
    =&~
    \frac{N_k}{\Vol_\CY(\Qt)}
    \int_\Qt     
    \frac{
      S \, \bar{S}'
    }{
      \sum_{\alpha,\betabar}  h^{\alpha\betabar} s_\alpha \sbar_\betabar
    }
    \dVol_\CY 
    .
  \end{split}
\end{equation}
Since $\langle~,~\rangle$ is again sesquilinear, it is uniquely
determined by its value on the basis sections $s_\alpha$, that is, by
the hermitian matrix
\begin{equation}
  H_{\alpha\betabar} = \big\langle s_\alpha, s_\beta \big\rangle
  .
\end{equation}
In general, the matrices $h^{\alpha\betabar}$ and $H_{\alpha\betabar}$
are completely different. However, for special metrics, they might
coincide:
\begin{definition}
  Suppose that
  \begin{equation}
    \label{eq:balanced}
    h^{\alpha\betabar} = \big( H_{\alpha\betabar} \big)^{-1}
    .
  \end{equation}
Then the metric $h$ on the line bundle $\Osheaf_\Qt(k)$ is called
balanced.
\end{definition}
We note that, in the balanced case, one can find a new basis of
sections $\{ \tilde{s}_\alpha \}_{\alpha=0}^{N_k-1}$ which
simultaneously diagonalizes
$\tilde{H}_{\alpha\betabar}=\delta_{\alpha\betabar}$ and
$\tilde{h}^{\alpha\betabar}=\delta^{\alpha\betabar}$. The interesting
thing about balanced metrics is that they have special curvature
properties, in particular
\begin{theorem}[Donaldson~\cite{MR1916953}]
  For each $k\geq 1$ the balanced metric $h$ exists and is unique. As
  $k\rightarrow \infty$, the sequence of metrics
  \begin{equation}
  \label{eq:bmetric} 
    g_{i\jbar}^{(k)}
    =
    \frac{1}{k \pi} \partial_i \bar\partial_\jbar 
    \ln \sum_{\alpha ,\betabar=0}^{N_k-1}
    h^{\alpha \betabar} s_\alpha \sbar_\betabar    
  \end{equation}
  on $\Qt$  converges to the unique Calabi-Yau metric for the given \Kahler{}
  class and complex structure.
\end{theorem}
Hence, the problem of finding the Calabi-Yau metric boils down to
finding the balanced metric for each $k$. Unfortunately, since
$H_{\alpha\betabar}$ depends non-linearly on $h^{\alpha\betabar}$ one
can not simply solve eq.~\eqref{eq:balanced} defining the balanced
condition.  However, iterating eq.~\eqref{eq:balanced} turns out
to converge quickly. That is, let
\begin{equation}
  \label{eq:T-operator}
  T(h)_{\alpha\betabar} 
  \eqdef
  H_{\alpha\betabar} 
  =
  \frac{N_k}{\Vol_\CY\big( \Qt \big)}
  \int_\Qt
  \frac{s_\alpha\sbar_{\bar\beta}}
  {\sum_{\gamma\bar\delta} h^{\gamma\bar\delta} s_\gamma \sbar_{\bar\delta}}
  \dVol_\CY
\end{equation}
be Donaldson's T-operator. Then
\begin{theorem}[Donaldson,~\cite{DonaldsonNumerical}]
  For any initial metric $h_0$, the sequence\footnote{At this point, it
    is crucial to work with a \emph{basis} of sections $s_0$, $\dots$,
    $s_{N_k-1}$. For if there were a linear relation between them
    then the matrix $T(h)$ would be singular.}
  \begin{equation}
    h_{n+1} = \big( T(h_n) \big)^{-1}
  \end{equation}
  converges to the balanced metric as $n\rightarrow \infty$ .
\end{theorem}
In practice, only very few ($\leq 10$) iterations are necessary to get
very close to the fixed point. Henceforth, we will also refer to
$g_{i\jbar}^{(k)}$ in eq.~\eqref{eq:bmetric}, the approximating metric
for fixed $k$, as a balanced metric.

\subsection{Integrating over the Calabi-Yau threefold}
\label{sec:integrating}

We still need to be able to integrate over the manifold in order to
evaluate the T-operator. Luckily, we know the exact Calabi-Yau volume
form,
\begin{equation}
  \dVol_\CY = \Omega \wedge \bar\Omega,
\end{equation}
since we can express the holomorphic volume form $\Omega$ as a
Griffiths residue. To do this, first note that the hypersurface
$\Qt\subset \CP^4$ has complex codimension one, so we can encircle any
point in the transverse direction. The corresponding residue integral
\begin{equation}
  \label{eq:OmegaDef}
  \Omega = \oint \frac{ \diff^4 z}{ \Qt(z) \strut}
\end{equation}
is a nowhere vanishing holomorphic $(3,0)$-form and, hence, must be the
holomorphic volume form $\Omega$. As an example, consider the Fermat
quintic defined by eq.~\eqref{eq:FermatQuintic}. In a patch where we can
use the homogeneous rescaling to set $z_0=1$ and where $z_2$, $z_3$,
and $z_4$ are good local coordinates,
\begin{equation}
  \Omega = \int 
  \frac
  { \diff z_1 \wedge \cdots \wedge \diff z_4 }
  {1 + z_1^5 + z_2^5 + z_3^5 + z_4^5 }
  =
  \frac{ \diff z_2 \wedge \diff z_3 \wedge \diff z_4 }{ 5 z_1^4 }  
  .
\end{equation}

To apply, for example, Simpson's rule to numerically integrate over
the Calabi-Yau threefold one would need local coordinate
charts. However, there is one integration scheme that avoids having to
go into these details: approximate the integral by $\Npoints$ random
points $\{p_i\}$,
\begin{equation}
  \frac{1}{\Npoints}
  \sum_{i=1}^{\Npoints} f(p_i) 
  \longrightarrow
  \int f \dVol
  .
\end{equation}
Of course, we have to define which ``random'' distribution the points
lie on, which in turn determines the integration measure $\dVol$. In
practice, we will only be able to generate points with the
\emph{wrong} random distribution, leading to some auxiliary
distribution $\diff A$. However, one can trivially account for this
by weighting the points with $w_i = \Omega\wedge\bar\Omega /  \diff A$,
\begin{equation}
  \frac{1}{\Npoints}
  \sum_{i=1}^{\Npoints} f(p_i)  w_i
  =
  \frac{1}{\Npoints}
  \sum_{i=1}^{\Npoints} f(p_i)  
  \frac{\Omega\wedge\bar\Omega}{\diff A}
  \longrightarrow
  \int f \;
  \frac{\Omega\wedge\bar\Omega}{\diff A}
  \diff A
  =
  \int f \; \dVol_\CY
  .
\end{equation}
Note that taking $f=1$ implies 
\begin{equation}
\frac{1}{\Npoints}
  \sum_{i=1}^{\Npoints} w_i
  =\Vol_\CY(\Qt).
\end{equation}

\subsubsection*{Points from Patches}

We start out with what is probably the most straightforward way to
pick random points. This method only works on the Fermat quintic, to which 
we now restrict. Let us split
$\CP^4$ into $5\cdot 4 = 20$ closed sets
\begin{multline}
  U_{\ell m} = 
  \Big\{
  \big[z_0:z_1:z_2:z_3:z_4\big] \Big|
  \\
  ~
  |z_\ell| = \max(|z_0|,\dots, |z_4|)
  ,~
  |z_m| = \max(|z_0|,\dots, \widehat{|z_\ell|},\dots, |z_4|)
  \Big\}
  .
\end{multline}
In other words, $z_\ell$ has the largest absolute value and $z_m$ has
the second-largest absolute value. They intersect in real
codimension-1 boundaries where the absolute values are the same and
induce the decomposition
\begin{equation}
  \QtF = \bigcup_{\ell,m} \QtFsub{\ell m}
\end{equation}
with $\QtFsub{\ell m} = \QtF \cap U_{\ell m}$. Since permuting coordinates
is a symmetry of the Fermat quintic, it suffices to consider
$\QtFsub{01}$. We define ``random'' points by
\begin{itemize}
\item Pick $x$, $y$, $z\in \C_{\leq 1}$ on the complex unit disk with
  the standard ``flat'' distribution.
\item Test whether
  \begin{equation}
    |x|, |y|, |z| \leq  \big| 1+x^5+y^5+z^5 \big|^{\frac{1}{5}} \leq 1
    .
  \end{equation}
  If this is not satisfied, start over and pick new $x$, $y$, and
  $z$. Eventually, the above inequality will be satisfied.
\item The ``random'' point is now
  \begin{equation}
    \left[ 
      1: - \big( 1+x^5+y^5+z^5 \big)^{\frac{1}{5}} 
      : x : y : z
    \right]
    \in \QtFsub{01}
    ,
  \end{equation}
  where one chooses a uniformly random phase for the fifth root of unity.
\end{itemize}
By construction, the auxiliary measure is then independent of the
position $(x,y,z)\in \QtFsub{\ell m}$. Hence,
\begin{equation}
  \diff A = 
  \frac{1}{20}
  \mathrm{d}^2\! x \wedge
  \mathrm{d}^2\! y \wedge
  \mathrm{d}^2\! z 
  .
\end{equation}

\subsubsection*{Points From Intersecting Lines With The Quintic}

The previous definition only works on the Fermat quintic, but not on
arbitrary quintics. A much better algorithm~\cite{Douglas:2006rr} is 
to pick random lines
\begin{equation}
  L \simeq \CP^1 \subset \CP^4
  .
\end{equation}
Any line $L$ determines $5$ points by the intersection $L\cap\Qt=\{5
\text{pt.}\}$ whose coordinates can be found by solving a quintic
polynomial (in one variable) numerically. Explicitly, a line can be
defined by two distinct points 
\begin{equation}
  p=[p_0:p_1:p_2:p_3:p_4]
  ,\;
  q=[q_0:q_1:q_2:q_3:q_4]
  ~\in \CP^4
\end{equation}
as
\begin{equation}
  L: \C\cup\{\infty\} \to \CP^4
  ,~
  t \mapsto 
  [ p_0+t q_0: p_1+t q_1: p_2+t q_2: p_3+t q_3: p_4+t q_4]
  .
\end{equation}
The $5$ intersection points $L\cap\Qt$ are then given by the $5$
solutions of
\begin{equation}
  \Qt \circ L (t) = 
  \Qt\big(
  p_0+t q_0, p_1+t q_1, p_2+t q_2, p_3+t q_3, p_4+t q_4
  \big)  = 0  
  .
\end{equation}

Clearly, the auxiliary measure will depend on how we pick ``random''
lines. The easiest way is to choose lines uniformly distributed with
respect to the $SU(5)$ action on $\CP^4$. Note that a line $L$ is
Poincar\'e dual to a $(3,3)$-current, that is, a $(3,3)$-form whose
coefficients are delta-functions supported on the line $L$. For the
expected distribution of lines, we then average over all ``random''
configurations of lines. Because of this averaging procedure, the
Poincar\'e dual\footnote{By the usual abuse of notation, we will not
  distinguish Poincar\'e dual quantities in the following.}  of the
expected distribution of lines $\langle L \rangle$ is a smooth $(3,3)$
form. Since there is (up to scale) only one $SU(5)$-invariant $(3,3)$
form on $\CP^4$, the expected distribution of lines must be
\begin{equation}
  \langle L \rangle \sim \omega_\FS^3 
  ,
\end{equation}
where $\omega_\FS$ is the \Kahler{} form defined by the unique
$SU(5)$-invariant Fubini-Study \Kahler{} potential
eq.~\eqref{eq:K_FS}. Restricting both sides to an embedded quintic
$i:\Qt\hookrightarrow \CP^4$, we obtain the auxiliary measure as the
expected distribution of the intersection points,
\begin{equation}
  \diff A 
  =
  \big\langle \Qt \cap L \big\rangle
  \sim 
  i^* \big( \omega_\FS^3 \big)
  .
\end{equation}
As a final remark, note that the the symmetry of the ambient space is,
in general, not enough to unambiguously determine the auxiliary
measure. It is, as we just saw, sufficient for any quintic
hypersurface $\Qt$. However, for more complicated threefolds one needs
a more general theory. We will have to come back to this point in
\autoref{sec:SchoenPoints}.

\subsection{Results}
\label{sec:QuinticPlot}

\begin{figure}[htbp]
  \centering
  \include{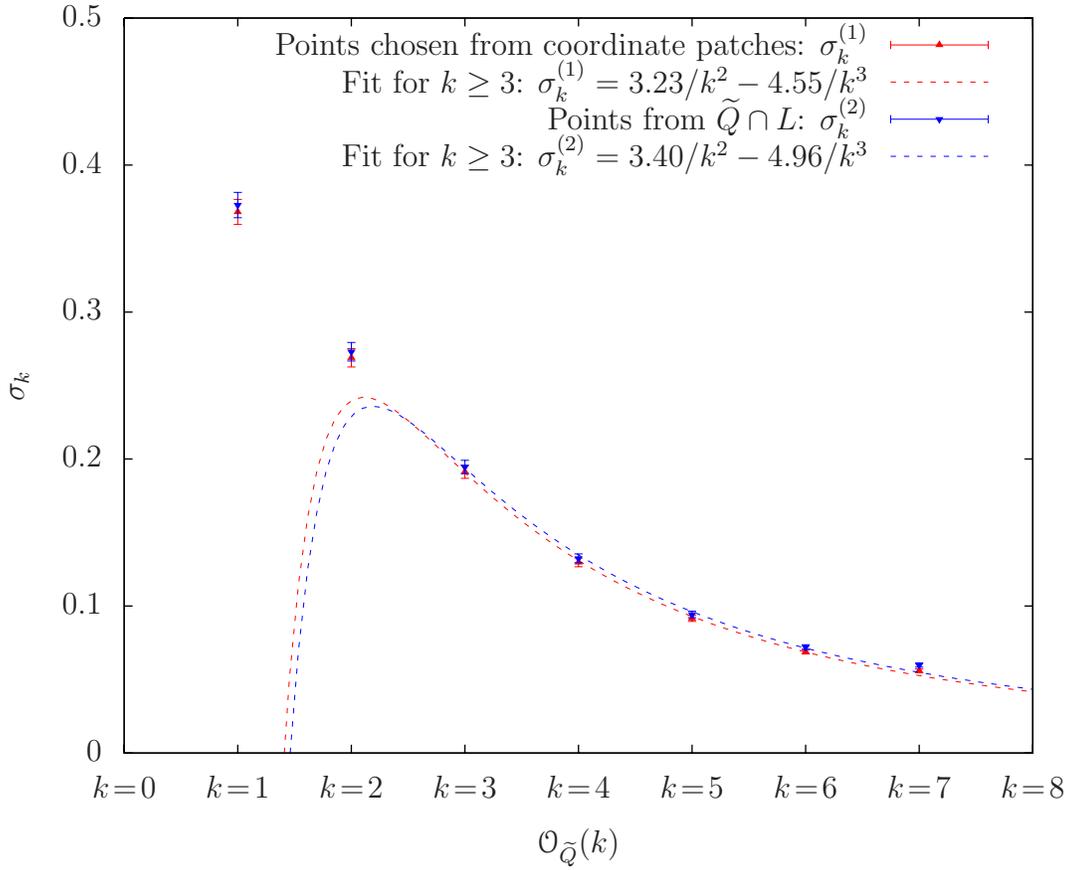}
  \caption{The error measure $\sigma_k$ for the metric on the Fermat
    quintic, computed with the two different point generation
    algorithms described in \autoref{sec:integrating}. In each case we
    iterated the T-operator $10$ times, numerically integrating over
    $\Npoints=\comma{200000}$ points. Then we evaluated $\sigma_k$
    using \comma{10000} different test points. The error bars are the
    numerical errors in the $\sigma_k$ integral.}
  \label{fig:FermatQuinticPoints}
\end{figure}
Following the algorithm laid out in this section, we can now compute
the successive approximations to the Calabi-Yau metric on $\Qt$. In
order to test the result, we need some kind of measure for how close
the approximate metric is to the Calabi-Yau metric. Douglas et
al.~\cite{Douglas:2006rr} proposed the following: First, remember that
the \Kahler{} form $\omega$ eq.~\eqref{eq:KahlerForm} is the
Calabi-Yau \Kahler{} form if and only if its associated volume form
$\omega^3$ is proportional to the Calabi-Yau volume form
eq.~\eqref{eq:OmegaDef}. That is,
\begin{equation}
  \label{eq:w3OObarProp}
  \omega^3 (p)= 
  (\text{const.})\times \Big(\Omega(p) \wedge \bar{\Omega} (p)\Big)
  \quad \forall p \in \Qt
\end{equation}
(the Monge-Amp\'ere equation) with a non-vanishing proportionality constant
independent of $p\in \Qt$\footnote{And varying over the moduli space.
However, this will not concern us here.}. Let us define 
\begin{subequations}
  \begin{equation}
    \label{eq:VolK}
    \Vol_\text{K}\big(\Qt\big) 
    = 
    \int_\Qt \omega^3
  \end{equation}
  and recall that
  \begin{equation}
    \label{eq:VolCY}
    \Vol_\CY\big(\Qt\big)
    = 
    \int_\Qt \Omega\wedge\bar\Omega
    .
  \end{equation}
\end{subequations}
The ratio of these two constants determines the proportionality
factor in eq.~\eqref{eq:w3OObarProp}. This equation can now
be rewritten 
\begin{equation}
  \label{eq:w3OObar}
  \frac{
    \omega^3 (p)
  }{
    \Vol_\text{K}\big(\Qt\big)     
  }
  = 
  \frac{
    \Omega(p) \wedge \bar{\Omega} (p)
  }{
    \Vol_\CY\big(\Qt\big)     
  }
  \qquad \forall p \in \Qt
  .
\end{equation}
Note that one often demands that the two constants,
eqns.~\eqref{eq:VolK} and~\eqref{eq:VolCY}, are unity by rescaling
$\omega$ and $\Omega$ respectively. However, this would be cumbersome
later on and we will not impose this normalization. Then the integral
\begin{equation}
  \label{eq:sigmaQtDef}
  \sigma\big( \Qt \big) = 
  \frac{1}{\Vol_\CY\big(\Qt\big)}
  \int_\Qt \left|
    1 - 
    \frac{
      \omega^3 \Big/ \Vol_\text{K}\big(\Qt\big)
    }{
      \Omega \wedge \bar\Omega \Big/ \Vol_\CY\big(\Qt\big)
    }
  \right| \dVol_\CY
\end{equation}
vanishes if and only if $\omega$ is the Calabi-Yau \Kahler{} form. In
practice, Donaldson's algorithm determines successive approximations
to the Calabi-Yau metric. Since we know the exact Calabi-Yau volume
form $\Omega\wedge\bar\Omega$, only $\omega$ is approximate and
depends on the degree $k$. We define $\sigma_k$ to be the above
integral evaluated with this degree-$k$ approximation to the
Calabi-Yau \Kahler{} form.

Let us quickly summarize the steps necessary to compute the
metric. To do that, one has to
\begin{enumerate}
\item Choose a degree $k$ at which to compute the balanced metric
  which will approximate the Calabi-Yau metric. 
\item Choose the number \Npoints{} of points, and generate this many
  points $\{p_i\}_{i=1}^{\Npoints}$ on $\Qt$. Although $k$ and
  \Npoints{} can be chosen independently, we will argue below that
  $N_p$ should be sufficiently larger than $N_k^2$ for accuracy.
\item For each point $p_i$, compute its weight $w_i=\diff
  A(p_i)/(\Omega\wedge\bar\Omega)$. 
\item Calculate a basis $\{ s_\alpha \}_{\alpha=0}^{N_k-1}$ for the
  quotient eq.~\eqref{eq:QuinticCoordRing} at degree $k$.
\item At each point $p_i$, calculate the (complex)
  numbers $\{ s_\alpha(p_i) \}_{\alpha=0}^{N_k-1}$ and, hence, the
  integrand of the T-operator.
\item \label{l:goto}Choose an initial invertible, hermitian matrix for 	
  $h^{\gamma\bar{\delta}}$. Now perform the numerical integration
  \begin{equation}
    T(h)_{\alpha\betabar} = 
    \frac{N_k}{\sum_{j=1}^{\Npoints} w_j}
    \sum_{i=1}^{\Npoints}
    \frac{ 
      s_\alpha(p_i) \, \overline{ s_\beta(p_i) } w_i
    }{ 
      \sum_{\gamma\bar\delta} h^{\gamma\bar\delta}  \,
      s_\gamma(p_i) \, \overline{ s_\delta(p_i) } 
    }.
  \end{equation}
\item Set the new $h^{\alpha\betabar}$ to be $h^{\alpha\betabar} =
  \big(T_{\alpha\betabar}\big)^{-1}$.
\item Return to \autoref{l:goto} and repeat until $h^{\alpha\betabar}$
  converges close to its fixed point. In practice, this procedure is insensitive 
  to the initial choice of $h^{\alpha\betabar}$ and fewer than $10$
  iterations suffice.
\end{enumerate}
Having determined the balanced $h^{\alpha\betabar}$, we can evaluate
the metric $g_{i\jbar}^{(k)}$ using eq.~\eqref{eq:bmetric} and, hence,
the \Kahler{} form $\omega(p)$ at each point $p$, see
eq.~\eqref{eq:KahlerForm}. Now form $\omega^3(p)$. This lets us
compute $\sigma_k$ by the following steps:
\begin{enumerate}
\item The $\sigma_k$ integral requires much less accuracy, so one may
  pick a smaller number $\Npoints$ of points $\{p_i\}_{i=1}^{\Npoints}$.
\item Compute 
  \begin{equation}
    \Vol_\CY = \frac{1}{\Npoints}\sum_{i=1}^{\Npoints} w_i
    ,
    \qquad 
    \Vol_\text{K} = 
    \frac{1}{\Npoints}
    \sum_{i=1}^{\Npoints} 
    \frac{ 
      \omega^3(p_i) 
    }{ 
      \Omega(p_i) \wedge \overline{\Omega(p_i)}  
    }
    w_i
    ,
  \end{equation}
  which numerically approximate $\int_\Qt \Omega\wedge\bar\Omega$ and
  $\int_\Qt \omega^3$, respectively.
\item The numerical integral approximating $\sigma_k$ is
  \begin{equation}
    \sigma_k = 
    \frac{1}{\Npoints \Vol_\CY}
    \sum_{i=1}^{\Npoints} 
    \left| 
      1 - 
      \frac{ 
        \omega(p_i)^3 
        \big/ \Vol_\text{K} 
      }{
        \Omega(p_i) \wedge \overline{\Omega(p_i)} 
        \big/ \Vol_\CY
      }
    \right|
    w_i
    .
  \end{equation}
\end{enumerate}

As a first application, we apply this procedure to compute the
Calabi-Yau metric for the simple Fermat quintic $\QtF$ defined by
eq.~\eqref{eq:FermatQuintic}. In this case, there are two point selection
algorithms, both given in \autoref{sec:integrating}. We do the
calculation for each and show the results in
\autoref{fig:FermatQuinticPoints}. One can immediately see that both
point selection strategies give the same result, as they should. In
fact, there is a theoretical prediction for how fast $\sigma_k$
converges to $0$, see~\cite{Douglas:2006rr, MR1064867,
  MR1916953}. Expanding in $\tfrac{1}{k}$, the error goes to zero at
least as fast as
\begin{equation}
  \label{eq:SigmaEstim}
  \sigma_k = \frac{S_2}{k^2} + \frac{S_3}{k^3} + \cdots
  , \qquad
  S_i \in \R
  .
\end{equation}
In particular, the coefficient of $\frac{1}{k}$ is proportional to the
scalar curvature and vanishes on a Calabi-Yau manifold. In
\autoref{fig:FermatQuinticPoints}, we fit $\sigma_k = \frac{S_2}{k^2}
+ \frac{S_3}{k^3}$ for $k\geq 3$ and find good agreement with the data
points. 

An important question is how many points are necessary to approximate
the Calabi-Yau threefold in the numerical integration for any given
$k$. The problem is that we really are trying to compute the
$N_k\times N_k$-matrix $h^{\alpha\betabar}$, whose dimension increases
quickly with $k$, see \autoref{tab:N_k_quintic}. Hence, to have more
equations than indeterminates, we expect to need
\begin{equation}
  \Npoints > N_k^2
\end{equation}
points to evaluate the integrand of the T-operator on.
\begin{figure}[htbp]
  \centering
  \include{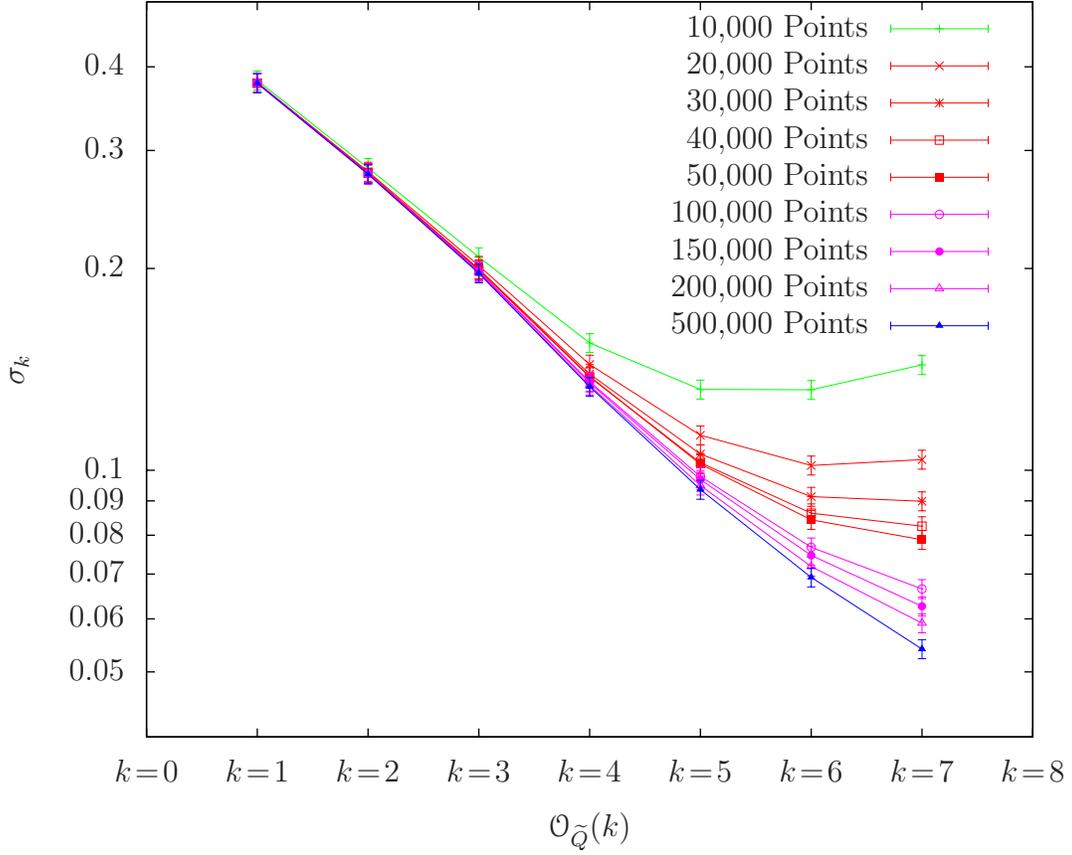}
  \caption{The error measure $\sigma_k$ for the balanced metric on the
    Fermat quintic as a function of $k$, computed by numerical
    integration with different numbers of points $\Npoints$. In each
    case, we iterated the T-operator $10$ times and evaluated
    $\sigma_k$ on \comma{5000} different test points. Note that we use
    a logarithmic scale for the $\sigma_k$ axis.}
  \label{fig:NumberOfPoints1}
\end{figure}
To numerically test this, we compute $\sigma_k$ using
different numbers  of points $\Npoints$. The result is displayed in
\autoref{fig:NumberOfPoints1}, where we used the more convenient
logarithmic scale for $\sigma_k$. Clearly, the error measure $\sigma_k$
starts out decreasing with $k$. However, at some $\Npoints$-dependent 
point it
\begin{figure}[htbp]
  \centering
  \include{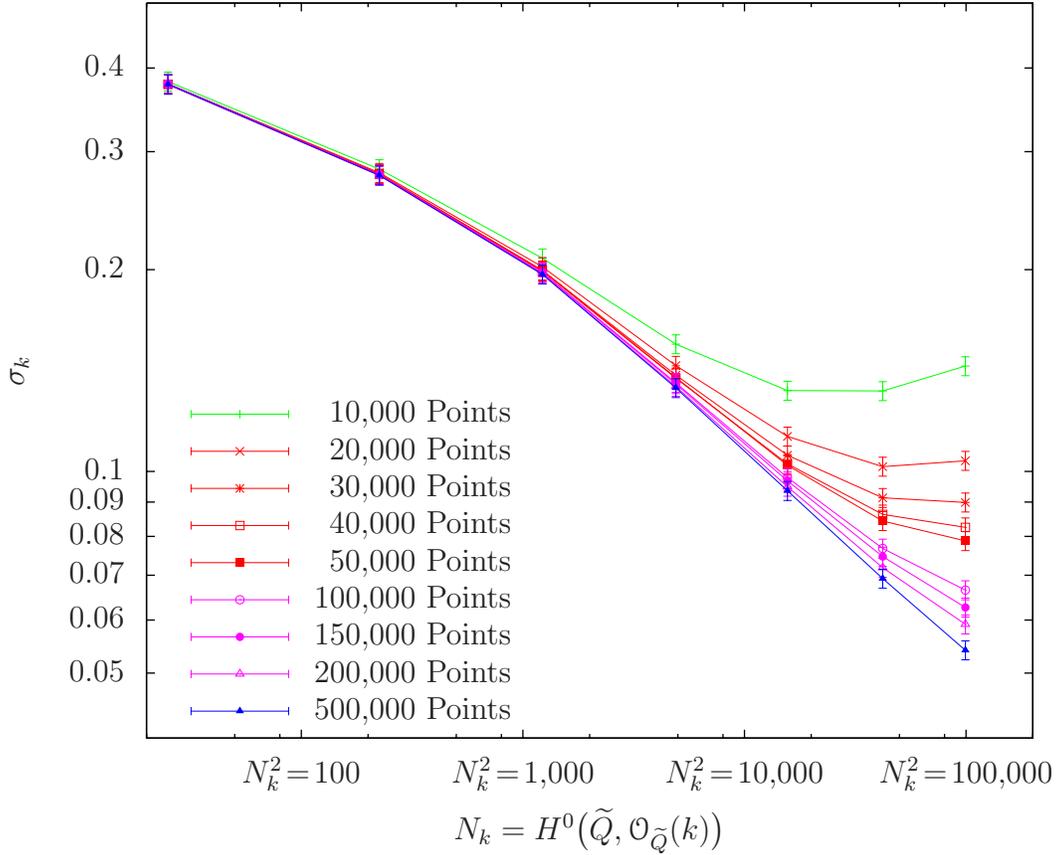}
  \caption{The error measure $\sigma_k$ for the balanced metric on the 
    Fermat quintic as a function of $N_k^2 =$ number of entries in
    $h^{\alpha\betabar}\in \Mat_{N_k\times N_k}$. In other words,
    evaluating the T-operator requires $N_k^2$ scalar integrals.  In
    each case, we iterated the T-operator $10$ times and finally
    evaluated $\sigma_k$ using \comma{5000} different test points.
    We use a logarithmic scale for both axes.}
  \label{fig:NumberOfPoints2}
\end{figure}
reaches a minimum and then starts to increase. In
\autoref{fig:NumberOfPoints2}, we plot the same $\sigma_k$ as a
function of $N_k^2$. This confirms our guess that we need $\Npoints >
N_k^2$ points in order to accurately perform the numerical
integration. One notes that the data points in
\autoref{fig:NumberOfPoints1} seem to approach a straight line as we
increase \Npoints. This would suggest an exponential fall-off
\begin{equation}
  \sigma_k \approx 0.523 e^{ -0.324  k}
  .
\end{equation}
It is possible, therefore, that the theoretical error estimate
eq.~\eqref{eq:SigmaEstim} could be improved upon.

\begin{figure}
  \centering
  \include{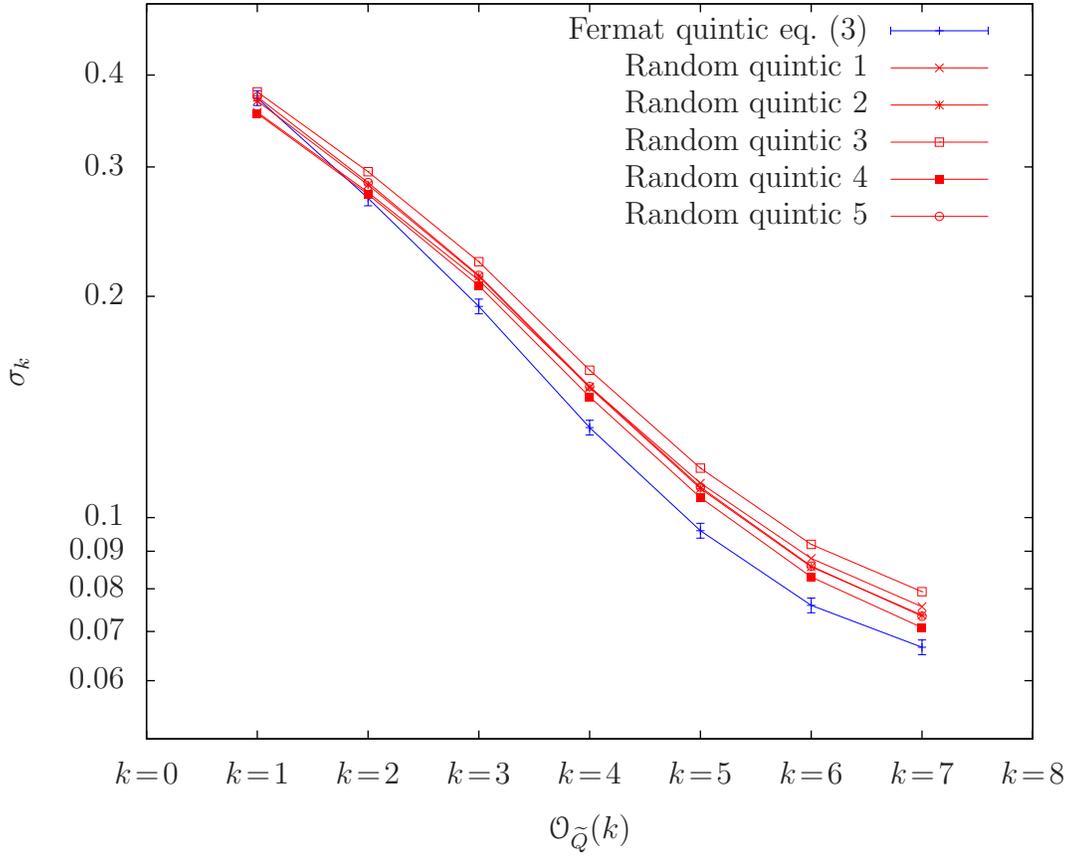}
  \caption{The error measure $\sigma_k$ as a function of $k$ for five 
    random quintics, as well as for the Fermat quintic. The random
    quintics are the sum over the $126$ quintic monomials in $5$
    homogeneous variables with coefficients random on the unit disk.
    We use a logarithmic scale for $\sigma_k$.}
  \label{fig:MultipleQuintics}
\end{figure}
So far, we have applied our procedure to the Fermat quintic $\QtF$ for
simplicity. However, our formalism applies equally well to any quintic
$\Qt$ in the $101$-dimensional complex structure moduli space with the
proviso that, for a non-Fermat quintic, one must use the $L\cap\Qt$
method of choosing points. An important property of the programs that
implement our procedure is that they make no assumptions about the
form of the quintic polynomial eq.~\eqref{eq:generic_quintic}.  We
proceed as follows. First, fix a quintic by randomly (in the usual
flat distribution) choosing each coefficient
$c_{(n_0,n_1,n_2,n_3,n_4)}$ on the unit disk, see
eq.~\eqref{eq:generic_quintic}. Then approximate the Calabi-Yau metric
via Donaldson's algorithm and compute the error measure $\sigma_k$. In
\autoref{fig:MultipleQuintics}, we present the results for $\sigma_k$
for five randomly chosen quintics, and compare them to the Fermat
quintic. We observe that the convergence to the Calabi-Yau metric does
not strongly depend on the complex structure parameters.

\section{Group Actions and Invariants}
\label{sec:Invariants}

\subsection{Quotients and Covering Spaces}
\label{sec:quotients}

Thus far, we have restricted our formalism to quintic Calabi-Yau
threefolds $\Qt\subset \CP^4$. These are, by construction,
 simply connected. However, for applications in heterotic string theory we 
 are particularly interested in non-simply connected Calabi-Yau manifolds 
 where one can reduce the number of quark/lepton generations
 and turn on discrete Wilson
lines~\cite{Witten:1985xc, Sen:1985eb, Evans:1986nq, Breit:1985ud, 
Breit:1985ns, Braun:2004xv, Ovrut:2002jk}. Therefore,
it is of obvious interest to compute the metrics in such
cases. However, these manifolds are more complicated than hypersurfaces in
projective spaces. In fact, any complete intersection in a smooth
toric variety will be simply connected\footnote{Note, however, that
there are $16$ cases of smooth, non-simply connected hypersurfaces
in singular toric varieties~\cite{Batyrev:2005jc:bat}.}. Therefore,
we are usually forced to study non-simply connected Calabi-Yau
threefolds $Y$,
\begin{equation}
  \pi_1(Y)
  =\Pi\not=1  
  ,
\end{equation}
via their universal covering space $\Yt$ and the free group action
$\Pi:\Yt\to\Yt$.

In order to carry through Donaldson's algorithm on $Y$, we now need to
generalize the notion of ``homogeneous polynomials'' to arbitrary
varieties. As mentioned previously, the homogeneous coordinates
on the quintic $\Qt\in\CP^4$ can be interpreted as the 
basis of sections of the line bundle
$\Osheaf_{\Qt}(1)$,
\begin{equation}
  \Span\{ z_0, z_1, z_2, z_3, z_4 \} 
  = 
  H^0\big( \Qt, \Osheaf_\Qt(1) \big)
  .
\end{equation}
The special property of $\Osheaf_{\Qt}(1)$ is that it is ``very
ample'', that is, its sections define an embedding
\begin{equation}
  \Phi_{\Osheaf_{\Qt}(1)}: 
  \Qt \to \CP^4, 
  ~
  x \mapsto \big[ z_0(x): z_1(x): z_2(x): z_3(x): z_4(x) \big]
  .
\end{equation}
Hence, we need to pick a ``very ample'' line bundle on $\Yt$ in order
to compute the metric there. Furthermore, to discuss $Y$, we will also 
need to ``mod out'' by the group action. It follows that the group $\Pi$ 
must act properly on the line bundle. In mathematical terms this is 
called an ``equivariant line bundle'', and there is a one-to-one 
correspondence
\begin{equation}
  \vcenter{\xymatrix@C=3cm{
      \framebox{\parbox{4cm}
        {$\strut$\centering 
          $\Pi$-equivariant\\line bundles on $\Yt$}}
      \ar@<1ex>[r]^{ \Pi}
      & 
      \ar@<1ex>[l]^{\Pi^*}
      \framebox{\parbox{3cm}
        {$\strut$\centering 
          Line bundles\\on $Y$.}}
    }}
\end{equation}
Let us denote such a line bundle on $\Yt$ by $\Lsheaf$.  
We are specifically interested in the sections of this line bundle,
since they generalize the homogeneous coordinates. The important
observation here is that the sections of a $\Pi$-equivariant line
bundle on $\Yt$ themselves form a representation of $\Pi$.
Furthermore, the $\Pi$-invariant sections correspond to the sections
on the quotient. That is,
\begin{equation}
  H^0\big( \Yt, \Lsheaf \big)^\Pi 
  = 
  H^0\big( Y, \Lsheaf/\Pi \big)
  .
\end{equation}
Hence, in order to compute the metric on the quotient $Y=\Yt/\Pi$, we
can work on the covering space $\Yt$ if we simply replace all sections
by the $\Pi$-invariant sections.

In this paper, we will always consider the case where $\Yt$ is a
hypersurface or a complete intersection in (products of) projective
spaces. Then
\begin{itemize}
\item The sections on the ambient projective space are homogeneous
  polynomials.
\item The sections on $\Yt$ are the quotient of these polynomials by 
  the defining equations. 
\item The invariant sections on $\Yt$ are the invariant homogeneous
  polynomials modulo the invariant polynomials generated by the
  defining equations. 
\end{itemize}
The mathematical framework for counting and finding these
invariants is provided by Invariant Theory~\cite{MR1255980}, which we
review in the remainder of this section.

\subsection{Poincar\'e and Molien}
\label{sec:Molien}

Let $\C[\vec{x}]$ be a polynomial ring in $n$ commuting variables 
\begin{equation}
  \vec{x}=\big(  x_1, \dots, x_n \big)
  .
\end{equation}
As a vector space over the ground field $\C$, it is generated by all
monomials
\begin{equation} 
  \label{eq:PolynomialRing}
  \C[\vec{x}]
  =
  \C 1 \oplus 
  \C x_1 \oplus \cdots \oplus
  \C x_n \oplus 
  \C x_1^2 \oplus \cdots
  .
\end{equation}
Clearly, $\C[\vec{x}]$ is an infinite dimensional vector space. However, at each
degree $k$ we have a finite dimensional vector space of homogeneous
degree-$k$ polynomials. A concrete basis for the degree-$k$
polynomials would be all distinct monomials of that degree. 

By definition, the Poincar\'e series is the generating function for
the dimensions of the vector subspaces of fixed degree, that is,
\begin{equation} 
  P \big( \C[\vec{x}], t \big)
  = 
  \sum_{k=0}^\infty 
  \Big( \dim_{\C} \C[\vec{x}]_k \Big)
  \,
  t^k
\end{equation}
where $\C[\vec{x}]_k$ is the vector subspace of $\C[\vec{x}]$ of
degree $k$. The monomials of the polynomial ring in $n$ commuting
variables $x_1$, $\dots$, $x_n $ can be counted just like $n$ species
of bosons, and one obtains
\begin{equation}
\label{eq:P4Poincare} 
  P \big( \C[\vec{x}], t \big)
  = 
  \prod_n \frac{1}{1-t}
  =
  \sum_{k=0}^\infty
  \left(\begin{array}{c}n+k-1 \\ k\end{array}\right)
  t^k.
\end{equation} 
We have already mentioned that the homogeneous degree-$k$ polynomials in
$n$ variables are just the sections of $\Osheaf_{\CP^{n-1}}(k)$. Hence, the
number of degree-$k$ polynomials is the same as the dimension of the
space of sections of the line bundle $\Osheaf_{\CP^{n-1}}(k)$,
\begin{equation}
  \dim_{\C} \C[\vec{x}]_k
  =
  h^0\big( \CP^{n-1}, \Osheaf_{\CP^{n-1}}(k) \big)
  . 
\end{equation}
Acknowledging this geometric interpretation, we also write
\begin{equation} 
  P \big( \Osheaf_{\CP^{n-1}}, t \big)
  \eqdef 
  \sum_{k=0}^\infty 
  h^0 \big( \CP^{n-1}, \Osheaf_{\CP^{n-1}}(k) \big)
  \,
  t^k
  = 
  P \big( \C[\vec{x}], t \big)  
  .
\end{equation}
Furthermore, note that
\begin{equation}
 \label{eq:molienadd}
  P(M \oplus M' ,t)=P(M,t)+P(M',t)
\end{equation}
for any rings $M$ and $M'$.

A $n$-dimensional representation of a finite group $G$ generates a
group action on the polynomials eq.~\eqref{eq:PolynomialRing}. One is
often interested in the invariant polynomials under this group action,
which again form a ring $\C\left[ \vec{x}\right]^G$. Clearly, the
invariant ring is a subring of $\C\left[ \vec{x}\right]$. Since the
group action preserves the degree of a polynomial, one can again
define the Poincar\'e series of the invariant ring,
\begin{equation} 
 \label{eq:poincareexp} 
  P \big( \C[\vec{x}]^G, t \big)= \sum_{k=0}^\infty 
  \Big( \dim_{\C} \C[\vec{x}]_k ^G\Big)
  \,
  t^k
  .
\end{equation}
The coefficients in eq.~\eqref{eq:poincareexp} can be obtained using
\begin{theorem}[Molien]
  Let $G\subset GL(n,\C)$ be a finite matrix group acting linearly on
  the $n$ variables $\vec{x}=(x_1,\dots,x_n)$. Then the Poincar\'e
  series of the ring of invariant polynomials, that is, the generating
  function for the number of invariant polynomials of each degree, is
  given by
  \begin{equation}
    \label{eq:molien}
    P \big( \C[\vec{x}]^G ,t \big)
    =
    \frac{1}{|G|} \sum_{g\in G}\frac{1}{\det(1-gt)}
    .
  \end{equation}
\end{theorem}
Equation~\eqref{eq:molien} is called the Molien formula.

\subsection{Hironaka Decomposition}
\label{sec:Hironaka}

Although eq.~\eqref{eq:molien} contains important information about
$\C[x_1,\dots,x_n]^G$, the most detailed description is provided by
the Hironaka decomposition, which we discus next. To construct
 this, one first needs to find $n$ homogeneous polynomials
$\theta_1,\dots,\theta_n$, invariant under the group action, such that
the quotient
\begin{equation}
  \C[x_1,\dots,x_n]\big/
  \left< \theta_1,\dots,\theta_n \right> 
\end{equation}
is zero-dimensional. The above condition is
equivalent~\cite{Sem:MR1690810} to demanding that the system
$\theta_i=0$, $i=1,\dots n$ has only the trivial solution. This
guaranties that the $\theta_i$ are algebraically independent. Then
\begin{theorem}[Hironaka decomposition]
  With respect to $\theta_1$, $\dots$, $\theta_n$ chosen as above, the 
  ring of $G$-invariant polynomials can be decomposed as
  \begin{equation} 
    \C[\vec{x}]^G
    =
    \eta_1  \C[\theta_1, \dots, \theta_n] 
    \oplus 
    \eta_2  \C[\theta_1, \dots, \theta_n] 
    \oplus \cdots \oplus 
    \eta_{s} \C[\theta_1, \dots, \theta_n]
    .
  \end{equation}
\end{theorem}
Clearly, the $\eta_i$ are themselves $G$-invariant polynomials in
$\C[\vec{x}]$. Thus any $G$-invariant polynomial is a unique
linear combination of $\eta_i$'s, where the coefficients are
polynomials in $\theta_i$. The polynomials $\theta_i$ are called the
``primary'' invariants and $\eta_{j}$ the ``secondary''
invariants. Note that, while the number of primary invariants is fixed
by the number of variables $x_1$, $\dots$, $x_n$, the number $s$ of
secondary polynomials depends on our choice of primary invariants. Using
eq.~\eqref{eq:P4Poincare} with each $x_i$ replaced by $\theta_i$, we 
find that the Poincar\'e series for $\C[\theta_1, \dots, \theta_n]$ is given by
\begin{equation} 
  P \big( \C[\theta_1, \dots, \theta_n], t \big)
  = \frac{1}
  {(1-t^{\deg(\theta_1)})\dots(1-t^{\deg(\theta_n)})}
  .
\end{equation}
Moreover, multiplication by $\eta_i$ shifts all degrees by
$\deg(\eta_i)$. Therefore, applying eq.~\eqref{eq:molienadd} we
obtain the Poincar\'e series for the Hironaka decomposition,
\begin{equation}
  \label{eq:moldec}
  \begin{split}
    P \big( \C[\vec{x}]^G, t \big)
    =&~
    \frac{t^{D_1}}{(1-t^{d_1})\cdots (1-t^{d_n})} 
    + \cdots 
    + \frac{t^{D_{s}}}{(1-t^{d_1})\cdots (1-t^{d_n})} 
    \\     =&~
    \frac{t^{D_1} + \cdots + t^{D_{s}}}{(1-t^{d_1})\cdots (1-t^{d_n})} 
    ,
  \end{split}
\end{equation}
where $D_{j}=\deg(\eta_j)$ and $d_i=\deg(\theta_i)$. Each term in the
numerator of eq.~\eqref{eq:moldec} corresponds to a secondary
invariant.

\section{Four-Generation Quotient of Quintics}
\label{sec:Z5Z5}

\subsection{Four Generation Models}
\label{sec:fourgen}

Were one to compactify the heterotic string on a generic quintic $\Qt$
using the standard embedding, then the four-dimensional effective
theory would contain $\frac{1}{2}\chi(\Qt)=100$ net generations. A
well known way to reduce this number~\cite{Candelas:1985en} is to compactify on
quintics that admit a fixed point free $\Z_5 \times \Z_5$ action. In
that case, the quotient manifold $Q=\Qt\big/ (\Z_5\times\Z_5)$ has
only $\frac{1}{2}\chi(Q)=\frac{100}{| \Z_5\times \Z_5
  |}=4$ generations. In this section, these special quintics and their
$\mathbb{Z}_5\times \mathbb{Z}_5$ quotient will be described. We then
compute the Calabi-Yau metrics directly on these quotients $Q$ using a
generalization of our previous formalism.
 
Recall from \autoref{sec:quintic} that a generic quintic $\Qt\subset \CP^4$
is defined as the zero locus of a degree-$5$ polynomial of the
form eq.~\eqref{eq:generic_quintic}. In general, it is the sum
of $126$ degree-$5$ monomials, leading to $126$
coefficients $ c_{(n_0,n_1,n_2,n_3,n_4)}\in\C $. However, not
all of these quintic threefolds admit a fixed point free
$ \mathbb{Z}_5\times \mathbb{Z}_5$ action. To be explicit,
we will consider the following two actions on the five
homogeneous variables defining $\CP^4$,
\begin{equation}
  \label{eq:Z5Z5action}
  \begin{split}
    g_1
    \begin{pmatrix}
      z_0 \\ z_1 \\ z_2 \\ z_3 \\ z_4
    \end{pmatrix}
    =&\;
    \begin{pmatrix}
      0&0&0&0&1\\
      1&0&0&0&0\\
      0&1&0&0&0\\
      0&0&1&0&0\\
      0&0&0&1&0
    \end{pmatrix}
    \begin{pmatrix}
      z_0 \\ z_1 \\ z_2 \\ z_3 \\ z_4
    \end{pmatrix}
    \\
    g_2
    \begin{pmatrix}
      z_0 \\ z_1 \\ z_2 \\ z_3 \\ z_4
    \end{pmatrix}
    =&\;
    \begin{pmatrix}
      1&0&0&0&0\\
      0&e^{\frac{2\pi i}{5}}&0&0&0\\
      0&0&e^{2\frac{2\pi i}{5}}&0&0\\
      0&0&0&e^{3\frac{2\pi i}{5}}&0\\
      0&0&0&0&e^{4\frac{2\pi i}{5}}\\
    \end{pmatrix}
    \begin{pmatrix}
      z_0 \\ z_1 \\ z_2 \\ z_3 \\ z_4
    \end{pmatrix}
    .
  \end{split}
\end{equation}
Clearly $g_1^5=1=g_2^5$, but they do not quite commute: 
\begin{equation}
  \label{eq:Z5Z5commutant}
  g_1 g_2 = e^{\frac{2\pi i}{5}} g_2 g_1
  \quad \Leftrightarrow \quad
  g_1 g_2 g_1^{-1} g_2^{-1} = e^{\frac{2\pi i}{5}}
  .
\end{equation}
However, even though $g_1$ and $g_2$ do not form a matrix
representation of $\Z_5\times\Z_5$, they do generate a
$\Z_5\times\Z_5$ action on $\CP^4$ because on the level of homogeneous
coordinates we have to identify
\begin{equation}
  \begin{split}
    [z_0:z_1:z_2:z_3:z_4] \
    =&~
    [
    e^{\frac{2\pi i}{3}} z_0:
    e^{\frac{2\pi i}{3}} z_1:
    e^{\frac{2\pi i}{3}} z_2:
    e^{\frac{2\pi i}{3}} z_3:
    e^{\frac{2\pi i}{3}} z_4
    ] 
    \\
    =&~
    g_1 g_2 g_1^{-1} g_2^{-1}
    \Big( [z_0:z_1:z_2:z_3:z_4] \Big).
  \end{split}
\end{equation}
If the quintic polynomial $\Qt(z)$ is $\Z_5\times\Z_5$-invariant, then
the corresponding hypersurface will inherit this group action. One can
easily verify that the dimension of the space of invariant homogeneous
degree-$5$ polynomials is $6$, as we will prove in
eq.~\eqref{eq:Gquintics} below. Taking into account that one can
always multiply the defining equation by a constant, there are $5$
independent parameters $\phi_1$, $\dots$, $\phi_5\in \C$. Thus the
$\Z_5\times\Z_5$ symmetric quintics form a five parameter family
which, at a generic point in the moduli space, can be written as
\begin{equation}
  \label{eq:QuinticZ5Z5}
  \begin{split}
    \Qt(z) = &~
    \big( z_0^5+z_1^5+z_2^5+z_3^5+z_4^5 \big)    
    \\+&~ \phi_1
    \big( z_0 z_1 z_2 z_3 z_4 \big)
    \\+&~ \phi_2
    \big( 
    z_0^3 z_1 z_4+z_0 z_1^3 z_2+z_0 z_3 z_4^3+z_1 z_2^3 z_3+z_2
    z_3^3 z_4 
    \big)
    \\+&~ \phi_3
    \big( 
    z_0^2 z_1 z_2^2 + 
    z_1^2 z_2 z_3^2 + 
    z_2^2 z_3 z_4^2 + 
    z_3^2 z_4 z_0^2 + 
    z_4^2 z_0 z_1^2
    \big)
    \\+&~ \phi_4
    \big( 
    z_0^2 z_1^2 z_3 + 
    z_1^2 z_2^2 z_4 + 
    z_2^2 z_3^2 z_0 + 
    z_3^2 z_4^2 z_1 + 
    z_4^2 z_0^2 z_2
    \big)
    \\+&~ \phi_5
    \big( 
    z_0^3 z_2 z_3 + 
    z_1^3 z_3 z_4 + 
    z_2^3 z_4 z_0 + 
    z_3^3 z_0 z_1 + 
    z_4^3 z_1 z_2
    \big)  
    .
  \end{split}
\end{equation}
The explicit form of these invariant polynomials is derived in \autoref{sec:Z5Z5invariants}
and given in eq.~\eqref{eq:Gquintics}. Note that, even though the
$\Z_5\times\Z_5$ action on $\CP^4$ necessarily has fixed points, one
can check that a generic (that is, for generic $\phi_1$, $\dots$,
$\phi_5$) quintic threefold $\Qt$ is fixed-point free. 

Now choose any quintic defined by eq.~\eqref{eq:QuinticZ5Z5}. Since the
 $\Z_5 \times \Z_5$ action on it is fixed point free, the quotient
\begin{equation}
  Q 
  \eqdef 
  \Qt \Big/ \big(\Z_5\times\Z_5\big)
\end{equation}
is again a smooth Calabi-Yau threefold. Its Hodge diamond
is given by~\cite{Green:1987mn}
\begin{equation}
  \label{eq:QuinticQuotientHodge}
  h^{p,q}\big( Q \big)
  =
  h^{p,q}\Big( \Qt / \big(\Z_5\times\Z_5\big) \Big)
  =
  \vcenter{\xymatrix@!0@=7mm@ur{
    1 &  0 &  0 & 1 \\
    0 &  5 &  1 & 0 \\
    0 &  1 &  5 & 0 \\
    1 &  0 &  0 & 1 
  }}
  ,
\end{equation}
where we again see that there is a $h^{2,1}(Q)=5$-dimensional complex
structure moduli space parametrized by the coefficients $\phi_1$,
$\dots$, $\phi_5$.

\subsection{Sections on the Quotient}
\label{sec:Z5Z5sections}

We now extend Donaldson's algorithm to compute the Calabi-Yau
metric directly on the quotient $Q=\Qt\big/(\Z_5\times\Z_5)$. To do
this, we will need to count and then explicitly construct the 
$\Z_5 \times \Z_5$ invariant sections, that is, the $\Z_5 \times \Z_5$
invariant polynomials, on the covering space $\Qt \in \CP^4$,
as discussed in \autoref{sec:quotients}. These then descend to
the quotient $Q$ and can be used to parametrize the Kahler
potential and the approximating balanced metrics.

One technical problem, however, is that the two group generators $g_1$
and $g_2$ in eq.~\eqref{eq:Z5Z5action} do not commute; they only
commute up to a phase. Therefore, the homogeneous coordinates
\begin{equation}
  \Span\big\{z_0,z_1,z_2,z_3,z_4\big\} 
  = 
  H^0\big( \Qt, \Osheaf_\Qt(1) \big)
\end{equation}
do not carry a $\Z_5\times\Z_5$ representation. The
solution to this problem is to enlarge the group. Each
generator has order $5$ and, even though they do not quite generate
$\Z_5\times\Z_5$, they commute up to a phase. Hence, $g_1$ and $g_2$
generate the ``central extension''
\begin{equation}
  1 
  \longrightarrow
  \Z_5
  \longrightarrow
  G
  \longrightarrow
  \Z_5\times\Z_5
  \longrightarrow
  1
\end{equation}
with $|G|=125$ elements. This group $G$ is also called a Heisenberg
group since it is formally the same as $[x,p]=1$, only in this case
over $\Z_5$. It follows that $H^0( \Qt, \Osheaf_\Qt(1) )$ \emph{does}
carry a representation of $G$ and, hence, so does $H^0( \Qt,
\Osheaf_\Qt(k) )$ for any integer $k$.

Note that, when acting on degree-$k$ polynomials $p_k(z)$, the
commutant eq.~\eqref{eq:Z5Z5commutant} becomes
\begin{equation}
  g_1g_2g_1^{-1}g_2^{-1} \Big( p_k(z) \Big) 
  =
  e^{2\pi i\frac{k}{5}}
  \; p_k(z)
  .
\end{equation}
Therefore, if and only if $k$ is divisible by $5$ then the $G$
representation reduces to a true $\Z_5\times\Z_5$ representation on
$H^0( \Qt, \Osheaf_\Qt(k) )$. That is, $k$ must be of the form
\begin{equation}
  k = 5\ell
  ,\quad
  \ell \in \Z
  .
\end{equation}
The formal reason for this is that only the line bundles
$\Osheaf_\Qt(5\ell)$ are $\Z_5\times\Z_5$ equivariant. The invariant
subspaces of these $\Z_5\times\Z_5$ representations define the
invariant sections. Hence, we only consider homogeneous
polynomials of degrees divisible by $5$ which are invariant under the
action of $\Z_5\times\Z_5$ in the following.

\subsection{Invariant Polynomials}
\label{sec:Z5Z5invariants}

As a first step, determine the $\Z_5\times\Z_5$ invariant
sections on the ambient space $\CP^4$. That is, we must find
the invariant ring
\begin{equation}
   \C[z_0,z_1,z_2,z_3,z_4]^G
  \end{equation}
over $\CP^4$, where $G$ is the Heisenberg group defined in the previous
subsection. One can read off the number of invariants $\hat{N}^G_k$ at
each degree $k$ from the Molien series
\begin{multline}
  \label{eq:MolienZ5Z5}
    P \big( \C[z_0,z_1,z_2,z_3,z_4]^G ,t \big)
    =
    \sum_k \hat{N}^G_k \; t^k
    =
    \frac{1}{|G|}
    \sum_{g\in G} \frac{1}{ \det\big( 1-t g\big) } 
    = 
    \\
    =
    1 +
    6 t^5 +
    41 t^{10} + 
    156 t^{15} + 
    426 t^{20} + 
    951 t^{25} + 
    1856 t^{30} + 
    3291 t^{35} + 
    5431 t^{40} + 
    \\+
    8476 t^{45} + 
    12651 t^{50} +
    18206 t^{55} +
    25416 t^{60} +
    34581 t^{65} +
    \cdots
\end{multline}
We see that the only invariants are of degree $k=5\ell$, as
discussed in the previous subsection. To go further than just counting
the invariants, one uses the Hironaka decomposition which was
introduced in \autoref{sec:Hironaka}. For that, we need to choose $5$
primary invariants, the same number as homogeneous
coordinates. Unfortunately, any $5$ out of the $6$ quintic invariant
polynomials are never algebraically independent. Hence, picking five
degree-$5$ invariants never satisfies the requirements for them to be
primary invariants. It turns out that the primary invariants of
minimal degree consist of three degree-$5$ and two degree-$10$
invariants, which we will list in eq.~\eqref{eq:thetadef}
below. First, however, let us rewrite the Molien series as in
eq.~\eqref{eq:moldec},
\begin{equation}
  \label{eq:MolienZ5Z5rewrite}
  P \big( \C[z_0,z_1,z_2,z_3,z_4]^G ,t \big)
  = 
  \frac{ 1 + 3 t^5 + 24 t^{10} + 44 t^{15} + 24 t^{20} + 3 t^{25} + t^{30} }
  {\big(1-t^5\big)^3 \big(1-t^{10}\big)^2}
  .
\end{equation}
We see that this choice of primary invariants requires
\begin{equation}
  \frac{1}{|G|}
  \prod_{i=1}^5 \deg \theta_i
  =
  \frac{5^3 {10}^2}{|G|}
  =
  100
  =
  1 + 3 + 24 + 44 + 24 + 3 + 1
\end{equation}
secondary invariants in degrees up to $30$. We again note that this
decomposition is not unique, as one can always find different primary
and secondary invariants. However, our choice of primary invariants is
minimal, that is, leads to the least possible number ($=100$) of
secondary invariants.

Knowing the number of secondary invariants is not enough, however, and
we need the actual $G$-invariant polynomials. As will be explicitly
checked in \autoref{sec:thetaQuintic}, the five $G$-invariant
polynomials
\begin{equation}
  \label{eq:thetadef}
  \begin{array}{rcl}
    \theta_1 \eqdef&\; 
    z_0^5+z_1^5+z_2^5+z_3^5+z_4^5 
    \;&= z_0^5 + \cyclperm
    \\
    \theta_2 \eqdef&\; 
    z_0 z_1 z_2 z_3 z_4 
    \\
    \theta_3 \eqdef&\; 
    z_0^3 z_1 z_4+z_0 z_1^3 z_2+z_0 z_3 z_4^3
    +z_1 z_2^3 z_3+z_2 z_3^3 z_4 
    \;&= z_0^3 z_1 z_4 + \cyclperm
    \\
    \theta_4 \eqdef&\; 
    z_0^{10}+z_1^{10}+z_2^{10}+z_3^{10}+z_4^{10} 
    \;&= z_0^{10} + \cyclperm
    \\
    \theta_5 \eqdef&\; 
    z_0^8 z_2 z_3+z_0 z_1 z_3^8+z_0 z_2^8 z_4+z_1^8 z_3 z_4+z_1 z_2
    z_4^8
    \;&= z_0^8 z_2 z_3 + \cyclperm
  \end{array}
\end{equation}
satisfy the necessary criterion to be our primary invariants, where
$\cyclperm$ denotes the sum over the five different cyclic
permutations $z_0\to z_1\to \cdots\to z_4 \to z_0$. Next, we need a
basis for the corresponding secondary invariants, which must be of
degrees $0$, $5$, $10$, $15$, $20$, $25$, and $30$ according to
eq.~\eqref{eq:MolienZ5Z5rewrite}. In practice, these $100$ secondary
invariants can easily be found using \textsc{Singular}~\cite{GPS05,
  finvar}. They are
\begin{subequations}
\begin{gather}
  \begin{aligned}
    \eta_1 \eqdef&\; 1    ,\\ 
  \end{aligned}
  \displaybreak[2] \\[1ex]
  \begin{aligned}
    \eta_2 \eqdef&\; z_0^2 z_1 z_2^2 + \cyclperm    ,&
    \eta_3 \eqdef&\; z_0^2 z_1^2 z_3 + \cyclperm    ,&
    \eta_4 \eqdef&\; z_0^3 z_2 z_3 + \cyclperm      ,\\ 
  \end{aligned}
  \displaybreak[2] \\[1ex]
  \begin{aligned}
    \eta_5 \eqdef&\; z_0^5 z_2^5 + \cyclperm         ,&
    \eta_6 \eqdef&\; z_0^4 z_2^3 z_3^3 + \cyclperm    ,&
    \eta_7 \eqdef&\; z_0^4 z_1^3 z_4^3 + \cyclperm    ,\\ 
    \eta_8 \eqdef&\; z_0^4 z_1^2 z_2^4 + \cyclperm    ,&
    \eta_9 \eqdef&\; z_0^4 z_1^4 z_3^2 + \cyclperm    ,&
    \eta_{10} \eqdef&\; z_0^6 z_2^2 z_3^2 + \cyclperm   ,\\ 
    \eta_{11} \eqdef&\; z_0^6 z_1^2 z_4^2 + \cyclperm   ,&
    \eta_{12} \eqdef&\; z_0^6 z_1 z_3^3 + \cyclperm     ,& 
    \eta_{13} \eqdef&\; z_0^6 z_2^3 z_4 + \cyclperm     ,\\ 
    \eta_{14} \eqdef&\; z_0^6 z_1^3 z_2 + \cyclperm     ,& 
    \eta_{15} \eqdef&\; z_0^6 z_3 z_4^3 + \cyclperm     ,& 
    \eta_{16} \eqdef&\; z_0^7 z_1 z_2^2 + \cyclperm     ,\\ 
    \eta_{17} \eqdef&\; z_0^7 z_3^2 z_4 + \cyclperm     ,& 
    \eta_{18} \eqdef&\; z_0^7 z_1^2 z_3 + \cyclperm     ,& 
    \eta_{19} \eqdef&\; z_0^8 z_1 z_4 + \cyclperm       ,\\
    \eta_{20} \eqdef&\; z_0^3 z_1^2 z_2^2 z_3^3 + \cyclperm  ,& 
    \eta_{21} \eqdef&\; z_0^4 z_1^2 z_3^3 z_4 + \cyclperm    ,& 
    \eta_{22} \eqdef&\; z_0^4 z_1 z_2^3 z_4^2 + \cyclperm    ,\\ 
    \eta_{23} \eqdef&\; z_0^4 z_1^3 z_2^2 z_3 + \cyclperm    ,& 
    \eta_{24} \eqdef&\; z_0^4 z_1 z_2 z_3^4 + \cyclperm      ,&
    \eta_{25} \eqdef&\; z_0^5 z_1^2 z_2 z_3^2 + \cyclperm    ,\\ 
    \eta_{26} \eqdef&\; z_0^5 z_1^2 z_2^2 z_4 + \cyclperm    ,&
    \eta_{27} \eqdef&\; z_0^5 z_1 z_2^3 z_3 + \cyclperm      ,&
    \eta_{28} \eqdef&\; z_0^5 z_1^3 z_3 z_4 + \cyclperm      ,\\ 
  \end{aligned}
  \displaybreak[2] \\[1ex]
  \begin{aligned}
    \eta_{29} \eqdef&\; z_0^{15} + \cyclperm             ,& 
    \eta_{30} \eqdef&\; z_0^{10} z_2^5 + \cyclperm       ,&
    \eta_{31} \eqdef&\; z_0^{10} z_3^5 + \cyclperm       ,\\ 
    \eta_{32} \eqdef&\; z_0^{10} z_1^5 + \cyclperm       ,& 
    \eta_{33} \eqdef&\; z_0^6 z_1^3 z_2^6 + \cyclperm    ,& 
    \eta_{34} \eqdef&\; z_0^6 z_1^6 z_3^3 + \cyclperm    ,\\ 
    \eta_{35} \eqdef&\; z_0^7 z_2^4 z_3^4 + \cyclperm    ,& 
    \eta_{36} \eqdef&\; z_0^7 z_1^4 z_4^4 + \cyclperm    ,&
    \eta_{37} \eqdef&\; z_0^7 z_1^2 z_3^6 + \cyclperm    ,\\ 
    \eta_{38} \eqdef&\; z_0^7 z_2^6 z_4^2 + \cyclperm    ,& 
    \eta_{39} \eqdef&\; z_0^7 z_1^6 z_2^2 + \cyclperm    ,& 
    \eta_{40} \eqdef&\; z_0^8 z_1^3 z_3^4 + \cyclperm    ,\\ 
    \eta_{41} \eqdef&\; z_0^8 z_2^4 z_4^3 + \cyclperm    ,& 
    \eta_{42} \eqdef&\; z_0^8 z_1^4 z_2^3 + \cyclperm    ,& 
    \eta_{43} \eqdef&\; z_0^7 z_1^7 z_3 + \cyclperm      ,\\ 
    \eta_{44} \eqdef&\; z_0^8 z_2^6 z_3 + \cyclperm      ,& 
    \eta_{45} \eqdef&\; z_0^8 z_2 z_3^6 + \cyclperm      ,& 
    \eta_{46} \eqdef&\; z_0^8 z_1^6 z_4 + \cyclperm      ,\\ 
    \eta_{47} \eqdef&\; z_0^9 z_1^2 z_2^4 + \cyclperm    ,& 
    \eta_{48} \eqdef&\; z_0^9 z_1^4 z_3^2 + \cyclperm    ,& 
    \eta_{49} \eqdef&\; z_0^11 z_1^2 z_4^2 + \cyclperm   ,\\ 
    \eta_{50} \eqdef&\; z_0^11 z_1 z_3^3 + \cyclperm     ,& 
    \eta_{51} \eqdef&\; z_0^11 z_1^3 z_2 + \cyclperm     ,& 
    \eta_{52} \eqdef&\; z_0^11 z_3 z_4^3 + \cyclperm     ,\\ 
    \eta_{53} \eqdef&\; z_0^12 z_1 z_2^2 + \cyclperm     ,& 
    \eta_{54} \eqdef&\; z_0^12 z_1^2 z_3 + \cyclperm     ,& 
    \eta_{55} \eqdef&\; z_0^5 z_1^3 z_2^4 z_3^3 + \cyclperm    ,\\ 
    \eta_{56} \eqdef&\; z_0^5 z_1^3 z_2^3 z_4^4 + \cyclperm    ,& 
    \eta_{57} \eqdef&\; z_0^5 z_1^4 z_2^2 z_3^4 + \cyclperm    ,& 
    \eta_{58} \eqdef&\; z_0^5 z_1^2 z_3^4 z_4^4 + \cyclperm    ,\\ 
    \eta_{59} \eqdef&\; z_0^6 z_1^2 z_2^3 z_3^4 + \cyclperm    ,& 
    \eta_{60} \eqdef&\; z_0^6 z_2^4 z_3^3 z_4^2 + \cyclperm    ,& 
    \eta_{61} \eqdef&\; z_0^6 z_1^4 z_2^2 z_4^3 + \cyclperm    ,\\ 
    \eta_{62} \eqdef&\; z_0^6 z_1^4 z_3^4 z_4 + \cyclperm      ,&
    \eta_{63} \eqdef&\; z_0^6 z_1^4 z_2^4 z_3 + \cyclperm      ,& 
    \eta_{64} \eqdef&\; z_0^6 z_1 z_2^5 z_3^3 + \cyclperm      ,\\ 
    \eta_{65} \eqdef&\; z_0^7 z_1^3 z_2^3 z_3^2 + \cyclperm    ,& 
    \eta_{66} \eqdef&\; z_0^7 z_1^4 z_2 z_3^3 + \cyclperm      ,& 
    \eta_{67} \eqdef&\; z_0^7 z_2^3 z_3 z_4^4 + \cyclperm      ,\\ 
    \eta_{68} \eqdef&\; z_0^7 z_1^3 z_2^4 z_4 + \cyclperm      ,& 
    \eta_{69} \eqdef&\; z_0^7 z_1 z_2^2 z_3^5 + \cyclperm      ,& 
    \eta_{70} \eqdef&\; z_0^8 z_1^2 z_2^2 z_3^3 + \cyclperm    ,\\ 
    \eta_{71} \eqdef&\; z_0^8 z_1 z_3^5 z_4 + \cyclperm        ,& 
    \eta_{72} \eqdef&\; z_0^9 z_1 z_2 z_3^4 + \cyclperm        ,& 
  \end{aligned}
  \displaybreak[2] \\[1ex]
  \begin{aligned}
    \eta_{73} \eqdef&\; z_0^{20} + \cyclperm               ,&
    \eta_{74} \eqdef&\; z_0^{10} z_2^{10} + \cyclperm       ,& 
    \eta_{75} \eqdef&\; z_0^{15} z_2^5 + \cyclperm         ,\\ 
    \eta_{76} \eqdef&\; z_0^{15} z_1^5 + \cyclperm         ,& 
    \eta_{77} \eqdef&\; z_0^7 z_1^7 z_3^6 + \cyclperm      ,& 
    \eta_{78} \eqdef&\; z_0^7 z_1^6 z_2^7 + \cyclperm      ,\\ 
    \eta_{79} \eqdef&\; z_0^8 z_2^6 z_3^6 + \cyclperm      ,& 
    \eta_{80} \eqdef&\; z_0^8 z_1^6 z_4^6 + \cyclperm      ,& 
    \eta_{81} \eqdef&\; z_0^8 z_1^4 z_2^8 + \cyclperm      ,\\ 
    \eta_{82} \eqdef&\; z_0^8 z_1^8 z_3^4 + \cyclperm      ,& 
    \eta_{83} \eqdef&\; z_0^9 z_1^4 z_3^7 + \cyclperm      ,& 
    \eta_{84} \eqdef&\; z_0^9 z_2^7 z_4^4 + \cyclperm      ,\\ 
    \eta_{85} \eqdef&\; z_0^9 z_1^7 z_2^4 + \cyclperm      ,& 
    \eta_{86} \eqdef&\; z_0^9 z_3^4 z_4^7 + \cyclperm      ,& 
    \eta_{87} \eqdef&\; z_0^9 z_2^8 z_3^3 + \cyclperm      ,\\ 
    \eta_{88} \eqdef&\; z_0^9 z_2^3 z_3^8 + \cyclperm      ,& 
    \eta_{89} \eqdef&\; z_0^9 z_1^3 z_4^8 + \cyclperm      ,& 
    \eta_{90} \eqdef&\; z_0^9 z_1^2 z_2^9 + \cyclperm      ,\\ 
    \eta_{91} \eqdef&\; z_0^{11} z_1^3 z_2^6 + \cyclperm    ,& 
    \eta_{92} \eqdef&\; z_0^{11} z_1^6 z_3^3 + \cyclperm    ,& 
    \eta_{93} \eqdef&\; z_0^{11} z_2^3 z_4^6 + \cyclperm    ,\\ 
    \eta_{94} \eqdef&\; z_0^{11} z_2^7 z_3^2 + \cyclperm    ,& 
    \eta_{95} \eqdef&\; z_0^{11} z_2^2 z_3^7 + \cyclperm    ,& 
    \eta_{96} \eqdef&\; z_0^{11} z_1^7 z_4^2 + \cyclperm    ,\\ 
  \end{aligned}
  \displaybreak[2] \\[1ex]
  \begin{aligned}
     \eta_{97} \eqdef&\; z_0^9 z_2^8 z_3^8 + \cyclperm    ,&
     \eta_{98} \eqdef&\; z_0^9 z_1^8 z_4^8 + \cyclperm    ,& 
     \eta_{99} \eqdef&\; z_0^9 z_1^9 z_3^7 + \cyclperm ,    \\ 
  \end{aligned}
  \displaybreak[2] \\[1ex]
  \begin{aligned}
     \eta_{100} \eqdef&\; z_0^{30} + \cyclperm
     = z_0^{30} + z_1^{30} + z_2^{30} + z_3^{30} + z_4^{30}
     .
  \end{aligned}
\end{gather}
\end{subequations}
The Hironaka decomposition of the ring of $G$-invariant 
homogeneous polynomials is then
\begin{equation}
  \label{eq:HironakaZ5Z5}
  \C[z_0,z_1,z_2,z_3,z_4]^G =
  \bigoplus_{i=1}^{100}
  \eta_i 
  \, 
  \C[\theta_1,\theta_2,\theta_3,\theta_4,\theta_5 ]
  .
\end{equation}
As a simple application, we can read off a basis for the invariant
degree-$5$ polynomials,
\begin{equation}
  \label{eq:Gquintics}
  \begin{split}
    \C\big[ z_0,z_1,z_2,z_3,z_4 \big]^G_5 
    =&~
    \Span\Big\{ 
    \eta_1 \theta_1, \;
    \eta_1 \theta_2, \;
    \eta_1 \theta_3, \;
    \eta_2 ,\;
    \eta_3 ,\;
    \eta_4 
    \Big\}
    \\
    =&~
    \Span\Big\{ 
    \theta_1, \;
    \theta_2, \;
    \theta_3, \;
    \eta_2 ,\;
    \eta_3 ,\;
    \eta_4 
    \Big\}
    .
  \end{split}
\end{equation}
Note that this is the basis of invariant quintic polynomials used
in eq.~\eqref{eq:QuinticZ5Z5} to define $\Qt(z)$.

\subsection{Invariant Sections on the Quintic}

The next step is to restrict the $G$-invariant sections on $\CP^4$ to
the hypersurface $\Qt$. In \autoref{sec:quintic}, we showed how to
accomplish this for all sections on generic quintics
$\Qt\in\CP^4$. Since the sections on the ambient space are nothing but
homogeneous polynomials, the restricted sections were the quotient of
the homogeneous polynomials by the hypersurface equation $\Qt=0$,
\begin{equation}
  \vcenter{\xymatrix@C=2cm{
    H^0\big( \CP^4, \Osheaf_{\CP^4}(k) \big) 
    \ar[r]^-{\text{restrict}}
    \ar@{=}[d]
    & 
    H^0\big( \Qt, \Osheaf_\Qt(k) \big)     
    \ar@{=}[d]
    \\
    \C[z_0,z_1,z_2,z_3,z_4]_k
    \ar[r]^-{\Qt=0}
    & 
    \Big( 
    \C[z_0,z_1,z_2,z_3,z_4] \Big/ \big\langle \Qt \big\rangle
    \Big)_k
    .
  }}
\end{equation}
Now consider the quintics defined by eq.~\eqref{eq:QuinticZ5Z5}, which
allow a $\Z_5\times\Z_5$ action. Here, one only wants to know the
$G$-invariant sections on $\Qt$, since these correspond to the sections
on the $\Z_5\times\Z_5$ quotient $Q=\Qt\big/(\Z_5\times\Z_5)$. Moreover, since the 
$G$-invariant polynomials are of degree $5\ell$, we only consider
this case. Hence, the $G$-invariant sections are
\begin{equation}
  \vcenter{\xymatrix@C=2cm{
    H^0\big( \CP^4, \Osheaf_{\CP^4}(5\ell) \big)^G
    \ar[r]^-{\text{restrict}}
    \ar@{=}[d]
    & 
    H^0\big( \Qt, \Osheaf_\Qt(5\ell) \big)^G
    \ar@{=}[d]
    \\
    \C[z_0,z_1,z_2,z_3,z_4]^G_{5\ell}
    \ar[r]^-{\Qt=0}
    & 
    \left(
      \C[z_0,z_1,z_2,z_3,z_4] \Big/ \big\langle \Qt \big\rangle
    \right)^G_{5\ell}
    .
  }}
\end{equation}
Finally, we identify the invariant sections on $\Qt$ with sections on
the quotient manifold $Q$, as discussed in
\autoref{sec:quotients}. Therefore, the sections on $Q$ are
\begin{equation}
  H^0\Big( Q, \Osheaf_\Qt(5\ell)\big/(\Z_5\times\Z_5) \Big)
  =
  H^0\big( \Qt, \Osheaf_\Qt(5\ell) \big)^G
  =
  \left(
    \C[z_0,z_1,z_2,z_3,z_4] \Big/ \big\langle \Qt \big\rangle
  \right)^G_{5\ell}
  .
\end{equation}
By unravelling the definitions and using eq.~\eqref{eq:HironakaZ5Z5},
the invariant subspace of the quotient ring is given by
\begin{multline}
  \Big(
  \C[z_0,z_1,z_2,z_3,z_4] 
  \Big/ 
  \big\langle \Qt(z) \big\rangle
  \Big)^G
  \\ =
  \C[z_0,z_1,z_2,z_3,z_4]^G
  \Big/ 
  \big\langle \Qt(z) \big\rangle^G
   \\ =
  \left(
    \bigoplus_{i=1}^{100}
    \eta_i 
    \, 
    \C\big[\theta_1,\theta_2,\theta_3,\theta_4,\theta_5\big]
  \right)
  \Big/
  \left(
    \bigoplus_{i=1}^{100}
    \Qt \, \eta_i \, 
    \C\big[\theta_1,\theta_2,\theta_3,\theta_4,\theta_5\big]
  \right)
  .
\end{multline}
Using eq.~\eqref{eq:QuinticZ5Z5}, the hypersurface equation is
\begin{equation}
  \begin{aligned}
    \Qt(z) 
    =&~
    0
    &\Leftrightarrow
    \\
    z_0^5+z_1^5+z_2^5+z_3^5+z_4^5 
    =&~
    - \phi_1 \big( z_0 z_1 z_2 z_3 z_4 \big) - \cdots
    &\Leftrightarrow
    \\
    \theta_1
    =&~ -\phi_1\theta_2 - \phi_2 \theta_3 
    - \phi_3\eta_2 - \phi_4 \eta_3 -\phi_5 \eta_4
    ,
  \end{aligned}
\end{equation}
and, hence, we can simply eliminate $\theta_1$. Therefore, forming the
quotient is particularly easy, and we obtain
\begin{table}
  \centering
  \renewcommand{\arraystretch}{1.3}
  \begin{tabular}{c|cccccccc}
    $5\ell$ &
    $5$ &
    $10$ &
    $15$ &
    $20$ &
    $25$ &
    $30$ &
    $35$ &
    $40$
    \\ \hline \strut
    $\hat{N}^G_{5\ell}$ &
    $6$ &
    $41$ & 
    $156$ &
    $426$ &
    $951$ &
    $1856$ &
    $3291$ &
    $5431$
    \\
    $N^G_{5\ell}$ &
    $5$ &
    $35$ &
    $115$ &
    $270$ &
    $525$ &
    $905$ &
    $1435$ &
    $2140$
  \end{tabular}
  \caption{The number of $G$-invariant degree $5\ell$-homogeneous 
    polynomials $\hat{N}^G_{5\ell}$, eq.~\eqref{eq:MolienZ5Z5}, and 
    the number of remaining invariant polynomials $N^G_{5\ell}$ 
    after imposing the 
    hypersurface equation $\Qt(z)=0$, 
    see eq.~\eqref{eq:QuinticZ5Z5}.}
  \label{tab:N_k_quinticZ5Z5}
\end{table}
\begin{equation}
  \label{eq:Z5Z5quotient}
  \Big(
  \C[z_0,z_1,z_2,z_3,z_4] 
  \Big/ 
  \big\langle \Qt(z)\big\rangle
  \Big)^G
  =
  \bigoplus_{i=1}^{100}
  \eta_i \, 
  \C[\theta_2,\theta_3,\theta_4,\theta_5 ]
  .
\end{equation}
We list the number $\hat{N}_{5\ell}^G$ of $G$-invariant degree-$5\ell$
polynomials on $\CP^4$ as well as the number of invariant polynomials
after restricting to $\Qt$, $N_{5\ell}^G$, in
\autoref{tab:N_k_quinticZ5Z5}. Since we know the homogeneous degrees
of the primary and secondary invariants, $\theta$ and $\eta$
respectively, it is a simple combinatorial problem to list all
$N_{5\ell}^G$ monomials in eq.~\eqref{eq:Z5Z5quotient} of fixed degree
$5\ell$. They then form a basis for the sections on $Q$,
\begin{multline}
  \label{eq:sections5k}
  H^0\Big( Q, \Osheaf_\Qt(5\ell)\big/(\Z_5\times\Z_5) \Big)
  =
  \Span\big\{ s_\alpha  \big\}_{\alpha=0}^{N_{5\ell}^G-1}
  \\
  =
  \left(
    \bigoplus_{i=1}^{100}
    \eta_i \, 
    \C[\theta_2,\theta_3,\theta_4,\theta_5]
  \right)_{5\ell}
  = 
  \bigoplus_{i=1}^{100}
  \eta_i \, 
  \C[\theta_2,\theta_3,\theta_4,\theta_5]_{5\ell-\deg \eta_i}
  .
\end{multline}

\subsection{Results}
\label{sec:fourgenmetric}

\begin{figure}[htbp]
  \centering
  \include{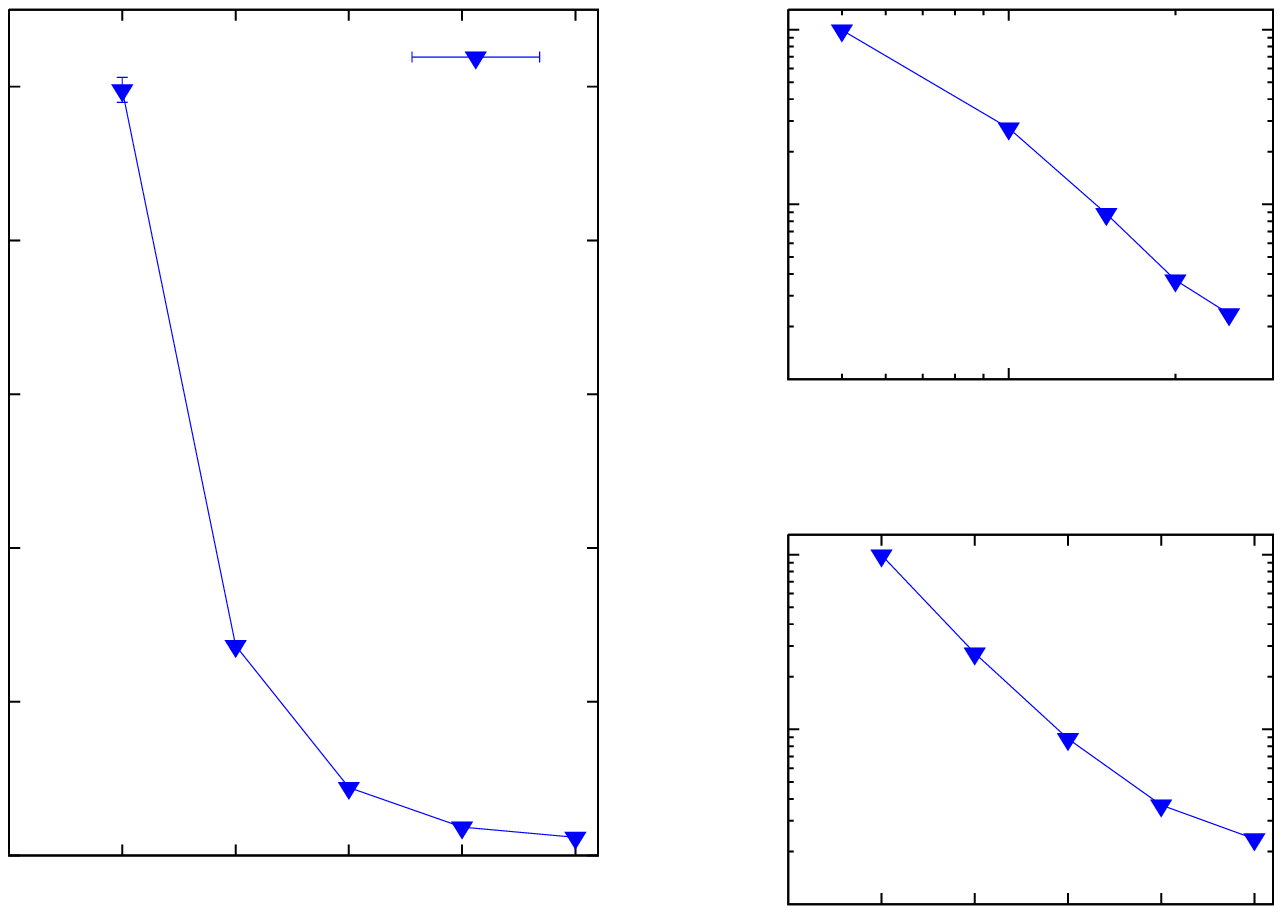}
  \caption{The error measure $\sigma_{5\ell}(Q_F)$ on the non-simply
    connected threefold $Q_F=\QtF\big/(\Z_5\times\Z_5)$. For each
    $\ell\in\Z_>$ we iterated the T-operator $10$ times, numerically
    integrating using $\Npoints=\comma{1000000}$ points. Then we
    evaluated $\sigma_{5\ell}(Q_F)$ using \comma{20000} different test
    points. Note that all three plots show the same data, but with
    different combinations of linear and logarithmic axes.}
  \label{fig:FermatQuinticZ5Z5_k}
\end{figure}
We have now computed an explicit basis of invariant sections of
$\Osheaf_\Qt(5\ell)$, which can be identified with a basis of sections
on the quotient manifold $Q=\Qt\big/(\Z_5\times\Z_5)$.  This is all we
need to extend Donaldson's algorithm to $Q$.  Literally the only
difference in the computer program used in \autoref{sec:QuinticPlot}
is that now
\begin{itemize}
\item the degree of the polynomials must be $k = 5\ell$, $\ell\in\Z_>$,
  and
\item the sections are given in eq.~\eqref{eq:sections5k}.
\end{itemize}
Hence, one can compute the balanced metrics on $Q$. As
$\ell\to\infty$, these will approach the unique Calabi-Yau metric. We
write $\sigma_{5\ell}(Q)$ for the error measure computed directly for
the balanced metrics on the non-simply connected threefold $Q$. Note
that there is still a $5$-dimensional complex structure moduli space
of such threefolds. However, as we have seen in
\autoref{fig:MultipleQuintics}, the details of the complex structure
essentially play no role in how fast the balanced metrics converge to
the Calabi-Yau metric. Therefore, as an example, in
\autoref{fig:FermatQuinticZ5Z5_k} we plot $\sigma_{5\ell}$ for the
quotient $Q_F=\QtF\big/(\Z_5\times\Z_5)$ of the Fermat quintic. Note
that the error measure tends to zero as $\ell\to\infty$, as it should.

\subsubsection*{Comparison With the Covering Space}

\begin{figure}[htbp]
  \centering
    \parbox{7cm}{\include{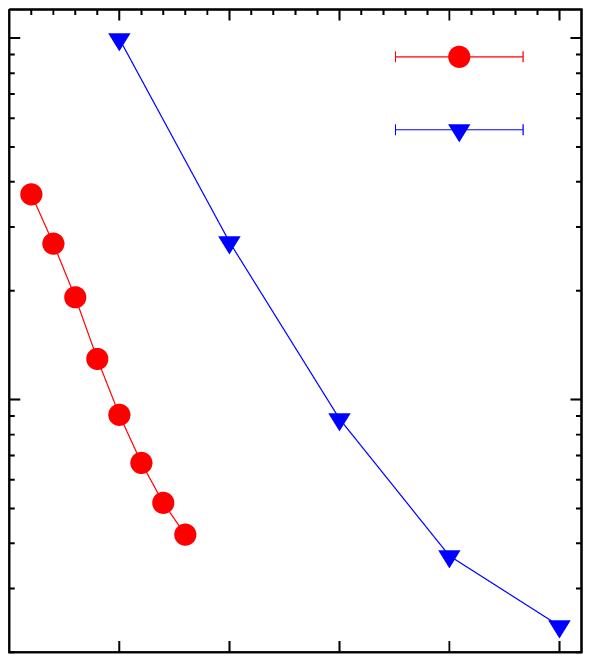}}
  \hfill
    \parbox{7cm}{\include{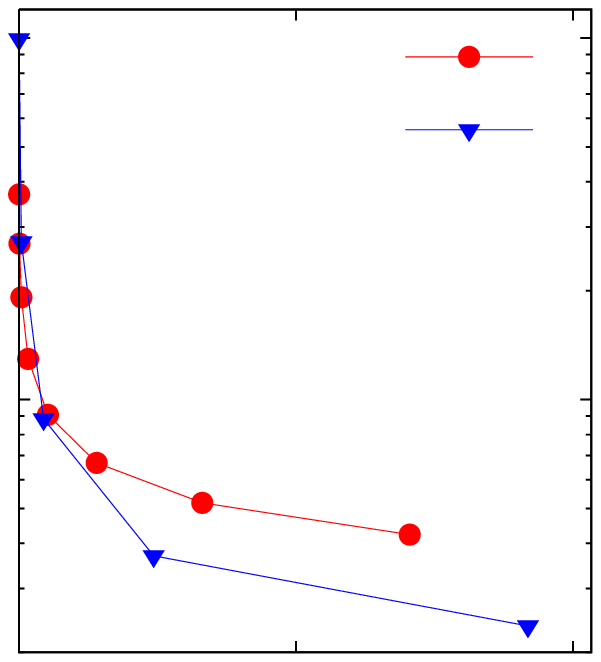}}
    \caption{The metric pulled back from
      $Q_F=\QtF\big/(\Z_5\times\Z_5)$ compared with the metric
      computation on $\QtF$. The error measures are
      $\tilde\sigma_{5\ell}(\QtF)$ and $\sigma_k(\QtF)$,
      respectively. On the left, we plot them by the degree of the
      homogeneous polynomials. On the right, we plot them as a
      function of $N^2$, the number of sections squared. On $Q_F$, the
      number of sections is $N^G_{5\ell}$; on $\QtF$ the number of
      sections is $N_k$. The $\sigma$-axis is logarithmic.}
  \label{fig:FermatQuinticZ5Z5_N}
\end{figure}
We have now extended Donaldson's algorithm so as to compute the
successive approximations to the Calabi-Yau metric directly on the
quotient manifold $Q$. Clearly, these metrics can be pulled back to
$\Z_5\times\Z_5$ symmetric metrics on the covering space $\Qt$,
thus approximating the Calabi-Yau metric on $\Qt$. Let us denote by
$\omega_{5\ell}$ the balanced \Kahler{} form on $Q$ computed at
degree-$5\ell$, and by $q^* \omega_{5\ell}$ its pull-back to $\Qt$. We
define $\tilde\sigma_{5\ell}(\Qt)$ to be the error measure evaluated
using the pull-back metric, that is,
\begin{multline}
  \tilde{\sigma}_{5\ell}\big( \Qt \big) = 
  \frac{1}{\Vol_\CY\big(\Qt\big)}
  \int_\Qt \left|
    1 - 
    \frac{
      q^*  \omega_{5\ell}^3  \Big/ \Vol_\text{K}\big(\Qt\big)
    }{
      \Omega \wedge \bar\Omega \Big/ \Vol_\CY\big(\Qt\big)
    }
  \right| \dVol_\CY
  = \\ =
  \frac{1}{\Vol_\CY(Q)}
  \int_Q \left|
    1 - 
    \frac{
      \omega_{5\ell}^3  \Big/ \Vol_\text{K}(Q)
    }{
      \Omega \wedge \bar\Omega \Big/ \Vol_\CY(Q)
    }
  \right| \dVol_\CY
  = 
  \sigma_{5\ell}(Q)
  .
\end{multline}
Now recall that in \autoref{sec:quintic} it was shown how to
determine the Calabi-Yau metric on $\textit{any}$ quintic
threefold. This, of course, includes the $\Z_5\times\Z_5$ quintics
$\Qt$ defined by eq.~\eqref{eq:QuinticZ5Z5}. However, since most
quintics do not admit a finite group action, the procedure specified
in \autoref{sec:quintic} finds the Calabi-Yau metric using
\emph{generic} homogeneous polynomials. That is, it finds an explicit
polynomial basis for $H^0( \Qt, \Osheaf_\Qt(k) )$, computes the
balanced metric and determines the Calabi-Yau metric as the
$k\to\infty$ limit. When applied to our $\Z_5\times\Z_5$ quintics,
this second method will also compute the unique $\Z_5\times\Z_5$
symmetric Calabi-Yau metric.  However, it does so as the limit of
balanced metrics constructed from sections of $\Osheaf_\Qt(k)$ which
do not share this symmetry, rather than from invariant sections of
$\Osheaf_\Qt(5\ell)$ as above. That is, this second method does
\emph{not} exploit the $\Z_5\times\Z_5$ symmetry. The associated error
measure $\sigma_k$ is evaluated using eq.~\eqref{eq:sigmaQtDef} for
$\Z_5\times\Z_5$-symmetric quintics. It is of some interest to compare
these two methods for calculating the Calabi-Yau metric on
$\Qt$. Specifically, in the left plot of
\autoref{fig:FermatQuinticZ5Z5_N} we compare the error measure
$\tilde\sigma_{5\ell}$ to $\sigma_k$ on the Fermat quintic
$\QtF$. Interestingly, for fixed degrees $k=5\ell$ the pull-back
metric is a worse approximation to the Calabi-Yau metric on $\Qt$ than
the metric computed on $\Qt$ without taking the symmetry into
account. The reason is that, in addition to the $\Z_5\times\Z_5$
invariant polynomials on $\Qt$, there are many more that transform
with some character of $\Z_5\times\Z_5$. These polynomials provide
extra degrees of freedom at fixed degree $5\ell$, which allow the
balanced metric to be a better fit to the Calabi-Yau metric.

However, a more just comparison is by the amount of the numerical
effort, that is, the number $(N_{5\ell}^G)^2$ and $(N_k)^2$,
respectively, of entries in the $h^{\alpha\betabar}$ matrix. We plot
$\tilde\sigma_{5\ell}$ and $\sigma_k$ as a function of $N^2$ in
\autoref{fig:FermatQuinticZ5Z5_N}. We see that, except for the two
lowest-degree cases $5\ell=k=5$ and $5\ell=k=10$, 
the pull-back metric computation
(that is, using invariant sections) is more efficient.

\section{Schoen Threefolds}
\label{sec:Schoen}

\subsection{As Complete Intersections}
\label{sec:SchoenCICY}

By definition, Schoen type Calabi-Yau threefolds are the fiber product
of two $\dP9$ surfaces, $B_1$ and $B_2$, fibered over $\CP^1$. Recall
that a $\dP9$ surface is defined as a blow-up of $\CP^2$ at $9$
points. In principle, these points can be ``infinitesimally close'',
that is, one of the blow-up points lies within a previous blow-up, but
we will only consider the generic case where all $9$ points are
distinct. Moreover, we are going to restrict ourselves to the case
where ``no Kodaira fibers collide''. In that case, the Hodge diamond of
the Schoen threefold $\Xt$ is~\cite{Donagi:2000zf, Braun:2004xv, MR923487}
\begin{equation}
  \label{eq:SchoenHodge}
  h^{p,q}\big( \Xt \big)
  =
  \vcenter{\xymatrix@!0@=7mm@ur{
    1 &  0 &  0 & 1 \\
    0 & 19 & 19 & 0 \\
    0 & 19 & 19 & 0 \\
    1 &  0 &  0 & 1 
  }}
  \,.
\end{equation}
These generic Schoen Calabi-Yau threefolds can be written as a
complete intersection as follow~\cite{Braun:2005nv, Braun:2005ux,
   Ovrut:2002jk, Braun:2005zv}. First, consider the ambient variety
$\CP^2 \times \CP^1 \times \CP^2$ with coordinates
\begin{equation}
  \Big( 
  [x_0:x_1:x_2],~ 
  [t_0:t_1],~
  [y_0:y_1:y_2]
  \Big)
  \in 
  \CP^2 \times \CP^1 \times \CP^2
  .
\end{equation}
The Calabi-Yau threefold $\Xt$ is then cut out as the zero-set of two
equations of multi-degrees $(3,1,0)$ and $(0,1,3)$, respectively. The
two equations are of the form
\begin{subequations}
\begin{align}
  \label{eq:P}
  \Pt(x,t,y) ~=&~ 
  t_0 \Pt_1\big(x_0, x_1, x_2\big) + 
  t_1 \Pt_2\big(x_0, x_1, x_2\big)
  = 0 
  ,
  \\
  \label{eq:R}
  \Rt(x,t,y) ~=&~ 
  t_1 \Rt_1\big(y_0, y_1, y_2\big) + 
  t_0 \Rt_2\big(y_0, y_1, y_2\big)  
  = 0 
\end{align}
\end{subequations}
where $\Pt_1$, $\Pt_2$, $\Rt_1$, and $\Rt_2$ are cubic
polynomials. The ambient space $\CP^2 \times \CP^1 \times \CP^2$ is a
toric variety and $\Xt$ is a toric complete intersection Calabi-Yau
threefold~\cite{Braun:2007xh, Braun:2007vy, Braun:2007tp}.

\subsection{Line Bundles and Sections}
\label{sec:SchoenLb}

The first Chern classes of line bundles on $\Xt$ form a
\begin{equation}
  h^{1,1}\big(\Xt\big)=19 
\end{equation}
dimensional lattice. Note, however, that most of them do not come from
the ambient space which has
\begin{equation}
  h^{1,1}\big(\CP^2\times\CP^1\times\CP^2\big)=3  
  .
\end{equation}
In other words, most of the divisors $D$ and their associated line
bundles $\Lsheaf(D)$ are not toric; that is, they cannot be described
by toric methods. We could embed $\Xt$ in a much more complicated
toric variety~\cite{Braun:2007vy} where all divisors are
toric. However, for now\footnote{This will be partially justified in
  \autoref{sec:Z3Z3}, where we investigate a certain $\ZZZ$-quotient
  of $\Xt$. There, only the toric line bundles will be relevant.}  we
will simply ignore the non-toric divisors and restrict ourselves to line
bundles on $\Xt$ that are induced from $\CP^2\times\CP^1\times\CP^2$.

The line bundles on $\CP^2\times\CP^1\times\CP^2$ are classified by
their first Chern class
\begin{equation}
  c_1\Big( 
  \Osheaf_{\CP^2\times\CP^1\times\CP^2}(a_1,b,a_2) \Big) = 
  (a_1,b,a_2) \in \Z^3 
  = H^2\big(\CP^2\times\CP^1\times\CP^2,\Z\big)
  .
\end{equation}
Just as in the $\CP^4$ case previously, their sections are homogeneous
polynomials of the homogeneous coordinates. Now, however, there are
three independent degrees, one for each factor. That is, the sections
of $\Osheaf_{\CP^2\times\CP^1\times\CP^2}(a_1,b,a_2)$ are homogeneous
polynomials of
\begin{itemize}
\item degree $a_1$ in $x_0$, $x_1$, $x_2$,
\item degree $b$ in $t_0$, $t_1$, 
\item degree $a_2$ in $y_0$, $y_1$, $y_2$.
\end{itemize}
The number of such polynomials (that is, the dimension of the linear
space of polynomials) is counted by the Poincar\'e series
\begin{equation}
  \begin{split}
    P\Big( \Osheaf_{\CP^2\times\CP^1\times\CP^2}, (x,t,y) \Big)
    =&~
    \sum_{a_1,b,a_2}
    h^0\Big(\CP^2\times\CP^1\times\CP^2, \Osheaf(a_1,b,a_2) \Big)
    x^{a_1} t^b y^{a_2}
    \\
    =&~
    \frac{1}{(1-x)^3}
    \frac{1}{(1-t)^2}
    \frac{1}{(1-y)^3}
    .
  \end{split}
\end{equation}
We now want to restrict the sections to the complete intersection
$\Xt\subset\CP^2\times\CP^1\times\CP^2$; that is, find the image
\begin{equation}
  \vcenter{\xymatrix{
      H^0\Big( \CP^2\times\CP^1\times\CP^2, \Osheaf(a_1,b,a_2) \Big)
      \ar[rr]^-{\text{restrict}} 
      &&
      H^0\Big( \Xt, \Osheaf_\Xt(a_1,b,a_2) \Big)
      \ar[r] 
      &
      0
   }}
\end{equation}
for\footnote{Note that $c_1\big( \Osheaf_\Xt(a_1,b,a_2)\big) \in
  H^2(X,\Z)$ is in the interior of the \Kahler{} cone if and only if
  $a_1,b,a_2>0$, see ~\cite{Gomez:2005ii}.} $a_1, b, a_2 >0$. As
discussed previously, this amounts to finding a basis for the quotient
space
\begin{equation}
  \label{eq:SchoenCoordRing}
  H^0\big( \Xt, \Osheaf_\Xt(a_1,b,a_2) \big) 
  = 
  \Big(
    \C\big[x_0,x_1,x_2,t_0,t_1,y_0,y_1,y_2\big]
  \Big/ 
    \big\langle \Pt, \Rt \big\rangle
  \Big)_{(a_1,b,a_2)}
\end{equation}
of degree $(a_1,b,a_2)$. Note that this quotient by more than one
polynomial is much more difficult than the case where one quotients
 out a single polynomial, as we did for quintics in \autoref{sec:quintic}. 
 In general, this requires the technology of
Gr\"obner bases~\cite{MR1189133}. Suffices to say that we are in a
very advantageous position here.

By a suitable coordinate change, we can assume that the $t_0 y_0^3$
term in $\Rt$ is absent. That is,
\begin{equation}
  \begin{split}
    \Pt =&~ t_0 x_0^3 + \cdots
    \\
    \Rt =&~ 0 \cdot t_0 y_0^3 + t_0 y_0^2 y_1 + \cdots
    .
  \end{split}
\end{equation}
Then, for otherwise generic polynomials $\Pt$ and
$\Rt$ and lexicographic monomial order
\begin{equation}
  x_0 \prec
  y_0 \prec
  t_0 \prec
  x_1 \prec
  y_1 \prec
  t_1 \prec
  x_2 \prec
  y_2
  ,
\end{equation}
the two polynomials generating 
\begin{equation}
  \big\langle \Pt, \Rt \big\rangle
  \subset 
  \C[x_0,x_1,x_2,t_0,t_1,y_0,y_1,y_2 ]
\end{equation}
already form a Gr\"obner basis. This means that the quotient in
eq.~\eqref{eq:SchoenCoordRing} can be implemented simply by
eliminating the leading monomials $t_0 x_0^3$ and $t_0 y_0^2 y_1$ in
the polynomial ring $\C[x_0,x_1,x_2,t_0,t_1,y_0,y_1,y_2]$.

\subsection{The Calabi-Yau Volume Form}
\label{sec:CI_Omega}

As in the case of a hypersurface, one can express the
$(3,0)$-form of the complete intersection as a Griffiths residue. By
definition, the zero loci $\Pt=0$ and $\Rt=0$
intersect transversally, so one can encircle each in an independent
transverse direction. The double residue integral
\begin{equation}
  \Omega = \oint\!\oint 
  \frac{\diff^2x \diff t \diff^2 y}{\Pt \cdot \Rt}
\end{equation}
is again independent of the chosen inhomogeneous coordinate
chart. Hence, it defines a holomorphic $(3,0)$-form which must
be the holomorphic volume form.

\subsection{Generating Points}
\label{sec:SchoenPoints}

Since the defining Equations~\eqref{eq:P},~\eqref{eq:R} are at most
cubic in the x and y coordinates, there is a particularly 
nice way to pick points. 
This is a generalization of the $L\cap \Qt$ method presented in
\autoref{sec:integrating} to generate points in generic quintics. In
the present case, select a specific $\CP^1\times\CP^1$ in the ambient
space, namely,
\begin{equation}
  \label{eq:hyperplanes}
  \CP^1 \times \ptset \times \CP^1
  ~\subset~
  \CP^2\times\CP^1\times\CP^2
  .
\end{equation}
This can easily be done with an $SU(3)\times SU(2)\times
SU(3)$-invariant probability density of such configurations. The
intersection 
\begin{equation}
  \Big( \CP^1 \times \ptset \times \CP^1 \Big) 
  \cap 
  \Xt
  = 
  \{ 9\text{ points}\}
\end{equation}
consists of nine points. To compute the coordinates of the nine
points, one needs to solve two cubic equations, which can be done
analytically\footnote{Recall that, to generate points on the quintic,
  we had to solve a quintic polynomial. This can only be done
  numerically.}.

We still need the distribution of these ``random'' points. First, note
that there are three obvious $(1,1)$-forms. These are the pull-backs
\begin{equation}
  \label{eq:pullback_i}
  \pi_1^*\big( \omega_{\CP^2} \big) 
  ,\quad
  \pi_2^*\big( \omega_{\CP^1} \big) 
  ,\quad
  \pi_3^*\big( \omega_{\CP^2} \big) 
\end{equation}
of the standard ($SU(m+1)$ symmetric) Fubini-Study \Kahler{} forms on
$\CP^m$, where $\pi_i$ is the projection on the $i$-th factor of the
ambient space. However, here the $SU(3)\times SU(2)\times SU(3)$
symmetry of the ambient space is not enough to determine the
distribution of points uniquely.

In general, the question about the distribution of zeros was answered
by Shifman and Zelditch~\cite{Zelditch:Shif}. Let us quickly review
the result. Let $\Lsheaf$ be a line bundle on a complex manifold $Y$
and pick a basis $s_0$, $\dots$, $s_{N-1}$ of sections
\begin{equation}
  \Span\big\{ s_0, \dots, s_{N-1} \big\} = 
  H^0 (Y,\Lsheaf)
  .
\end{equation}
Moreover, let $\Lsheaf$ be base-point free, that is, the sections do
not have a common zero. In other words, 
\begin{equation}
  \Phi_\Lsheaf: Y \rightarrow \CP^{N-1}
  ,~
  x \mapsto \big[ s_0(t) : s_1(t) : \cdots : s_{N-1}(t) \big]
\end{equation}
is a well-defined map. The sections generate the $N$-dimensional
vector space $H^0(Y,\Lsheaf)$ which contains the unit sphere
$SH^0(Y,\Lsheaf)$. In other words, if we define $s_0,\dots,s_{N-1}$ to
be an orthonormal basis, then $SH^0(Y,\Lsheaf)$ is the common
$SU(N)$-orbit of the basis sections. We take a random section $s\in
SH^0(Y,\Lsheaf)$ to be uniformly distributed with respect to the
usual ``round'' measure, that is, $SU(N)$-uniformly distributed.

Finally, switch from each such section $s$ to its zero locus $Z_s$ in
$Y$, and consider the expected distribution of the random zero
loci. Then
\begin{theorem}[Shifman, Zelditch]
  Under the above assumptions (in particular, that $\Phi_\Lsheaf$ is
  well-defined) the expected distribution of zero loci $Z_s$ is
  \begin{equation}
     \big\langle Z_\Lsheaf  \big\rangle
     = 
     \frac{1}{N} \Phi_\Lsheaf^* \omega_\FS
    ,
  \end{equation}
  where $\omega_\FS$ is the standard Fubini-Study \Kahler{} form on
  $\CP^{N-1}$.
\end{theorem}
Note that, in our case, the embedding $\Xt\subset
\CP^2\times\CP^1\times\CP^2$ is generated by the three line bundles
\begin{equation}
  \renewcommand{\arraystretch}{1.3}
  \begin{array}{r@{=}ll}
    H^0\big( \Xt, \Osheaf_\Xt(1,0,0) \big) 
    & \Span\{ x_0, x_1, x_2 \}
    & \quad \Rightarrow 
    \Phi_{\Osheaf_\Xt(1,0,0)}: \Xt \to \CP^2
    ,
    \\
    H^0\big( \Xt, \Osheaf_\Xt(0,1,0) \big) 
    & \Span\{ t_0, t_1 \}
    & \quad \Rightarrow 
    \Phi_{\Osheaf_\Xt(0,1,0)}: \Xt \to \CP^1
    ,
    \\
    H^0\big( \Xt, \Osheaf_\Xt(0,0,1) \big) 
    & \Span\{ y_0, y_1, y_2 \}
    & \quad \Rightarrow 
    \Phi_{\Osheaf_\Xt(0,0,1)}: \Xt \to \CP^2
    .
  \end{array}
\end{equation}
Although none of the three $\Phi$ maps is an embedding, they are all
well-defined. This is sufficient for the theorem of Shifman and
Zelditch. We point out that the $\Phi$ maps are nothing but the
restriction of the projections $\pi$ to
$\Xt\subset\CP^2\times\CP^1\times\CP^2$,
\begin{equation}
  \Phi_{\Osheaf_\Xt(1,0,0)} = \pi_1|_\Xt
  ,\quad
  \Phi_{\Osheaf_\Xt(0,1,0)} = \pi_2|_\Xt
  ,\quad
  \Phi_{\Osheaf_\Xt(0,0,1)} = \pi_3|_\Xt
  .
\end{equation}
Hence, the expected distribution of a zero-loci of sections on $\Xt$
is
\begin{equation}
  \left\langle Z_{\Osheaf_\Xt(1,0,0)} \right\rangle
  \sim
  \pi_1^* \big(\omega_{\CP^2} \big)\big|_\Xt 
  ,\quad
  \left\langle Z_{\Osheaf_\Xt(0,1,0)} \right\rangle
  \sim 
  \pi_2^* \big(\omega_{\CP^1} \big)\big|_\Xt 
  ,\quad
  \left\langle Z_{\Osheaf_\Xt(0,0,1)} \right\rangle
  \sim 
  \pi_3^* \big(\omega_{\CP^2} \big)\big|_\Xt 
  .
\end{equation}
These are precisely the three $(1,1)$-forms we introduced previously
in eq.~\eqref{eq:pullback_i}. Therefore, if we independently pick the
two $\CP^1$ factors and the point in eq.~\eqref{eq:hyperplanes}, then
the distribution of simultaneous zero loci is
\begin{equation}
  \diff A \sim 
  \pi_1^*\big( \omega_{\CP^2} \big)
  \wedge
  \pi_2^*\big( \omega_{\CP^1} \big)
  \wedge
  \pi_3^*\big( \omega_{\CP^2} \big)\Big|_\Xt 
  .
\end{equation}
In other words, the points generated by the above algorithm are
randomly distributed with respect to the auxiliary measure $\diff A$.

\subsection{Results}
\label{sec:SchoenResult}

\begin{table}
  \centering
  \renewcommand{\arraystretch}{1.3}
  \begin{tabular}{c|cccccccc}
    $(a_1,b,a_2)$ &
    $(1,1,1)$ &
    $(2,2,2)$ &
    $(3,3,3)$ &
    $(4,4,4)$ &
    $(5,5,5)$ &
    $(6,6,6)$ &
    \\ \hline \strut
    $\hat{N}_{(a_1,b,a_2)}$ &
    $18$ &
    $108$ & 
    $400$ & 
    $1125$ & 
    $2646$ &
    $5488$
    \\
    $N_{(a_1,b,a_2)}$ &
    $18$ &
    $108$ & 
    $343$ &
    $801$ & 
    $1566$ & 
    $2728$
  \end{tabular}
  \caption{The number of degree $(a_1,b,a_2)$-homogeneous 
    polynomials $\hat{N}_{(a_1,b,a_2)}$ over $\CP^2 \times \CP^1 \times \CP^2$
    and the number of remaining polynomials $N_{(a_1,b,a_2)}$ on $\Xt$ after 
    imposing the two equalities $\Pt=0=\Rt$ 
    defining the complete intersection.}
  \label{tab:N_k_Schoen}
\end{table}
The new feature of the Schoen Calabi-Yau threefold, as opposed to the
quintic, is that one now has different directions in the \Kahler{}
moduli space. On quintic threefolds there is only one \Kahler{}
modulus, which is just the overall volume. Now, however, there is a
$19=h^{1,1}(\Xt)$ dimensional \Kahler{} moduli space of which we
parametrize $3$ directions by the toric line bundles
$\Osheaf_\Xt(a_1,b,a_2)$. Note that, here as elsewhere in algebraic
geometry, one has to work with integral \Kahler{} classes that are the
first Chern classes of some line bundle. This is not a real restriction,
however, since any irrational slope direction in the \Kahler{} moduli 
space can be approximated by a rational slope. A line with rational 
slope always intersects points in $H^2(\Xt,\Z)$.

\begin{figure}[htbp]
  \centering
  \include{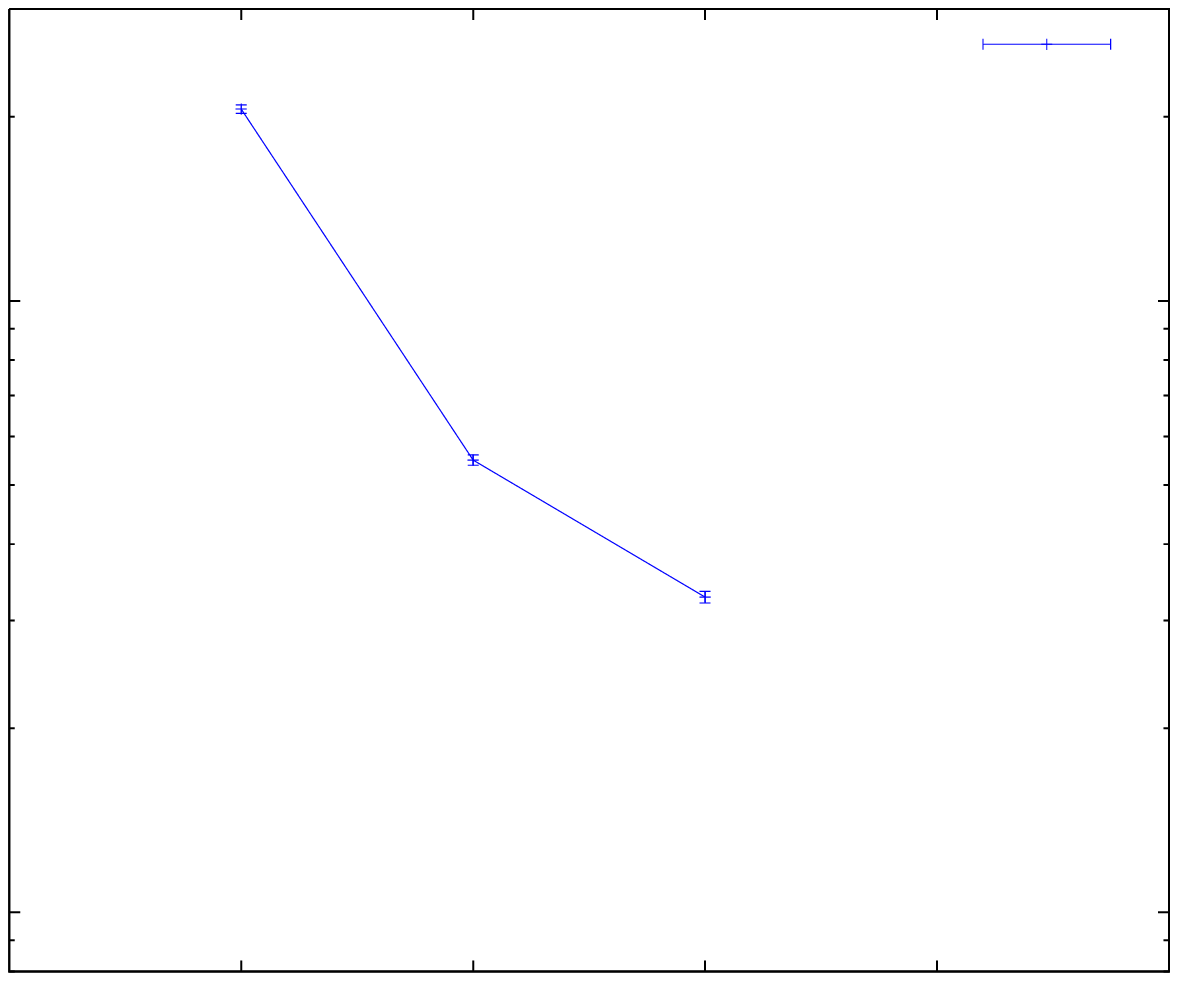}
  \caption{The error measure $\sigma_{(k,k,k)}$ for the metric on
    a $\Z_3\times\Z_3$ Schoen threefold $\Xt$. 
    We iterated the T-operator $5$ times,
    numerically integrating using $\Npoints=\comma{1000000}$
    points. Finally, we integrated $\sigma_{(k,k,k)}$ using
    \comma{10000} points.  $N_{(k,k,k)}$ is the number of sections
    $h^0\big(\Xt,\Osheaf_\Xt(k,k,k)\big)$.}
  \label{fig:Schoen}
\end{figure}
By way of an example, choose the direction $(1,1,1)\Z_> \subset
H^2(\Xt,\Z)$ in the \Kahler{} moduli space; that is, the line bundles
of the form $\Osheaf_\Xt(k,k,k)$ for $k\in\Z$, $k>0$. We list in
\autoref{tab:N_k_Schoen} the number of sections in both $\CP^2 \times
\CP^1 \times \CP^2$ and in its restriction to the Schoen manifold
$\Xt$. Note that they grow very fast with $k$, and quickly grow
outside of the range amenable to computation. However, the degree of
accuracy of the metric on $\Xt$ is essentially determined by
$N_{(k,k,k)}^2$, the number of metric parameters that we fit to
approximate the Calabi-Yau metric. Recall from the Hodge diamond
eq.\eqref{eq:SchoenHodge} that the complex structure moduli space is
$19=h^{2,1}(\Xt)$-dimensional. However, as in
\autoref{fig:MultipleQuintics}, the convergence of the balanced
metrics is essentially independent of the choice of complex
structure. Hence, as an example, we choose a specific $\Z_3\times\Z_3$
symmetric Schoen threefold ($\lambda_1=\lambda_2=0$, $\lambda_3=1$)
defined in the next section. In \autoref{fig:Schoen}, we plot the
error measure $\sigma_{(k,k,k)}$ vs.~$k$ for this manifold and find
very fast convergence. Note how the $k=3$ data point already
approaches to within $10\%$ of the limit
$\Npoints(=10^6)>N_{(k,k,k)}^2(=117,649)$, but still yields a quite
small value of $\sigma_{(3,3,3)}\approx 4\times 10^{-2}$.

\section
[The $\mathbf{\ZZZ}$ Manifold]
[The Z3 x Z3 Manifold]
{The $\mathbf{\ZZZ}$ Manifold}
\label{sec:Z3Z3}

\subsection{A Symmetric Schoen Threefold}   
\label{sec:SymmSchoen}

For special complex structures, the Schoen Calabi-Yau threefold has a
free $\ZZZ$ group action~\cite{Braun:2005zv, Candelas:2007ac}, which
we now describe. Recall that the Schoen threefolds can be
written as complete intersections in
\begin{equation}
  \Big( 
  [x_0:x_1:x_2],~ 
  [t_0:t_1],~
  [y_0:y_1:y_2]
  \Big)
  \in 
  \CP^2 \times \CP^1 \times \CP^2
  ,
\end{equation}
as discussed in \autoref{sec:Schoen}. Let us start by defining the
$\ZZZ$ group action on the ambient space~\cite{Braun:2004xv}, where it
is generated by ($\omega=e^{\frac{2\pi i}{3}}$)
\begin{subequations}
  \begin{equation}
    \label{eq:gamma1}
    \gamma_1:  
    \begin{cases}
      [x_0:x_1:x_2] \mapsto
      [x_0:\omega x_1:\omega^2 x_2]
      \\
      [t_0:t_1] \mapsto
      [t_0:\omega t_1] 
      \\
      [y_0:y_1:y_2] \mapsto
      [y_0:\omega y_1:\omega^2 y_2]
    \end{cases}
  \end{equation}
  and
  \begin{equation}
    \label{eq:gamma2}
    \gamma_2:  
    \begin{cases}
      [x_0:x_1:x_2] \mapsto
      [x_1:x_2:x_0]
      \\
      [t_0:t_1] \mapsto
      [t_0:t_1] 
      ~\text{(no action)}
      \\
      [y_0:y_1:y_2] \mapsto
      [y_1:y_2:y_0]
      .
    \end{cases}
  \end{equation}
\end{subequations}
The two generators commute up to phases on each of the two $\CP^2$
factors and, hence, define a $\ZZZ$ group action on the ambient
space. Note that $\gamma_2$ acts non-torically, that is, not by a
phase rotation. In order to define a $\ZZZ$-symmetric Calabi-Yau
threefold, we have to ensure that the zero locus $\Pt=0=\Rt$ is mapped
to itself by the group action. For that to be the case, one must  restrict the
polynomials $\Pt$ and $\Rt$ to have a special form. It was shown
in~\cite{Braun:2004xv} that one need only constrain the cubic
polynomials $\Pt_1$, $\Pt_2$, $\Rt_1$, $\Rt_2$ in eqns.~\eqref{eq:P}
and~\eqref{eq:R}. Specifically, the $\ZZZ$-symmetric Schoen Calabi-Yau
threefolds are defined by the simultaneous vanishing of the two
polynomials
\begin{equation}
  \begin{split}
    \label{eq:manifold}
    \Pt(x,t,y) =&~ 
    t_0 \Pt_1\big(x_0, x_1, x_2\big) + t_1 \Pt_2\big(x_0, x_1, x_2\big)
    \\
    \Rt(x,t,y) =&~ 
    t_1 \Rt_1\big(y_0, y_1, y_2\big) + t_0 \Rt_2\big(y_0, y_1,
    y_2\big)  
    ,
  \end{split}
\end{equation}
where
\begin{equation}
  \begin{split}
    \Pt_1\big(x_0, x_1, x_2\big)
    =&~
    x_0^3 + x_1^3 + x_2^3 + \lambda_1 x_0 x_1 x_2
    \\
    \Pt_2\big(x_0, x_1, x_2\big)
    =&~
    \lambda_3  \big(  x_0^2 x_2 + x_1^2 x_0 + x_2^2 x_1  \big)
    \\
    \Rt_1\big(y_0, y_1, y_2\big)
    =&~
    y_0^3 + y_1^3 + y_2^3 + \lambda_2 y_0 y_1 y_2
    \\
    \Rt_2\big(y_0, y_1, y_2\big)  
    =&~
    y_0^2 y_1 + y_1^2 y_2 + y_2^2 y_0
    .
  \end{split}
\end{equation}
In the following, we will always take $\Pt$, $\Rt$ to be of this form.
Note that, up to coordinate changes, the polynomials depend on $3$
complex parameters $\lambda_1$, $\lambda_2$, and $\lambda_3$.

One can easily check that $\Pt$ is completely invariant under the
$\ZZZ$ group action, as one naively expects. However, $\Rt$ is not
quite invariant. Rather, it transforms like a character of
$\ZZZ$. That is,
\begin{align} 
  \Pt(\gamma_1 x,\gamma_1 t,\gamma_1 y) =& \Pt(x,t,y) &
  \Pt(\gamma_2 x,\gamma_2 t,\gamma_2 y) =& \Pt(x,t,y) \\
  \label{eq:RtXform}
  \Rt(\gamma_1 x,\gamma_1 t,\gamma_1 y) =& e^{\frac{2\pi i}{3}} \Rt(x,t,y) &
  \Rt(\gamma_2 x,\gamma_2 t,\gamma_2 y) =& \Rt(x,t,y) 
  .
\end{align}
Nevertheless, the zero set $\Pt=0=\Rt$ \emph{is} invariant under the
group action. Moreover, the fixed point sets of $\gamma_1$ and
$\gamma_2$ on the ambient space $\CP^2\times\CP^1\times\CP^2$ are
\begin{equation}
  \begin{gathered}
  \big\{{\scriptstyle [1:0:0],~[0:1:0],~[0:0:1]}\big\} \times
  \big\{{\scriptstyle[0:1],~[1:0]}\big\} \times
  \big\{{\scriptstyle[1:0:0],~[0:1:0],~[0:0:1]}\big\}
  ,
  \\
  \big\{{\scriptstyle [1:1:1],~[1:\omega:\omega^2],~[1:\omega^2:\omega]}\big\} \times
  \CP^1 \times
  \big\{{\scriptstyle [1:1:1],~[1:\omega:\omega^2],~[1:\omega^2:\omega]}\big\}
  , 
\end{gathered}
\end{equation}
respectively. For generic\footnote{Note, however, that
  $\lambda_1=\lambda_2=\lambda_3=0$ is singular. A non-singular choice
  of complex structure is, for example, $\lambda_1=\lambda_2=0$ and
  $\lambda_3=1$.} $\lambda_i$, the Calabi-Yau threefold $\Xt$ misses
the $\ZZZ$-fixed points. Therefore, the quotient
\begin{equation}
  X = \Xt \Big/ \big(\ZZZ\big) 
  = \Big\{ \Pt=0=\Rt \Big\} \Big/ \big(\ZZZ\big)
\end{equation}
is a smooth Calabi-Yau threefold with fundamental group
$\pi_1(X)=\ZZZ$. Its Hodge diamond is given by~\cite{Braun:2004xv} 
\begin{equation}
  \label{eq:SchoenZ3Z3Hodge}
  h^{p,q}\big( X \big)
  =h^{p,q}\big( \Xt \Big/ \big(\ZZZ\big)\big)=
  \vcenter{\xymatrix@!0@=7mm@ur{
    1 &  0 &  0 & 1 \\
    0 & 3 & 3 & 0 \\
    0 & 3 & 3 & 0 \\
    1 &  0 &  0 & 1 
  }}
  \,.
\end{equation}
The complex structure moduli space is $h^{2,1}(X)=3$-dimensional and parametrized
by $\lambda_1$, $\lambda_2$, and $\lambda_3$.

\subsection{Invariant Polynomials}   

As discussed in \autoref{sec:SchoenLb}, sections of line bundles
on $\Xt$ are homogeneous polynomials in $[x_0:x_1:x_2]$, $[t_0:t_1]$
and $[y_0:y_1:y_2]$, modulo the ideal $\langle \Pt, \Rt\rangle$. We
now want to consider the quotient $X=\Xt\big/(\ZZZ)$. Therefore, we
are only interested in polynomials that are invariant under our group
action. Let us start with the group action on the homogeneous
coordinates $(x_0, x_1, x_2, t_0, t_1, y_0, y_1, y_2)$ of
$\CP^2\times\CP^1\times\CP^2$. The two generators defined in
eqns.~\eqref{eq:gamma1} and~\eqref{eq:gamma2} can be represented by
the $8\times 8$ matrices
\begin{equation}
\label{eq:gamma}
  \gamma_1=
  \begin{pmatrix}
    1 & 0 & 0 & 0 & 0 & 0 & 0 & 0 \\
    0 & \omega & 0 & 0 & 0 & 0 & 0 & 0 \\
    0 & 0 & \omega^2 & 0 & 0 & 0 & 0 & 0 \\
    0 & 0 & 0 & 1 & 0 & 0 & 0 & 0 \\
    0 & 0 & 0 & 0 & \omega & 0 & 0 & 0 \\
    0 & 0 & 0 & 0 & 0 & 1 & 0 & 0 \\
    0 & 0 & 0 & 0 & 0 & 0 & \omega & 0 \\
    0 & 0 & 0 & 0 & 0 & 0 & 0 & \omega^2
  \end{pmatrix}
  ,  ~
  \gamma_2= 
  \begin{pmatrix}
    0 & 1 & 0 & 0 & 0 & 0 & 0 & 0 \\
    0 & 0 & 1 & 0 & 0 & 0 & 0 & 0 \\
    1 & 0 & 0 & 0 & 0 & 0 & 0 & 0 \\
    0 & 0 & 0 & 1 & 0 & 0 & 0 & 0 \\
    0 & 0 & 0 & 0 & 1 & 0 & 0 & 0 \\
    0 & 0 & 0 & 0 & 0 & 0 & 1 & 0 \\
    0 & 0 & 0 & 0 & 0 & 0 & 0 & 1 \\
    0 & 0 & 0 & 0 & 0 & 1 & 0 & 0
  \end{pmatrix}
  .
\end{equation}
One can easily check that $\left[\gamma_1, \gamma_2\right]\neq 0$ and,
in fact, the $\gamma_1$ and $\gamma_2$ actions commute up to
multiplication by the central\footnote{Commuting with $\gamma_1$ and
  $\gamma_2$.} matrix
\begin{equation}
  \delta =
  \diag(\omega, \omega, \omega, 1, 1, \omega, \omega, \omega) 
  .
\end{equation}
In other words, the homogeneous coordinates 
\begin{multline}
  \Span\big\{ x_0, x_1, x_2, t_0, t_1, y_0, y_1, y_2 \big\}
  \\ =
  H^0\Big( \CP^2\times \CP^1 \times \CP^2, 
  \Osheaf(1,0,0) \oplus \Osheaf(0,1,0) \oplus \Osheaf(0,0,1) \Big)
\end{multline}
of $\CP^2\times \CP^1 \times \CP^2$ carry a representation of a
Heisenberg group $\Gamma$, which is the central extension
\begin{equation}
  0
  \longrightarrow 
  \Z_3 
  \longrightarrow 
  \Gamma
  \stackrel{\chi_1\times\chi_2}{\xrightarrow{\hspace{15mm}}}
  \Z_3 \times \Z_3 
  \longrightarrow 
  0
  .
\end{equation}
Note that the map $\chi_1\times\chi_2$ is defined in terms of
the two characters
\begin{equation}
  \label{eq:characters}
  \begin{aligned}
    \chi_1(\gamma_1) =& e^{\frac{2\pi i}{3}},  &
    \chi_1(\gamma_2) =& 1, &
    \chi_1(\delta) =& 1,
    \\
    \chi_2(\gamma_1) =& 1, &
    \chi_2(\gamma_2) =& e^{\frac{2\pi i}{3}}, &
    \chi_2(\delta) =& 1
  \end{aligned}
\end{equation}
of $\Gamma$, which will be important in the following. As discussed
previously for quintics, \autoref{sec:Z5Z5sections}, not all line
bundles are $\ZZZ$-equivariant. However, computing the polynomials
invariant under the Heisenberg group $\Gamma$ is sufficient for our purposes. 
The $\Gamma$-invariants are automatically the $\ZZZ$-invariant sections of
$\ZZZ$-equivariant line bundles. Their number
$\hat{N}^\Gamma_{(a_1,b,a_2)}$ in each multi-degree $(a_1,b,a_2)$ can
be read off from the multi-variable Molien series~\cite{Feng:2007ur},
\begin{multline}
  \label{eq:schmolienmulti} 
  P \Big( \C[x_0,x_1,x_2,t_0,t_1,y_0,y_1,y_2]^\Gamma , (x,t,y) \Big)
  = 
  \sum_{a_1,b,a_2}
  \hat{N}_{(a_1,b,a_2)}^\Gamma
  x^{a_1} t^b y^{a_2}
  \\=
  \frac{1}{|\Gamma|}\sum_{\gamma\in \Gamma}
  \frac{1}{\det\Big(1- \gamma \diag(x, x, x, t, t, y, y, y) \Big)}
  \\ =
  1+t+t^2+2t^3+2x^3+2y^3+ 2x^2y+2xy^2+2t^4
  + \cdots
  .
\end{multline}    
However, to construct the Hironaka decomposition it is sufficient
to determine the number of invariant linearly independent polynomials
of total degree $a_1+b+a_2$. The corresponding Poincar\'e series
can be obtained from eq.~\eqref{eq:schmolienmulti} by setting
$x=t=y=\tau$,
\begin{multline}
  \label{eq:schmolien} 
  P \Big( \C[x_0,x_1,x_2,t_0,t_1,y_0,y_1,y_2]^\Gamma ,\tau \Big)
  =
  \sum_{k}
  \hat{N}_{k}^\Gamma
  \tau^k
  \\=  
  \frac{1}{|\Gamma|}
  \sum_{\gamma\in \Gamma}\frac{1}{\det(1-\gamma\tau)}
   \\ =
  1+\tau+\tau^2+10\tau^3+16\tau^4+22\tau^5+ 85\tau^6+
  142\tau^7+199\tau^8+488\tau^9
  + \cdots
  .
\end{multline}    
Next, we need to choose $3+2+3=8$ primary invariants. Similarly to the
quintic case in \autoref{sec:Z5Z5invariants}, we choose our primary
invariants to be of the lowest possible degree. It is not hard to check
that homogeneous polynomials
\begin{subequations}
\label{eq:hopsschoen}
  \begin{align}
    \theta_1 &= t_0
    &
    \theta_2 &= t_1^3
    \\
    \theta_3 &= x_0x_1x_2
    &
    \theta_4 &= x_0^3+x_1^3+x_2^3
    \\
    \theta_5 &= y_0y_1y_2
    &
    \theta_6 &= y_0^3+y_1^3+y_2^3
    \\
    \theta_7 &= x_0^3x_1^3+ x_0^3x_2^3+x_1^3x_2^3
    &
    \theta_8 &= y_0^3y_1^3+ y_0^3y_2^3+y_1^3y_2^3
    .
  \end{align}
\end{subequations}
can be chosen as our primary invariants. They are, in fact, the choice
with the lowest degrees. Rewriting eq.~\eqref{eq:schmolien} as a
fraction with the denominator corresponding to our choice of the
primary invariants, we get
\begin{multline}
  P \Big( \C[x_0,x_1,x_2,t_0,t_1,y_0,y_1,y_2]^\Gamma ,\tau \Big)
  \\=
  \frac{1}{(1-\tau)(1-\tau^3)^5(1-\tau^6)^2} \Big( 
  1 + 4\tau^3 + 6\tau^4 + 6\tau^5 + 26\tau^6 + 27\tau^7 + 
  27\tau^8 +
  46\tau^9 +
  \\+   
  42\tau^{10} + 42\tau^{11} + 26\tau^{12} + 
  27\tau^{13} + 27 \tau^{14} + 4\tau^{15} + 
  6\tau^{16} + 6\tau^{17} + \tau^{18}
  \Big)
  .
\end{multline}    
Thus, the number of secondary invariants is 
\begin{equation}
  \begin{split}
    \frac{3^56^2}{|\Gamma|}=324= 
    1 + 4 &~+ 6+ 6 + 26 + 27 + 27+ 46+
    \\&~
    42+42+ 26 + 27+ 27 + 4 + 6 + 6+ 1
    .
  \end{split}
\end{equation}
Notice that the polynomials in eq.~\eqref{eq:hopsschoen} are
homogeneous of multi-degree $(a_1,b,a_2)$. Since the group action
eq.~\eqref{eq:gamma} does not mix the degrees, it follows that the
secondary invariants will also be homogeneous polynomials. They are,
moreover, separately homogeneous in the variables $[x_0:x_1:x_2]$,
$[t_0:t_1]$, and $[y_0:y_1:y_2]$. Here, we present the first few secondary
invariants
\begin{table}[t]
  \centering
  \begin{tabular}{c|c}
    $\deg(\eta)$ & \# of $\eta$\\
    \hline
    0&1\\
    3&4\\
    4&6\\
    5&6  \\
    6&26  \\
    7&27  \\
    8&27  \\
    9&46  \\
    10&42  \\ 
    11&42  \\     
    12&26 \\    
    13&27  \\    
    14&27  \\    
    15&4  \\  
    16&6  \\   
    17&6  \\      
    18&1 \\      
  \end{tabular}
  \hspace{2cm}
  \begin{tabular}{c|c}
    $\deg(\eta)$ & \# of $\eta$ \\  
    \hline
    $(0,0,0)$ & 1\\
    $(0,1,0)$ & 1 \\
    $(0,2,0)$ & 1 \\
    $(3,0,0)$ & 2 \\
    $(0,3,0)$ & 2 \\
    $(0,0,3)$ & 2 \\
    $(2,0,1)$ & 2 \\
    $(1,0,2)$ & 2 \\
    $(0,4,0)$ & 2 \\
    $(3,1,0)$ & 3 \\
    $(1,1,2)$ & 4 \\
    $(2,1,1)$ & 4 \\
    $(0,1,3)$ & 3 \\
    $\vdots$ & $\vdots$ 
  \end{tabular}
  \caption{Degrees of the $324$ secondary invariants
    $\eta_1,\dots,\eta_{324}$. On the left, we list the number of
    secondary invariants by total degree. On the right, we list
    some of invariants by their three individual $(a_1,b,a_2)$-degrees.}
  \label{tab:Z3Z3secondaryinvariants}
\end{table}   
\begin{subequations}
  \begin{gather}
    \begin{aligned}
      \eta_1 \eqdef \;1 ,\\
    \end{aligned}
    \displaybreak[2] \\[1ex]
    \begin{aligned}
      \eta_2 \eqdef&\; x_2y_0y_1+x_1y_0y_2+x_0y_1y_2 ,&
      \eta_3 \eqdef&\; x_2y_0y_1+x_1y_0y_2+x_0y_1y_2 ,\\
      \eta_4 \eqdef&\; x_0y_0^2+x_1y_1^2+x_2y_2^2 ,&
      \eta_5 \eqdef&\; x_1x_2y_0+x_0x_2y_1+x_0x_1y_2 ,\\
      \eta_6 \eqdef&\; x_0^2y_0+x_1^2y_1+x_2^2y_2 ,&
      \eta_7 \eqdef&\; t_1y_0y_1^2+t_1y_0^2y_2+t_1y_1y_2^2,\\
      \eta_8 \eqdef&\; x_1t_1y_0y_1+x_0t_1y_0y_2+x_2t_1y_1y_2,&
      \eta_9 \eqdef&\; x_2t_1y_0^2+x_0t_1y_1^2+x_1t_1y_2^2,\\
      \eta_{10} \eqdef&\; x_0x_2t_1y_0+x_0x_1t_1y_1+x_1x_2t_1y_2,&
      \eta_{11} \eqdef&\; x_1^2t_1y_0+x_2^2t_1y_1+x_0^2t_1y_2,\\
      \eta_{12} \eqdef&\; x_0x_1^2t_1+x_0^2x_2t_1+x_1x_2^2t_1,&
      \eta_{13} \eqdef&\; t_1^2y_0^2y_1+t_1^2y_1^2y_2+t_1^2y_0y_2^2,\\
      \eta_{14} \eqdef&\; x_1t_1^2y_0^2+x_2t_1^2y_1^2+x_0t_1^2y_2^2,&
      \eta_{15} \eqdef&\;
      x_0t_1^2y_0y_1+x_2t_1^2y_0y_2+x_1t_1^2y_1y_2,\\
      \vdots& & \vdots
      .
    \end{aligned}
  \end{gather}
\end{subequations}
We list the number of secondary invariants for a given degree in
\autoref{tab:Z3Z3secondaryinvariants}.
Thus we obtain the following Hironaka decomposition for the ring of
$\Gamma$-invariant polynomials,
\begin{equation}
  \label{eq:schoenhir}
  \C[x_0,x_1,x_2,t_0,t_1,y_0,y_1,y_2]^{\Gamma}
  =
  \bigoplus_{i=1}^{324}
  \eta_i
  \C[ \theta_1,\dots,\theta_8 ] 
  .
\end{equation}
Finally, we need to restrict the invariant ring
eq.~\eqref{eq:schoenhir} to the complete intersection threefold
$\Xt$. In other words, one must mod out the invariant ideal
\begin{equation}
  \big\langle \Pt,\Rt\big\rangle^\Gamma
  = 
  \big\langle \Pt,\Rt\big\rangle
  \cap
  \C[x_0,x_1,x_2,t_0,t_1,y_0,y_1,y_2]^{\Gamma}
\end{equation}
generated by the complete intersection equations $\Pt=0=\Rt$.

Since $\Pt$ is invariant, the ideal generated by $\Pt$ is just the
invariant ring multiplied by $\Pt$,
\begin{equation} 
  \label{eq:PidealG}
  \big\langle \Pt \big\rangle^\Gamma
  =
  \bigoplus_{i=1}^{324}
  \Pt
  \,
  \eta_i
  \,
  \C[ \theta_1,\dots,\theta_8 ] 
  .
\end{equation}
However, the ideal generated by $\Rt$ is not as simple. From
eqns.~\eqref{eq:RtXform} and~\eqref{eq:characters} we see that $\Rt$
transforms like the character $\chi_1$. Thus the elements of the
invariant ring that are divisible by $\Rt$ must also be divisible by a
$\chi_1^2$-transforming polynomial (like $t_1^2$, for example). One
can generalize the Molien formula eq.~\eqref{eq:schmolien} to count
these ``covariant'' polynomials transforming like
$\chi_1^2$~\cite{MR1709907}, namely
\begin{multline}
  \label{eq:modmolchi}
  P \Big( \C[x_0,x_1,x_2,t_0,t_1,y_0,y_1,y_2]^{\chi_1^2} ,\tau \Big)
  =
  \frac{1}{|\Gamma|}
  \sum_{\gamma\in \Gamma}
  \frac{\chi_1(\gamma)^2}{\det(1-\gamma\chi_1(\gamma)^2 \tau)}  
  \\
  =
 \tau^2+7\tau^3+13\tau^4+22\tau^5+79\tau^6+136\tau^7+199\tau^8+478\tau^9\dots
\end{multline}  
Choosing the same primary invariants as previously, eq.\eqref{eq:hopsschoen},
one can rewrite eq.\eqref{eq:modmolchi} as
 \begin{multline}
  \label{eq:modmol}
  P \Big( \C[x_0,x_1,x_2,t_0,t_1,y_0,y_1,y_2]^{\chi_1^2} ,\tau \Big)
  =\\
  \frac{1}{(1-\tau)(1-\tau^3)^5(1-\tau^6)^2} 
  \Big( 
  \tau^2 + 6\tau^3 + 6\tau^4 + 4\tau^5 
  + 27 \tau^6 + 27 \tau^7 + 26 \tau^8 +
  42\tau^9 
  +\\
  + 42 \tau^{10} 
  + 46\tau^{11} 
  + 27 \tau^{12} + 27 \tau^{13} + 26 \tau^{14} +
  6\tau^{15} + 6\tau^{16} + 4\tau^{17} + \tau^{20}
  \Big)
  .
\end{multline}  
Summing the coefficients in the numerator, we see that we again get the 
same number $(=324)$ of secondary $\chi_1^2$-covariant generators. 
This is expected since we are using the same primary invariants. 
The first few secondary $\chi_1^2$-covariants are:
\begin{subequations}
  \begin{gather}
    \begin{aligned}
      \eta_1^{\chi_1^2} \eqdef  t_1^2 ,\\
    \end{aligned}
    \displaybreak[2] \\[1ex]
    \begin{aligned}
      \eta_2^{\chi_1^2} \eqdef&\; x_0^2x_2+x_0x_1^2+x1x_2^2,&
      \eta_3^{\chi_1^2} \eqdef&\;  y_1y_2^2+y_0y_1^2+y_0^2y_2,\\
      \eta_4^{\chi_1^2} \eqdef&\; x_2y_1y_2+x_0y_0y_2+x_1y_0y_1,&
      \eta_5^{\chi_1^2} \eqdef&\;  x_2y_0^2+x_0y_1^2+x_1y_2^2,\\
      \eta_6^{\chi_1^2} \eqdef&\;  x_0^2y_2+x_2^2y_1+x_1^2y_0,&
      \eta_7^{\chi_1^2} \eqdef&\; x_0x_1y_1+x_0x_2y_0+x_1x_2y_2,\\
      \eta_8^{\chi_1^2} \eqdef&\;  y_0^2y_1+y_1^2y_2+y_0y_2^2t_1,&
      \eta_9^{\chi_1^2} \eqdef&\;  x_0x_1t_1y_0+x_0x_2t_1y_2+x_1x_2t_1y_1,\\
      \eta_{10}^{\chi_1^2} \eqdef&\;  x_1t_1y_0^2+x_2t_1y_1^2+x_0t_1y_2^2,&
      \eta_{11}^{\chi_1^2} \eqdef&\;  x_2t_1y_0y_2+x_0t_1y_0y_1+x_1t_1y_1y_2,\\
      \eta_{12}^{\chi_1^2} \eqdef&\;  x_0x_2^2t_1+x_0^2x_1t_1+x_1^2x_2t_1,&
      \eta_{13}^{\chi_1^2} \eqdef&\;  x_1^2t_1y_2+x_2^2t_1y_0+x_0^2t_1y_1,\\
      \eta_{14}^{\chi_1^2} \eqdef&\;  x_2t_1^2y_2^2+x_0t_1^2y_0^2+x_1t_1^2y_1^2,&
      \eta_{15}^{\chi_1^2} \eqdef&\;
      x_2t_1^2y_0y_1+x_0t_1^2y_1y_2+x_1t_1^2y_0y_2,\\
      \vdots& & \vdots.&
    \end{aligned}
  \end{gather}
\end{subequations}
Hence, the space of $\chi_1^2$-covariant polynomials, that is,
transforming like $\chi_1^2$, is given by the ``equivariant Hironaka
decomposition''~\cite{MR1709907} (compare with eq.~\eqref{eq:schoenhir})
\begin{equation}
  \C[ x_0,x_1,x_2,t_0,t_1,y_0,y_1,y_2 ]^{\chi_1^2}
  = 
  \bigoplus_{i=1}^{324}
  \eta_i^{\chi_1^2}
  \,
  \C[ \theta_1,\dots,\theta_8 ]   
  .
\end{equation}
To summarize, even though $\Rt$ is not invariant, it generates an
ideal which contains $\Gamma$-invariant polynomials. Using the above
generalization of the Hironaka decomposition, a basis for these
invariants is
\begin{equation} 
  \label{eq:RidealG}
  \big\langle \Rt \big\rangle^\Gamma
  =
  \bigoplus_{i=1}^{324}
  \Rt
  \,
  \eta_i^{\chi_1^2}
  \,
  \C\left[ \theta_1,\dots , \theta_8\right] 
  .
\end{equation}

\subsection{Quotient Ring}
\label{sec:ZZZquotient}

\begin{table}[htbp]
  \centering
  \begin{tabular}{c|cc}
    $(a_1,b,a_2)$ &  $\hat{N}^\Gamma$ &  $N^\Gamma$ 
    \\ \hline
        (2,1,1)	&   4	&   4	 \\
        (2,2,1)	&   6	&   6	 \\
        (2,3,1)	&   8	&   8	 \\
        (2,4,1)	&  10	&  10	 \\
        (2,5,1)	&  12	&  12	 \\
        (2,6,1)	&  14	&  14	 \\
        (2,7,1)	&  16	&  16	 \\
        (2,8,1)	&  18	&  18	 \\
        (2,9,1)	&  20	&  20	 \\
       (2,10,1)	&  22	&  22	 \\
       (2,11,1)	&  24	&  24	 \\
       (2,12,1)	&  26	&  26	 \\
       (2,13,1)	&  28	&  28	 \\
       (2,14,1)	&  30	&  30	 \\
       (2,15,1)	&  32	&  32	 \\
       (2,16,1)	&  34	&  34	 \\
       (2,17,1)	&  36	&  36	 \\
       (2,18,1)	&  38	&  38	 \\
       (2,19,1)	&  40	&  40	 \\
       (2,20,1)	&  42	&  42	 \\
       (2,21,1)	&  44	&  44	 \\
       (2,22,1)	&  46	&  46	 \\
       (2,23,1)	&  48	&  48	 \\
       (2,24,1)	&  50	&  50	 \\
       (2,25,1)	&  52	&  52	 \\
       (2,26,1)	&  54	&  54	 \\
  \end{tabular}
  \hfill
  \begin{tabular}{c|cc}
    $(a_1,b,a_2)$ &  $\hat{N}^\Gamma$ &  $N^\Gamma$ 
    \\ \hline
       (2,27,1)	&  56	&  56	 \\
       (2,28,1)	&  58	&  58	 \\
       (2,29,1)	&  60	&  60	 \\
       (2,30,1)	&  62	&  62	 \\
       (2,31,1)	&  64	&  64	 \\
       (2,32,1)	&  66	&  66	 \\
       (2,33,1)	&  68	&  68	 \\
       (2,34,1)	&  70	&  70	 \\
        (3,1,3)	&  23	&  20	 \\
        (3,2,3)	&  34	&  29	 \\
        (3,3,3)	&  46	&  38	 \\
        (3,4,3)	&  57	&  47	 \\
        (3,5,3)	&  68	&  56	 \\
        (3,6,3)	&  80	&  65	 \\
        (4,1,2)	&  20	&  18	 \\
        (4,2,2)	&  30	&  26	 \\
        (4,3,2)	&  40	&  34	 \\
        (4,4,2)	&  50	&  42	 \\
        (4,5,2)	&  60	&  50	 \\
        (4,6,2)	&  70	&  58	 \\
        (4,7,2)	&  80	&  66	 \\
        (5,1,1)	&  14	&  12	 \\
        (5,2,1)	&  21	&  17	 \\
        (5,3,1)	&  28	&  22	 \\
        (5,4,1)	&  35	&  27	 \\
        (5,5,1)	&  42	&  32	 \\
  \end{tabular}
  \hfill
  \begin{tabular}{c|cc}
    $(a_1,b,a_2)$ &  $\hat{N}^\Gamma$ &  $N^\Gamma$ 
    \\ \hline
        (5,6,1)	&  49	&  37	 \\
        (5,7,1)	&  56	&  42	 \\
        (5,8,1)	&  63	&  47	 \\
        (5,9,1)	&  70	&  52	 \\
       (5,10,1)	&  77	&  57	 \\
       (5,11,1)	&  84	&  62	 \\
       (5,12,1)	&  91	&  67	 \\
        (5,1,4)	&  70	&  53	 \\
        (6,1,3)	&  63	&  48	 \\
        (6,2,3)	&  94	&  66	 \\
        (7,1,2)	&  48	&  38	 \\
        (7,2,2)	&  72	&  52	 \\
        (7,3,2)	&  96	&  66	 \\
        (8,1,1)	&  30	&  23	 \\
        (8,2,1)	&  45	&  31	 \\
        (8,3,1)	&  60	&  39	 \\
        (8,4,1)	&  75	&  47	 \\
        (8,5,1)	&  90	&  55	 \\
        (8,6,1)	& 105	&  63	 \\
       (10,1,2)	&  88	&  64	 \\
       (11,1,1)	&  52	&  37	 \\
       (11,2,1)	&  78	&  48	 \\
       (11,3,1)	& 104	&  59	 \\
       (11,4,1)	& 130	&  70	 \\
       (14,1,1)	&  80	&  54	 \\
       (14,2,1)	& 120	&  68	 \\
  \end{tabular}
  \caption{All homogeneous degrees leading to
    few ($\leq 70$) invariant sections 
    $N^\Gamma = N^\Gamma_{(a_1,b,a_2)}$ on $\Xt$. For
    comparison, we also list the number
    $\hat{N}^\Gamma 
    = \hat{N}^\Gamma_{(a_1,b,a_2)} 
    = \dim \C[\vec{x},\vec{t},\vec{y}]^\Gamma_{(a_1,b,a_2)}$
    of invariant
    polynomials before quotienting out the relations generated by the
    complete intersection equations $\Pt=0=\Rt$.}
  \label{tab:Z3Z3NumOfSections}
\end{table}
By the results of the previous section, we know for any fixed
multi-degree $(a_1,b,a_2)$:
\begin{itemize}
\item A (finite) basis for the $\Gamma$-invariant polynomials
  \begin{equation}
    I \eqdef
    \C\big[\vec{x},\vec{t},\vec{y}\big]^\Gamma_{(a_1,b,a_2)}
    .
  \end{equation}
  In particular, the polynomials are linearly independent of each
  other.
\item Generators for the $\Gamma$-invariant ideal generated by the
  complete intersection eqns.~\eqref{eq:P}
  and~\eqref{eq:R},
  \begin{equation}
    \begin{split}
      J
      \eqdef&~ 
      \big\langle \Pt, \Rt \big\rangle^\Gamma_{(a_1,b,a_2)}
      =
      \big\langle \Pt \big\rangle^\Gamma_{(a_1,b,a_2)} 
      + 
      \big\langle \Rt \big\rangle^\Gamma_{(a_1,b,a_2)}
      \\
      =&~      
      \left\langle
        \Pt \cdot 
        \C\big[\vec{x},\vec{t},\vec{y}\big]^\Gamma_{(a_1-3,b-1,a_2)}
        ,~
        \Rt \cdot 
        \C\big[\vec{x},\vec{t},\vec{y}\big]^{\chi_2^2}_{(a_1,b-1,a_2-3)}
      \right\rangle_{(a_1,b,a_2)}
      .
    \end{split}
  \end{equation}
  The generating polynomials of $J$ are not automatically linearly
  independent.
\end{itemize}
It remains to find a basis for the quotient 
\begin{equation}
  \left(
  \C\big[\vec{x},\vec{t},\vec{y}\big]
  \Big/ 
  \big\langle \Pt, \Rt\big\rangle
  \right)^\Gamma_{(a_1,b,a_2)}
  =
  \C\big[\vec{x},\vec{t},\vec{y}\big]^\Gamma_{(a_1,b,a_2)}
  \Big/ 
  \big\langle \Pt, \Rt \big\rangle^\Gamma_{(a_1,b,a_2)}
  = I / J
  ,
\end{equation}
corresponding to the restriction of the invariant sections on
$\CP^2\times\CP^1\times\CP^2$ to the complete intersection $\Xt$.
This is technically more difficult than the previous quotients, where
we were able to use Gr\"obner bases or pick suitable primary
invariants to find the quotient. Here, we will resort to a numerical
computation of the quotient. To do this, note that the ideal elements
$J$ are linear combinations of invariants $I$. Hence, thinking of $I$,
$J$ as column vectors, there is a matrix
\begin{equation}
  M\in\Mat_{|J|\times |I|}(\C):
  \qquad
  M I = J
  .
\end{equation}
The kernel of $M$ is a basis for the quotient $I/J$. Of course, due to
floating-point precision limits, there are generally no exact
null-vectors. However, the singular value decomposition~\cite{laug} is
a well-behaved numerical algorithm to compute an orthonormal basis for
the kernel. In \autoref{tab:Z3Z3NumOfSections} we list the dimension
\begin{equation}
  N^\Gamma_{(a_1,b,a_2)} = \dim_\C\big( I/J \big)
\end{equation}
of the quotient space for various multi-degrees $(a_1,b,a_2)$.

\subsection{Results}
\label{sec:ZZZplots}

\begin{figure}[htbp]
  \centering
  \include{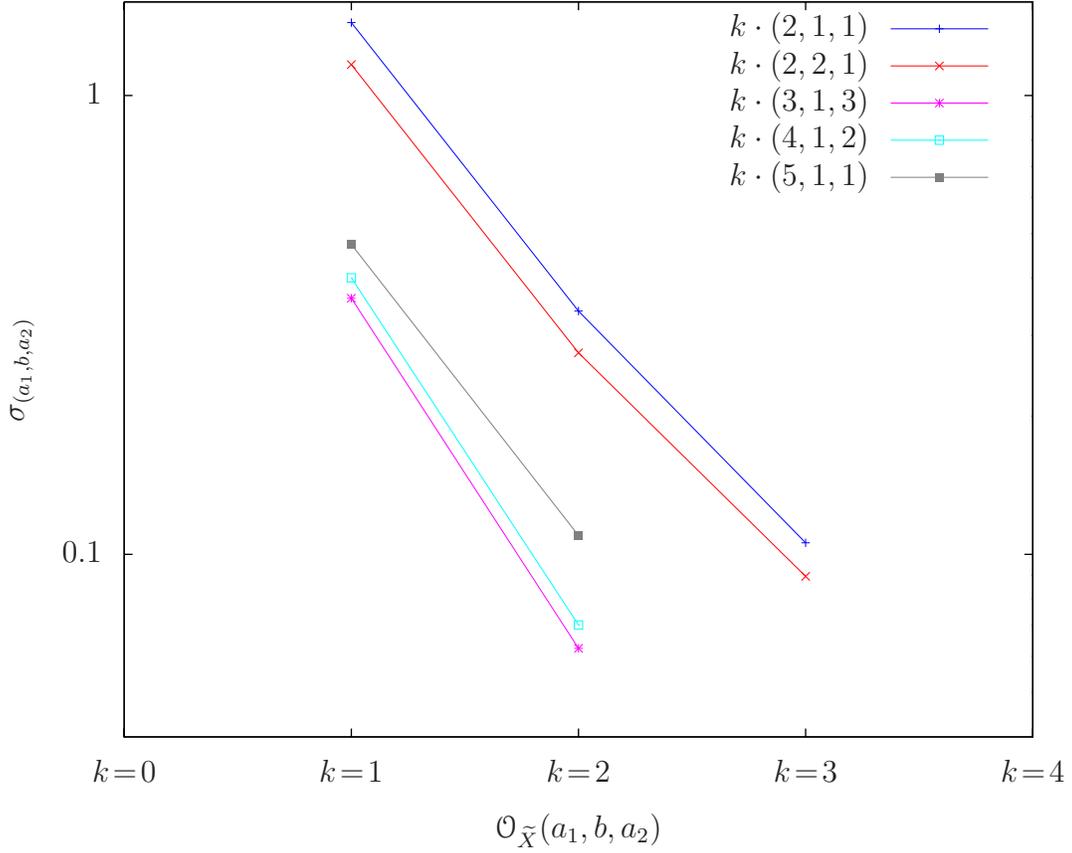}
  \caption{The error measure $\sigma_{(a_1,b,a_2)}(X)$ for the metric on
    the $\ZZZ$-quotient $X$, computed for different \Kahler{} moduli
    but common complex structure $\lambda_1=\lambda_2=0$,
    $\lambda_3=1$. Note that we chose $k=\gcd(a_1,b,a_2)$ as the
    independent variable, and stopped increasing $k$ as soon as
    $N^\Gamma$ exceeded $200$. In each case we iterated the T-operator
    $5$ times, numerically integrating using $\Npoints=\comma{50000}$
    points. Then we evaluated $\sigma_{(a_1,b,a_2)}(X)$ using
    \comma{5000} different test points.}
  \label{fig:sigma_k_Z3Z3}
\end{figure}
We implemented Donaldson's algorithm to compute the Calabi-Yau metric
on the threefold $X=\Xt \big/ (\ZZZ)$. As discussed earlier, the
convergence of the balanced metrics is essentially independent of the
complex structure. Hence, we will consider an explicit example where
$\lambda_1=\lambda_2=0$, $\lambda_3=1$. In \autoref{fig:sigma_k_Z3Z3}
we demonstrate that the numerical metric indeed approximates the
Calabi-Yau metric, as it should.

\begin{figure}[htbp]
  \centering
  \include{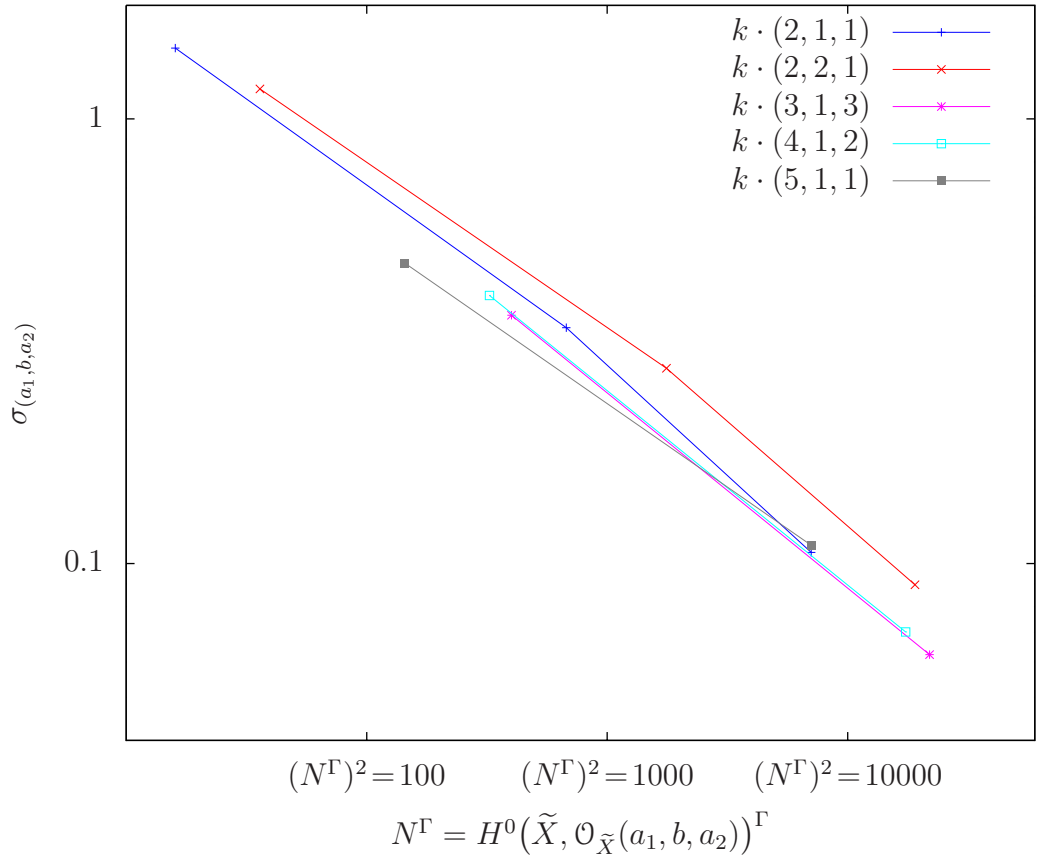}
  \caption{The same data as in \autoref{fig:sigma_k_Z3Z3}, but plotted
    as a function of the number of free parameters
    $(N^\Gamma_{(a_1,b,a_2)})^2$ in the ansatz for the \Kahler{}
    potential.}
  \label{fig:sigma_N2_Z3Z3}
\end{figure}
In contrast to the quintic, where the single \Kahler{} modulus is the
overall volume, the Schoen quotient threefold $X$ has a
$h^{1,1}(X)=3$-dimensional \Kahler{} moduli space, see
eq.~\eqref{eq:SchoenZ3Z3Hodge}.  The \Kahler{} moduli are determined
through the three independent degrees $(a_1,b,a_2)$. Note that the
integer $k=\gcd(a_1,b,a_2)$ in \autoref{fig:sigma_k_Z3Z3} serves only
to measure the refinements along a ray in the \Kahler{} moduli
space. In order to properly compare the metric convergence for
different rays in the \Kahler{} moduli space, we should consider
$(N^\Gamma_{(a_1,b,a_2)})^2$, which is the number of free parameters
in the ansatz for the \Kahler{} potential and, hence, measures the
numerical complexity of the whole algorithm. We do this in
\autoref{fig:sigma_N2_Z3Z3}, and see that the accuracy is essentially
determined by $(N^\Gamma_{(a_1,b,a_2)})^2$, and depends only slightly
on the details of the \Kahler{} moduli.

Finally, we note again that $\sigma_{(a_1,b,a_2)}(X)$ is also the error
measure for the metric pulled back to the covering space $\Xt$ of
$X$. It is useful to compare this result with the convergence of the
Calabi-Yau metric on $\Xt$ obtained directly as discussed in
\autoref{sec:Schoen}. We have numerically performed this comparison
and obtained results similar to those found in the quintic case, see
\autoref{fig:MultipleQuintics}. That is, when measured by the
numerical effort involved, the $\Z_3\times\Z_3$ symmetric method of
this section is far more efficient.

\section*{Acknowledgments}

This research was supported in part by the U.~S.~Department of Energy
grant DE-FG02-96ER40959, by the Department of Physics and the
Math/Physics Research Group at the University of Pennsylvania under
cooperative research agreement DE-FG02-95ER40893 with the
U.~S.~Department of Energy, and an NSF Focused Research Grant
DMS0139799 for ``The Geometry of Superstrings''.

\appendix
\makeatletter
\def\Hy@chapterstring{section}
\makeatother

\section{Primary Invariants}
\label{sec:thetaQuintic}

In this appendix, we check that the invariants in
eq.~\eqref{eq:thetadef} can be chosen to be the primary invariants,
that is, form a ``homogeneous system of parameters''. In fact, the
following criteria are equivalent, see~\cite{Sem:MR1690810}
Proposition 2.3:
\begin{itemize}
\item $\{\theta_1,\theta_2,\theta_3,\theta_4,\theta_5\}$ are a
  homogeneous system of parameters (h.s.o.p.).
\item $\dim \Big(\C[z_0,z_1,z_2,z_3,z_4]\big/\left<
    \theta_1,\theta_2,\theta_3,\theta_4,\theta_5 \right> \Big)= 0$
\item The only common solution to $\theta_i=0$, $i=1,\dots,5$
  is $z_0=z_1=z_2=z_3=z_4=0$.
\end{itemize}
Using \textsc{Singular}~\cite{GPS05}, we can test the dimension
criterion easily:
\begin{Verbatim}[fontsize=\small,frame=single,rulecolor=\color{red}]
                       SINGULAR                           /
 A Computer Algebra System for Polynomial Computations   /   version 3-0-1
                                                       0<
     by: G.-M. Greuel, G. Pfister, H. Schoenemann        \   October 2005
FB Mathematik der Universitaet, D-67653 Kaiserslautern    \
> ring r=0,(z0,z1,z2,z3,z4),dp;
> poly t1=z0*z1*z2*z3*z4;
> poly t2=z0^3*z1*z4+z0*z1^3*z2+z0*z3*z4^3+z1*z2^3*z3+z2*z3^3*z4;
> poly t3=z0^5+z1^5+z2^5+z3^5+z4^5;
> poly t4=z0^10+z1^10+z2^10+z3^10+z4^10;
> poly t5=z0^8*z2*z3+z0*z1*z3^8+z0*z2^8*z4+z1^8*z3*z4+z1*z2*z4^8;
> ideal i=t1,t2,t3,t4,t5;
> dim(std(i));
0
\end{Verbatim}
Hence, eq.~\eqref{eq:thetadef} is indeed a h.s.o.p.

\section{Implementation Details}
\label{sec:QuinticImplementation}

\subsection{Tensors}
\label{sec:blitz}

We use the \blitzpp~\cite{Blitz} library for all tensor computations. For
example, here is the ordinary (serial) computation of the T-operator:
\medskip

{\ttfamily \raggedright \small
001 template<{}\textbf{typename}\ Mfd>{}\\
002 Array<{}COMPLEX,2>{}\ Metric<{}Mfd>{}::Toperator\underline\ integrand\\
003 (\textbf{const}\ Point\ \&p)\ \textbf{const}\\
003 \{\\
004 \ \ using\ \textbf{namespace}\ blitz;\\
005 \ \\
006 \ \ Array<{}COMPLEX,1>{}\ s\underline\ val(N),\ sbar\underline\ val(N);\\
007 \ \ s\underline\ val\ =\ X.SectionsAt(p);\\
008 \ \ sbar\underline\ val\ =\ conj(s\underline\ val);\\
009 \ \\
010 \ \ COMPLEX\ D\ =\ s\underline\ h\underline\ sbar(s\underline\ val,\ sbar\underline\ val);\ \textsl{//\ =\ ||s||\underline\ h\textasciicircum 2}\\
011 \ \\
012 \ \ Array<{}COMPLEX,2>{}\ result(N,\ N);\\
013 \ \ firstIndex\ a;\ \ secondIndex\ b;\\
014 \ \ result\ =\ s\underline\ val(a)$\ast$sbar\underline\ val(b)/D;\\
015 \ \\
016 \ \ \textbf{return}(result);\\
017 \}\\
018 \ \\
019 \textbf{template}<{}\textbf{typename}\ Mfd>{}\\
020 Metric<{}Mfd>{}\ Metric<{}Mfd>{}::Toperator\underline\ slow()\ \textbf{const}\\
020 \{\\
021 \ \ Array<{}COMPLEX,2>{}\ tmp(N,N);\ tmp\ =\ 0;\\
022 \ \ \textbf{for}\ (\textbf{typename}\ Manifold::\textbf{const}\underline\ iterator\ p=X.begin();\ p!=X.end();\ p++)\ \\
023 \ \ \ \ tmp\ +=\ Toperator\underline\ integrand($\ast$p)\ $\ast$\ p-{}>{}weight;\\
024 \ \ tmp\ $\ast$=\ N\ /\ X.Volume();\\
025 \ \\
026 \ \ Metric<{}Manifold>{}\ result(X);\ \ result.h\underline\ ab\ =\ tmp;\\
027 \ \ result.compute\underline\ hinv();\ \textsl{//\ compute\ h\textasciicircum ab}\\
028 \ \ \textbf{return}(result);\\
029 \}
}
\normalfont\normalsize

\subsection{MPI}
\label{sec:MPI}

In order to speed up computations we use a cluster of ordinary PCs,
consisting of 11 machines connected via gigabit Ethernet. Each node
has one 2.2 GHz dual-core Opteron processor and 2 GiB of RAM.

The main task in computing the metrics is to compute the T-operator,
see eq.~\eqref{eq:T-operator}. Performing the numerical integration is
embarrassingly parallel, and does not even need any high-speed network
connection. For the stochastic integration one has to
\begin{itemize}
\item Compute the (weighted) integrand of the T-operator at each
  point, and 
\item Sum the resulting matrix. 
\end{itemize}
We solve this by a bag-of-tasks, where each node adds up the
contribution of a few points and then asks for another work set. In
the end, the partial sum computed at each node is reduced to the
master node. To effectively write distributed programs on the cluster,
we make use of the MPI standard implemented by
OpenMPI~\cite{gabriel04:_open_mpi}. For example, here is the parallel
implementation of the T-operator: \medskip

{\ttfamily \raggedright \small
001 template<{}\textbf{typename}\ Mfd>{}\\
002 \textbf{void}\ Metric<{}Mfd>{}::Toperator\underline\ slave(\textbf{int})\\
003 \{\\
004 \ \ WorkQueue<{}\textbf{typename}\ Mfd::\textbf{const}\underline\ iterator>{}\ work(X.begin(),\ X.end());\\
005 \ \ Array<{}COMPLEX,2>{}\ tmp(N,N);\ tmp\ =\ 0;\\
006 \ \ \textbf{while}\ (work.ReceiveMoreWork())\ \{\\
007 \ \ \ \ \textbf{for}\ (\textbf{typename}\ Mfd::\textbf{const}\underline\ iterator\ \\
008 \ \ \ \ \ p=work.begin();\ p!=work.end();\ p++)\\
009 \ \ \ \ \ \ tmp\ +=\ Toperator\underline\ integrand($\ast$p)\ $\ast$\ p-{}>{}weight;\\
010 \ \ \}\\
011 \ \ Cluster::SendSummandArray(tmp);\\
012 \}\\
013 \ \\
014 \textbf{template}<{}\textbf{typename}\ Mfd>{}\\
015 Metric<{}Mfd>{}\ Metric<{}Mfd>{}::Toperator()\ \textbf{const}\\
015 \{\\
016 \ \ ClusterExecMethod<{}\ Metric<{}Mfd>{},\ \&Metric<{}Mfd>{}::Toperator\underline\ slave\ >{}\\
017 \ \ \ \ ().Run($\ast$\textbf{this});\\
018 \ \\
019 \ \ WorkQueue<{}\textbf{typename}\ Mfd::\textbf{const}\underline\ iterator>{}\ work(X.begin(),\ X.end(),20);\\
020 \ \ work.Finish();\ \textsl{//\ main\ loop}\\
021 \ \\
022 \ \ Array<{}COMPLEX,2>{}\ tmp(N,N);\ \ tmp\ =\ 0;\\
023 \ \ Cluster::ReceiveSumArray(tmp);\\
024 \ \ tmp\ $\ast$=\ N\ /\ X.Volume();\\
025 \ \\
026 \ \ Metric<{}Manifold>{}\ result(X);\\
027 \ \ result.h\underline\ ab\ =\ tmp;\\
028 \ \ result.compute\underline\ hinv();\\
029 \ \ \textbf{return}\ result;\\
030 \}
}
\normalfont\normalsize

\subsection{Multivariate Polynomials}
\label{sec:MultivariatePolynomial}

Every section is, at the end of the day, some multivariate
polynomial. For that reason, we implemented a \CC{} library for sparse
multivariate polynomials. In addition to the usual arithmetic
operations, it supports differentiation and can copy polynomials to
remote nodes via MPI. Using this library, we can easily work with
arbitrary polynomials. For example, the program to compute the metric
on the Fermat quintic can, without change, also work with generic
quintics that are a non-trivial sum over all $126$ monomials, see
\autoref{fig:MultipleQuintics}.

\bibliographystyle{utcaps} \renewcommand{\refname}{Bibliography}
\addcontentsline{toc}{section}{Bibliography} 

\bibliography{Volker,Metric}

\providecommand{\href}[2]{#2}\begingroup\raggedright\begin{thebibliography}{10}

\bibitem{Candelas:1985en}
P.~Candelas, G.~T. Horowitz, A.~Strominger, and E.~Witten, ``Vacuum
  Configurations for Superstrings,'' {\em Nucl. Phys.} {\bf B258} (1985)
46--74.

\bibitem{Ovrut:2003zj}
B.~A. Ovrut, T.~Pantev, and R.~Reinbacher, ``Invariant homology on standard
  model manifolds,'' {\em JHEP} {\bf 01} (2004) 059,
\href{http://arXiv.org/abs/hep-th/0303020}{{\tt hep-th/0303020}}.

\bibitem{Buchbinder:2002pr}
E.~I. Buchbinder, R.~Donagi, and B.~A. Ovrut, ``Vector bundle moduli
  superpotentials in heterotic superstrings and M-theory,'' {\em JHEP} {\bf 07}
  (2002) 066,
\href{http://arXiv.org/abs/hep-th/0206203}{{\tt hep-th/0206203}}.

\bibitem{Donagi:2000zf}
R.~Donagi, B.~A. Ovrut, T.~Pantev, and D.~Waldram, ``Standard-model bundles on
  non-simply connected Calabi-Yau threefolds,'' {\em JHEP} {\bf 08} (2001) 053,
\href{http://arXiv.org/abs/hep-th/0008008}{{\tt hep-th/0008008}}.

\bibitem{Donagi:2000fw}
R.~Donagi, B.~A. Ovrut, T.~Pantev, and D.~Waldram, ``Spectral involutions on
  rational elliptic surfaces,'' {\em Adv. Theor. Math. Phys.} {\bf 5} (2002)
  499--561,
\href{http://arXiv.org/abs/math/0008011}{{\tt math/0008011}}.

\bibitem{Donagi:2004ia}
R.~Donagi, Y.-H. He, B.~A. Ovrut, and R.~Reinbacher, ``The particle spectrum of
  heterotic compactifications,'' {\em JHEP} {\bf 12} (2004) 054,
\href{http://arXiv.org/abs/hep-th/0405014}{{\tt hep-th/0405014}}.

\bibitem{Donagi:2004qk}
R.~Donagi, Y.-H. He, B.~A. Ovrut, and R.~Reinbacher, ``Moduli dependent spectra
  of heterotic compactifications,'' {\em Phys. Lett.} {\bf B598} (2004)
  279--284,
\href{http://arXiv.org/abs/hep-th/0403291}{{\tt hep-th/0403291}}.

\bibitem{Donagi:2004ub}
R.~Donagi, Y.-H. He, B.~A. Ovrut, and R.~Reinbacher, ``The spectra of heterotic
  standard model vacua,'' {\em JHEP} {\bf 06} (2005) 070,
\href{http://arXiv.org/abs/hep-th/0411156}{{\tt hep-th/0411156}}.

\bibitem{Donagi:2004su}
R.~Donagi, Y.-H. He, B.~A. Ovrut, and R.~Reinbacher, ``Higgs doublets, split
  multiplets and heterotic SU(3)C x SU(2)L x U(1)Y spectra,'' {\em Phys. Lett.}
  {\bf B618} (2005) 259--264,
\href{http://arXiv.org/abs/hep-th/0409291}{{\tt hep-th/0409291}}.

\bibitem{Braun:2005fk}
V.~Braun, Y.-H. He, B.~A. Ovrut, and T.~Pantev, ``Heterotic standard model
  moduli,'' {\em JHEP} {\bf 01} (2006) 025,
\href{http://arXiv.org/abs/hep-th/0509051}{{\tt hep-th/0509051}}.

\bibitem{Braun:2005bw}
V.~Braun, Y.-H. He, B.~A. Ovrut, and T.~Pantev, ``A standard model from the
  E(8) x E(8) heterotic superstring,'' {\em JHEP} {\bf 06} (2005) 039,
\href{http://arXiv.org/abs/hep-th/0502155}{{\tt hep-th/0502155}}.

\bibitem{Braun:2005nv}
V.~Braun, Y.-H. He, B.~A. Ovrut, and T.~Pantev, ``The exact MSSM spectrum from
  string theory,'' {\em JHEP} {\bf 05} (2006) 043,
\href{http://arXiv.org/abs/hep-th/0512177}{{\tt hep-th/0512177}}.

\bibitem{Braun:2005ux}
V.~Braun, Y.-H. He, B.~A. Ovrut, and T.~Pantev, ``A heterotic standard model,''
  {\em Phys. Lett.} {\bf B618} (2005) 252--258,
\href{http://arXiv.org/abs/hep-th/0501070}{{\tt hep-th/0501070}}.

\bibitem{Bouchard:2005ag}
V.~Bouchard and R.~Donagi, ``An SU(5) heterotic standard model,'' {\em Phys.
  Lett.} {\bf B633} (2006) 783--791,
\href{http://arXiv.org/abs/hep-th/0512149}{{\tt hep-th/0512149}}.

\bibitem{Braun:2006me}
V.~Braun, Y.-H. He, and B.~A. Ovrut, ``Yukawa couplings in heterotic standard
  models,'' {\em JHEP} {\bf 04} (2006) 019,
\href{http://arXiv.org/abs/hep-th/0601204}{{\tt hep-th/0601204}}.

\bibitem{Braun:2005xp}
V.~Braun, Y.-H. He, B.~A. Ovrut, and T.~Pantev, ``Moduli dependent mu-terms in
  a heterotic standard model,'' {\em JHEP} {\bf 03} (2006) 006,
\href{http://arXiv.org/abs/hep-th/0510142}{{\tt hep-th/0510142}}.

\bibitem{Candelas:1987rx}
P.~Candelas and S.~Kalara, ``Yukawa couplings for a three generation
  superstring compactification,'' {\em Nucl. Phys.} {\bf B298} (1988)
357.

\bibitem{Candelas:1990rm}
P.~Candelas, X.~C. De~La~Ossa, P.~S. Green, and L.~Parkes, ``A pair of
  Calabi-Yau manifolds as an exactly soluble superconformal theory,'' {\em
  Nucl. Phys.} {\bf B359} (1991)
21--74.

\bibitem{Greene:1993vm}
B.~R. Greene, D.~R. Morrison, and M.~R. Plesser, ``Mirror manifolds in higher
  dimension,'' {\em Commun. Math. Phys.} {\bf 173} (1995) 559--598,
\href{http://arXiv.org/abs/hep-th/9402119}{{\tt hep-th/9402119}}.

\bibitem{Donagi:2006yf}
R.~Donagi, R.~Reinbacher, and S.-T. Yau, ``Yukawa couplings on quintic
  threefolds,''
\href{http://arXiv.org/abs/hep-th/0605203}{{\tt hep-th/0605203}}.

\bibitem{DonaldsonNumerical}
S.~K. Donaldson, ``Some numerical results in complex differential geometry,''
  \href{http://arXiv.org/abs/math.DG/0512625}{{\tt math.DG/0512625}}.

\bibitem{Douglas:2006hz}
M.~R. Douglas, R.~L. Karp, S.~Lukic, and R.~Reinbacher, ``Numerical solution to
  the hermitian Yang-Mills equation on the Fermat quintic,''
\href{http://arXiv.org/abs/hep-th/0606261}{{\tt hep-th/0606261}}.

\bibitem{Douglas:2006rr}
M.~R. Douglas, R.~L. Karp, S.~Lukic, and R.~Reinbacher, ``Numerical Calabi-Yau
  metrics,''
\href{http://arXiv.org/abs/hep-th/0612075}{{\tt hep-th/0612075}}.

\bibitem{MR2161248}
S.~K. Donaldson, ``Scalar curvature and projective embeddings. {II},'' {\em Q.
  J. Math.} {\bf 56} (2005), no.~3, 345--356.

\bibitem{MR1064867}
G.~Tian, ``On a set of polarized {K}\"ahler metrics on algebraic manifolds,''
  {\em J. Differential Geom.} {\bf 32} (1990), no.~1, 99--130.

\bibitem{Headrick:2005ch}
M.~Headrick and T.~Wiseman, ``Numerical Ricci-flat metrics on K3,'' {\em Class.
  Quant. Grav.} {\bf 22} (2005) 4931--4960,
\href{http://arXiv.org/abs/hep-th/0506129}{{\tt hep-th/0506129}}.

\bibitem{Doran:2007zn}
C.~Doran, M.~Headrick, C.~P. Herzog, J.~Kantor, and T.~Wiseman, ``Numerical
  Kaehler-Einstein metric on the third del Pezzo,''
\href{http://arXiv.org/abs/hep-th/0703057}{{\tt hep-th/0703057}}.

\bibitem{MR2154820}
X.~Wang, ``Canonical metrics on stable vector bundles,'' {\em Comm. Anal.
  Geom.} {\bf 13} (2005), no.~2, 253--285.

\bibitem{MR1255980}
B.~Sturmfels, {\em Algorithms in invariant theory}.
\newblock Texts and Monographs in Symbolic Computation. Springer-Verlag,
  Vienna, 1993.

\bibitem{MR1916953}
S.~K. Donaldson, ``Scalar curvature and projective embeddings. {I},'' {\em J.
  Differential Geom.} {\bf 59} (2001), no.~3, 479--522.

\bibitem{Witten:1985xc}
E.~Witten, ``Symmetry Breaking Patterns in Superstring Models,'' {\em Nucl.
  Phys.} {\bf B258} (1985)
75.

\bibitem{Sen:1985eb}
A.~Sen, ``The Heterotic String in Arbitrary Background Field,'' {\em Phys.
  Rev.} {\bf D32} (1985)
2102.

\bibitem{Evans:1986nq}
M.~Evans and B.~A. Ovrut, ``Breaking the superstring vacuum degeneracy,''.
  Invited talk given at 21st Rencontre de Moriond, Les Arcs, France, Mar 9-16,
  1986.

\bibitem{Breit:1985ud}
J.~D. Breit, B.~A. Ovrut, and G.~C. Segre, ``E(6) Symmetry Breaking in the
  Superstring Theory,'' {\em Phys. Lett.} {\bf B158} (1985)
33.

\bibitem{Breit:1985ns}
J.~D. Breit, B.~A. Ovrut, and G.~Segre, ``The one loop effective Lagrangian of
  the superstring,'' {\em Phys. Lett.} {\bf B162} (1985)
303.

\bibitem{Braun:2004xv}
V.~Braun, B.~A. Ovrut, T.~Pantev, and R.~Reinbacher, ``Elliptic Calabi-Yau
  threefolds with Z(3) x Z(3) Wilson lines,'' {\em JHEP} {\bf 12} (2004) 062,
\href{http://arXiv.org/abs/hep-th/0410055}{{\tt hep-th/0410055}}.

\bibitem{Ovrut:2002jk}
B.~A. Ovrut, T.~Pantev, and R.~Reinbacher, ``Torus-fibered Calabi-Yau
  threefolds with non-trivial fundamental group,'' {\em JHEP} {\bf 05} (2003)
  040,
\href{http://arXiv.org/abs/hep-th/0212221}{{\tt hep-th/0212221}}.

\bibitem{Batyrev:2005jc:bat}
V.~Batyrev and M.~Kreuzer, ``Integral Cohomology and Mirror Symmetry for
  Calabi-Yau 3-folds,''
\href{http://arXiv.org/abs/math.AG/0505432}{{\tt math.AG/0505432}}.

\bibitem{Sem:MR1690810}
A.~V. Geramita, ed., {\em The {C}urves {S}eminar at {Q}ueen's. {V}ol. {XII}},
  vol.~114 of {\em Queen's Papers in Pure and Applied Mathematics}.
\newblock Queen's University, Kingston, ON, 1998.
\newblock Papers from the seminar held at Queen's University, Kingston, ON,
  1998.

\bibitem{Green:1987mn}
M.~B. Green, J.~H. Schwarz, and E.~Witten, ``Superstring Theory. Vol. 2: Loop
  Amplitudes, Anomalies and Phenomenology,''. Cambridge, Uk: Univ. Pr. (1987)
  596 P. (Cambridge Monographs On Mathematical Physics).

\bibitem{GPS05}
G.-M. Greuel, G.~Pfister, and H.~Sch\"onemann, ``{\sc Singular} 3.0,'' a
  computer algebra system for polynomial computations, Centre for Computer
  Algebra, University of Kaiserslautern, 2005.
\newblock {\tt http://www.singular.uni-kl.de}.

\bibitem{finvar}
A.~E. Heydtmann, ``\texttt{finvar.lib}. A {\sc Singular} 3.0 library,''
  {Invariant Rings of Finite Groups}, 2005.
\newblock {\tt http://www.singular.uni-kl.de}.

\bibitem{MR923487}
C.~Schoen, ``On fiber products of rational elliptic surfaces with section,''
  {\em Math. Z.} {\bf 197} (1988), no.~2, 177--199.

\bibitem{Braun:2005zv}
V.~Braun, Y.-H. He, B.~A. Ovrut, and T.~Pantev, ``Vector bundle extensions,
  sheaf cohomology, and the heterotic standard model,'' {\em Adv. Theor. Math.
  Phys.} {\bf 10} (2006) 4,
\href{http://arXiv.org/abs/hep-th/0505041}{{\tt hep-th/0505041}}.

\bibitem{Braun:2007xh}
V.~Braun, M.~Kreuzer, B.~A. Ovrut, and E.~Scheidegger, ``Worldsheet instantons
  and torsion curves. Part A: Direct computation,'' {\em JHEP} {\bf 10} (2007)
  022,
\href{http://arXiv.org/abs/hep-th/0703182}{{\tt hep-th/0703182}}.

\bibitem{Braun:2007vy}
V.~Braun, M.~Kreuzer, B.~A. Ovrut, and E.~Scheidegger, ``Worldsheet Instantons
  and Torsion Curves, Part B: Mirror Symmetry,'' {\em JHEP} {\bf 10} (2007)
  023,
\href{http://arXiv.org/abs/arXiv:0704.0449 [hep-th]}{{\tt arXiv:0704.0449
  [hep-th]}}.

\bibitem{Braun:2007tp}
V.~Braun, M.~Kreuzer, B.~A. Ovrut, and E.~Scheidegger, ``Worldsheet Instantons,
  Torsion Curves, and Non-Perturbative Superpotentials,'' {\em Phys. Lett.}
  {\bf B649} (2007) 334--341,
\href{http://arXiv.org/abs/hep-th/0703134}{{\tt hep-th/0703134}}.

\bibitem{Gomez:2005ii}
T.~L. Gomez, S.~Lukic, and I.~Sols, ``Constraining the Kaehler moduli in the
  heterotic standard model,''
\href{http://arXiv.org/abs/hep-th/0512205}{{\tt hep-th/0512205}}.

\bibitem{MR1189133}
D.~Cox, J.~Little, and D.~O'Shea, {\em Ideals, varieties, and algorithms}.
\newblock Undergraduate Texts in Mathematics. Springer-Verlag, New York, 1992.
\newblock An introduction to computational algebraic geometry and commutative
  algebra.

\bibitem{Zelditch:Shif}
B.~Shiffman and S.~Zelditch, ``Distribution of zeros of random and quantum
  chaotic sections of positive line bundles,'' {\em Comm. Math. Phys.} {\bf
  200} (1999), no.~3, 661--683.

\bibitem{Candelas:2007ac}
P.~Candelas, X.~de~la Ossa, Y.-H. He, and B.~Szendroi, ``Triadophilia: A
  Special Corner in the Landscape,''
\href{http://arXiv.org/abs/arXiv:0706.3134 [hep-th]}{{\tt arXiv:0706.3134
  [hep-th]}}.

\bibitem{Feng:2007ur}
B.~Feng, A.~Hanany, and Y.-H. He, ``Counting gauge invariants: The plethystic
  program,'' {\em JHEP} {\bf 03} (2007) 090,
\href{http://arXiv.org/abs/hep-th/0701063}{{\tt hep-th/0701063}}.

\bibitem{MR1709907}
K.~Gatermann and F.~Guyard, ``Gr\"obner bases, invariant theory and equivariant
  dynamics,'' {\em J. Symbolic Comput.} {\bf 28} (1999), no.~1-2, 275--302.
  Polynomial elimination---algorithms and applications.

\bibitem{laug}
E.~Anderson, Z.~Bai, C.~Bischof, S.~Blackford, J.~Demmel, J.~Dongarra,
  J.~Du~Croz, A.~Greenbaum, S.~Hammarling, A.~McKenney, and D.~Sorensen, {\em
  {LAPACK} Users' Guide}.
\newblock Society for Industrial and Applied Mathematics, Philadelphia, PA,
  third~ed., 1999.

\bibitem{Blitz}
T.~L. Veldhuizen and J.~Cummings, ``{\blitzpp},'' a {\CC{}} class library for
  scientific computing., Open Systems Laboratory at Indiana University
  Bloomington, 2007.
\newblock \texttt{http://www.oonumerics.org/blitz/}.

\bibitem{gabriel04:_open_mpi}
E.~Gabriel, G.~E. Fagg, G.~Bosilca, T.~Angskun, J.~J. Dongarra, J.~M. Squyres,
  V.~Sahay, P.~Kambadur, B.~Barrett, A.~Lumsdaine, R.~H. Castain, D.~J. Daniel,
  R.~L. Graham, and T.~S. Woodall, ``Open {MPI}: Goals, Concept, and Design of
  a Next Generation {MPI} Implementation,'' in {\em Proceedings, 11th European
  PVM/MPI Users' Group Meeting}, pp.~97--104.
\newblock Budapest, Hungary, September, 2004.

\end{thebibliography}\endgroup

\end{document}